\documentclass[a4paper,11pt]{article}
\pdfoutput=1 

\usepackage{jheppub} 
\usepackage{empheq}
\usepackage{bm}
\usepackage{slashed}
\usepackage[T1]{fontenc} 
\usepackage[all]{xy}
\usepackage{rotating}
\usepackage{framed}
\usepackage{mdframed}
\usepackage{longtable}

\usepackage{slashed}

\def\a{\alpha}

\def\CA{{\cal A}}
\def\CB{{\cal B}}

\def\CF{{\cal F}}

\def\CH{{\cal H}}
\def\CI{{\cal I}}

\def\CK{{\cal K}}
\def\CL{{\cal L}}
\def\CM{{\cal M}}
\def\CN{{\cal N}}
\def\CO{{\cal O}}

\def\CS{{\cal S}}
\def\CT{{\cal T}}

\def\CW{{\cal W}}

\def\CZ{{\cal Z}}

\def\U{\mathrm{U}}

\def\beq#1\eeq{\begin{align}#1\end{align}}

\preprint{KIAS-P22046}
\title{\boldmath Infrared Phases of 3D Class R Theories}

\author[a]{Sunjin Choi,} 
\author[b,c]{Dongmin Gang,}
\author[c,d]{Hee-Cheol Kim}

\affiliation[a]{School of Physics, Korea Institute for Advanced Study,\\
85 Hoegiro, Dongdaemun-gu, Seoul 02455, Republic of Korea.}

\affiliation[b]{
	Department of Physics and Astronomy $\&$ Center for Theoretical Physics,
	\\
	Seoul National University, 1 Gwanak-ro, Seoul 08826, Republic of Korea.}
\affiliation[c]{Asia Pacific Center for Theoretical Physics (APCTP),
	Pohang 37673, Republic of Korea.}
\affiliation[d]{Department of Physics, Pohang University of Science and Technology (POSTECH), \\Pohang 37673, Republic of Korea.}

\emailAdd{sunjinchoi@kias.re.kr}
\emailAdd{arima275@snu.ac.kr}
\emailAdd{heecheol1@gmail.com}

\abstract{We study the IR phases of 3D class R theories associated with closed non-hyperbolic 3-manifolds. Non-hyperbolic 3-manifolds can be obtained by performing Dehn fillings on 1-cusped  hyperbolic 3-manifolds along exceptional slopes. In 3D-3D correspondence, the `exceptional' Dehn filling corresponds to the gauging of an $SU(2)$ flavor symmetry in a superconformal field theory associated with a 1-cusped 3-manifold with `small' Chern-Simons levels. With several explicit examples, we analyze  various interesting non-perturbative IR phenomena (such as  spontaneous SUSY breaking, generation of mass gap and  supersymmetry enhancement) from the `exceptional' gaugings.   Interestingly,  distinguished features of the IR phases can be captured by simple topological properties of non-hyperbolic 3-manifolds. We also find that 3D class R theories associated with certain classes of atoroidal non-hyperbolic 3-manifolds always exhibit supersymmetry enhancement at low energy  and actually flow to 3D rank-0 $\mathcal{N}=4$ SCFTs with trivial vacuum moduli space.}

\begin{document} 
\maketitle
\flushbottom

\section{Introduction and Summary}

Recently, considerable progress has been made in understanding lower dimensional physics from compactifications of higher dimensional superconformal field theories (SCFTs). In particular, plenty of 4D $\mathcal{N}=2$ SCFTs, called class S theories, are engineered by compactifications of 6D $\mathcal{N}=(2,0)$ SCFTs on punctured Riemann surfaces, and non-perturbative physics of the 4D theories such as dualities and enhanced symmetries at infrared have been studied in great detail through the compactifications \cite{Gaiotto:2009we,Gaiotto:2009hg}. Similarly, the compactification of the 6D SCFTs living on M5-branes wrapped around hyperbolic three-manifolds $M$ leads to a family of 3D $\mathcal{N}=2$ theories called 3D class R theories \cite{Dimofte:2011ju,Dimofte:2011py}. This construction allows us to understand intricate networks of 3D dualities and also to predict new dualities based on geometric operations on the internal 3-manifolds.

More recently, we have revisited the class R construction of 3D theories and extended it to incorporate non-hyperbolic 3-manifolds into the construction \cite{Cho:2020ljj}. Unlike the compactification on hyperbolic 3-manifolds, which engineers gapless phases at low energy, the 3D field theories constructed from non-hyperbolic manifolds can give rise to a wide variety of strongly coupled IR phases including gapped phases, SUSY-broken phases as well as conformal phases. For example, a certain family of non-hyperbolic 3-manifolds, which we can use in the class R construction to engineer 3D topological field theories (TFTs), was proposed in \cite{Cho:2020ljj}. The proposal was verified by explicitly constructing modular S- and T-matrices for the TFTs in terms of topological invariants of the non-hyperbolic 3-manifolds, which can be computed using various partition functions of complex $SL(2,\mathbb{C})$ Chern-Simons theories on the 3-manifolds. This establishes a new correspondence between 3D TFTs and non-hyperbolic 3-manifolds. 

Especially, when the modular data from the internal 3-manifold is non-unitary, the infrared phase of the corresponding 3D theory is conjectured to become an SCFT embedding a non-unitary subsector described by that modular data  \cite{Cho:2020ljj}. Additionally, this class of IR SCFTs are believed to enjoy SUSY enhancement from  $\mathcal{N}=2$ to $\mathcal{N}=4$, and they have neither Coulomb nor Higgs branch of vacua, i.e. they are of rank-0 $\mathcal{N}=4$ SCFTs. The subsector equipped with the non-unitary modular structure can be isolated in a degenerate limit where supersymmetric partition functions compute contributions only from the Coulomb or Higgs branch chiral ring operators and their descendants, which are empty in this case since the SCFT is of rank-0. Therefore, in the degenerate limit, the relevant physical observables are loop operators and their correlation functions, which together define the non-unitary modular tensor category (MTC) associated with the non-unitary sub-sector.  This novel relation between 3D $\mathcal{N}=4$ rank-0 SCFTs and non-unitary TFTs  was extensively studied in \cite{Gang:2021hrd}.

In this paper, we study the relation between topological invariants of closed non-hyperbolic 3-manifolds and infrared phases of 3D class R theories. We focus our attention on the non-hyperbolic 3-manifolds constructed by so-called exceptional Dehn fillings from 1-cusped hyperbolic 3-manifolds. The 3-manifolds of this class have been  classified in mathematical literature (see for instance \cite{2018arXiv181211940D}) based on the topological properties of 2-cycles embedded in the 3-manifolds, which is summarized in Fig. \ref{fig: census of non-hyperbolics}. Surprisingly, we find that the mathematical classification can be repackaged in physics as a systematic characterization of low-energy phases of the class R theories.

Our main aim in this paper is to develop a geometric scheme to identify IR phases of 3D class R theories in terms of topological properties of the associated non-hyperbolic 3-manifolds. As we will see, this task will also formulate a concrete dictionary between topological invariants of the internal non-hyperbolic 3-manifolds and physical observables in the 3D class R theories. The dictionary provides a number of criteria to characterize IR phases of the 3D theories without detailed analysis of strongly coupled low-energy dynamics which in general is notoriously difficult. 

The main tools we will use are the 3D index and the irreducible $SL(2,\mathbb{C})$ flat connections of the non-hyperbolic 3-manifolds. These are defined as the partition function of a complex $SL(2,\mathbb{C})$ Chern-Simons (CS) theory on the non-hyperbolic 3-manifold at Chern-Simons level $k=0$ and irreducible flat connections in the complex CS theory respectively. The convergence of the 3D index and the reality of the irreducible flat connections turn out to capture the existence of certain topological 2-cycles (more precisely, separating essential 2-spheres or tori), which plays a central role in the classification of 3-manifolds in mathematics.
Field theoretically, the 3D index corresponds to a supersymmetric index counting the number of local BPS operators, and the irreducible flat connections are related to discrete vacua (or Bethe vacua) of the 3D field theory with supersymmetric massive deformations. These are protected under renormalization group (RG) flows by supersymmetry, so we can use them for a systematic exploration of rich structure of low-energy phases. 
We first discuss what are the possible IR phases from non-hyperbolic closed 3-manifolds via the class R construction, and then propose geometric criteria on how to catalog the IR phases in terms of the properties of the 3D index and irreducible flat connections of the 3-manifolds. This will give a novel field theoretic interpretation of the mathematical classification of closed 3-manifolds. Utilizing the geometric criteria, we will identify all possible IR phases of 3D class R theories constructed from non-hyperbolic 3-manifolds obtained by exceptional Dehn fillings on $S^3\backslash \mathbf{4}_1, S^3\backslash \mathbf{5}_2, m003,m006,m007$ and  $m009$. This provides a concrete confirmation for the geometric criteria we propose.

The organization for the rest of this paper is as follows: In Section \ref{sec:2}, we review how to obtain closed non-hyperbolic 3-manifolds by exceptional Dehn fillings. We also introduce a mathematical classification of closed non-hyperbolic 3-manifolds, and explain how to understand it in terms of properties of 3D indices and irreducible flat connections of the 3-manifolds. In Section \ref{sec : IR phases}, we discuss IR phases of 3D class R theories constructed from closed non-hyperbolic 3-manifolds. We explain how to identify precise IR phases using supersymmetric partition functions, which can be computed from topological data of the associated 3-manifolds. Section \ref{sec:examples} presents various examples of 3D class R theories and explicit analysis of their infrared phases.

\section{Space of non-hyperbolic 3-manifolds}\label{sec:2}
This section provides a brief introduction to closed non-hyperbolic 3-manifolds with a focus on the construction using exceptional Dehn fillings from  1-cusped hyperbolic 3-manifolds, and their topological features in terms of basic properties of the 3D index and $SL(2,\mathbb{C})$ flat connections.  We also give some background mathematical materials relevant to the discussions in this paper.  

\subsection{Exceptional Dehn fillings on 1-cusped hyperbolic 3-manifolds}
Let us first explain the notations and a method for constructing closed non-hyperbolic 3-manifolds. We start with the hyperbolic 3-manifolds with a single torus boundary, which are often called  1-cusped hyperbolic 3-manifolds. We shall use the following notation:
\begin{align}
\begin{split}
&N \;: \; \textrm{1-cusped hyperbolic 3-manifold}\;,
\\
& pA+qB \in H_1(\partial N, \mathbb{Z}) = \mathbb{Z} \oplus \mathbb{Z} \;:\; \textrm{primitive boundary 1-cycle}\;. 
\end{split}
\end{align}
In this paper, the 1-cusped hyperbolic 3-manifold $N$'s are labeled by the nomenclature used in SnapPy  \cite{SnapPy} as
\begin{align}
N = m003,\, m004,\, m006,\ldots,
\end{align}
There is no canonical basis choice for boundary 1-cycles $A$ and $B$ in $H_1(\partial N,\mathbb{Z})$ of a generic 3-manifold $N$. We find it convenient to choose the particular basis used in SnapPy which we denote by
\begin{align}
A_{\rm SP} , B_{\rm SP} \;:\; \textrm{a basis of $H_1 (\partial N, \mathbb{Z})$ chosen by SnapPy}\;.
\end{align}
There is however a choice of canonical basis for $H_1(\partial N,\mathbb{Z})$, which are called meridian $\mu$ and longitude $\lambda$, for a special type of 3-manifold $N=S^3/\mathcal{K}$ engineered by a complement of a knot $\mathcal{K}$ in $S^3$.
\begin{align}
\mu , \;\lambda \;:\; \textrm{a canonical basis of $H_1 (\partial N, \mathbb{Z})$ when $N= S^3\backslash \CK$}\;.
\end{align}

The closed 3-manifolds we are interested in can be constructed from the 1-cusped 3-manifolds using so-called Dehn fillings. Dehn filling is a natural topological operation on a 1-cusped 3-manifold that produces a new 3-manifold without boundary. It is a simple operation that glues a solid torus to the boundary torus in a cusped 3-manifold along a primitive cycle $pA+qB$. Here, the gluing can be performed by identifying the primitive cycle $pA+qB$ with the shrinkable boundary $S^1$ of the solid torus. So the boundary of the 1-cusped 3-manifold is closed after the operation. We denote the closed 3-manifold resulting from the Dehn filling as
\begin{align}
\begin{split}
N_{pA+qB} &:= (\textrm{3-manifold obtained by Dehn filling on $N$ along $pA+qB$})\;,
\\
&:= (N\cup (\textrm{solid-torus}))/\sim \;, 
\\
& \quad \textrm{where }  (pA+qB) \sim (\textrm{shrinkable boundary $S^1$ of solid-torus})\;.
\end{split}
\end{align}
The primitive cycle $pA+qB$ (up to an overall sign) is called a slope. 
We remark that according to the well-known Thurston's hyperbolic Dehn surgery theorem, $N_{pA+qB}$ is always hyperbolic except only finitely many slopes called {\it exceptional slopes}.
\begin{align}
\textrm{A slope $pA+qB \in H_1 (\partial N, \mathbb{Z})$ is called exceptional if $N_{pA+qB}$ is non-hyperbolic}.
\end{align}
The exceptional Dehn fillings drastically alter geometrical structure of the 3-manifolds and can produce a large class of closed 3-manifolds with rich geometrical/topological structures. Actually 7 geometries among Thurston's 8 geometries can be realized on non-hyperbolic 3-manifolds \cite{thurston2014three}. 

\subsection{Topological types of non-hyperbolic manifolds}
\begin{figure}[htbp]
	\begin{center}
		\includegraphics[width=.49\textwidth]{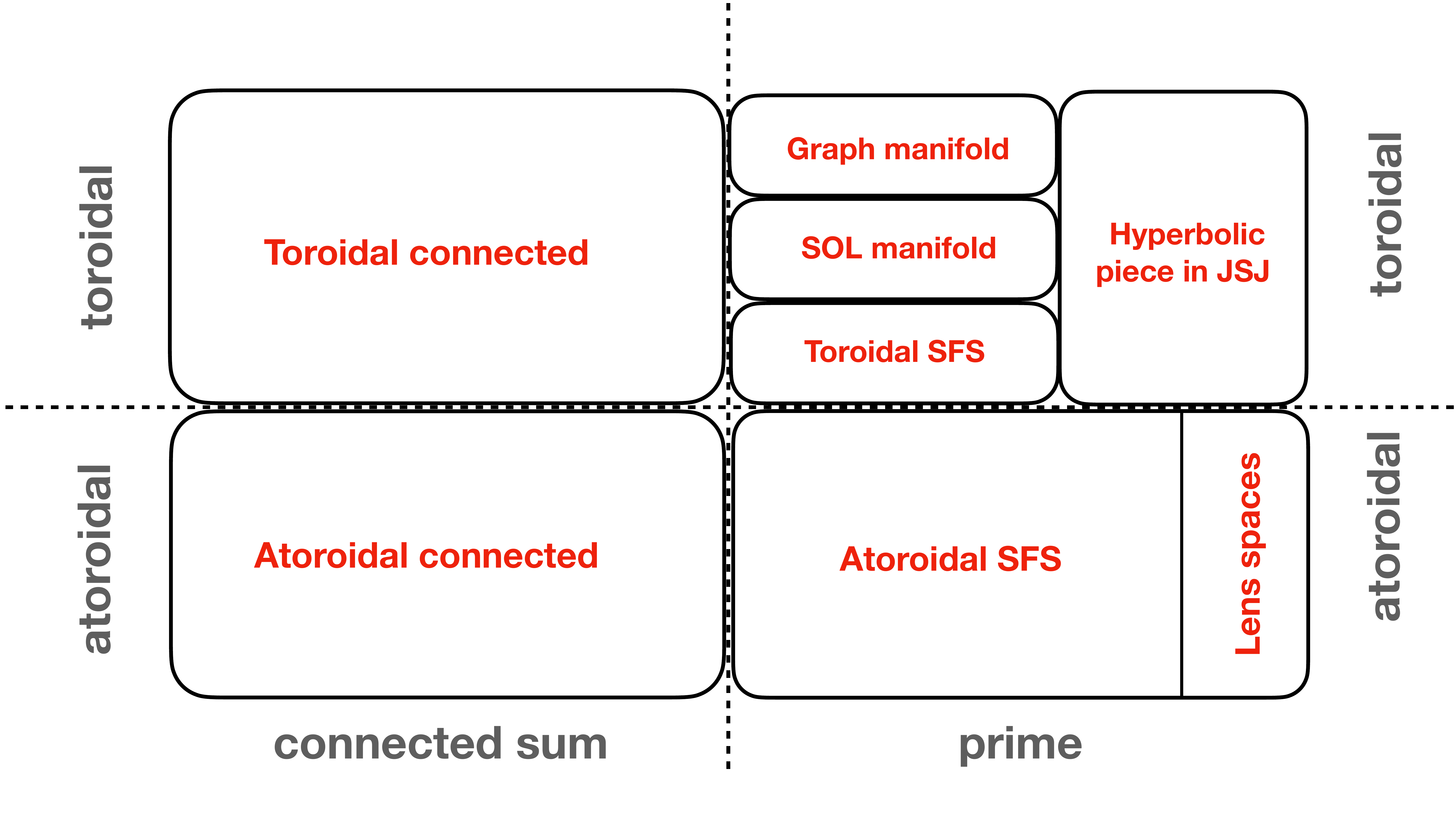}
	    \includegraphics[width=.49\textwidth]{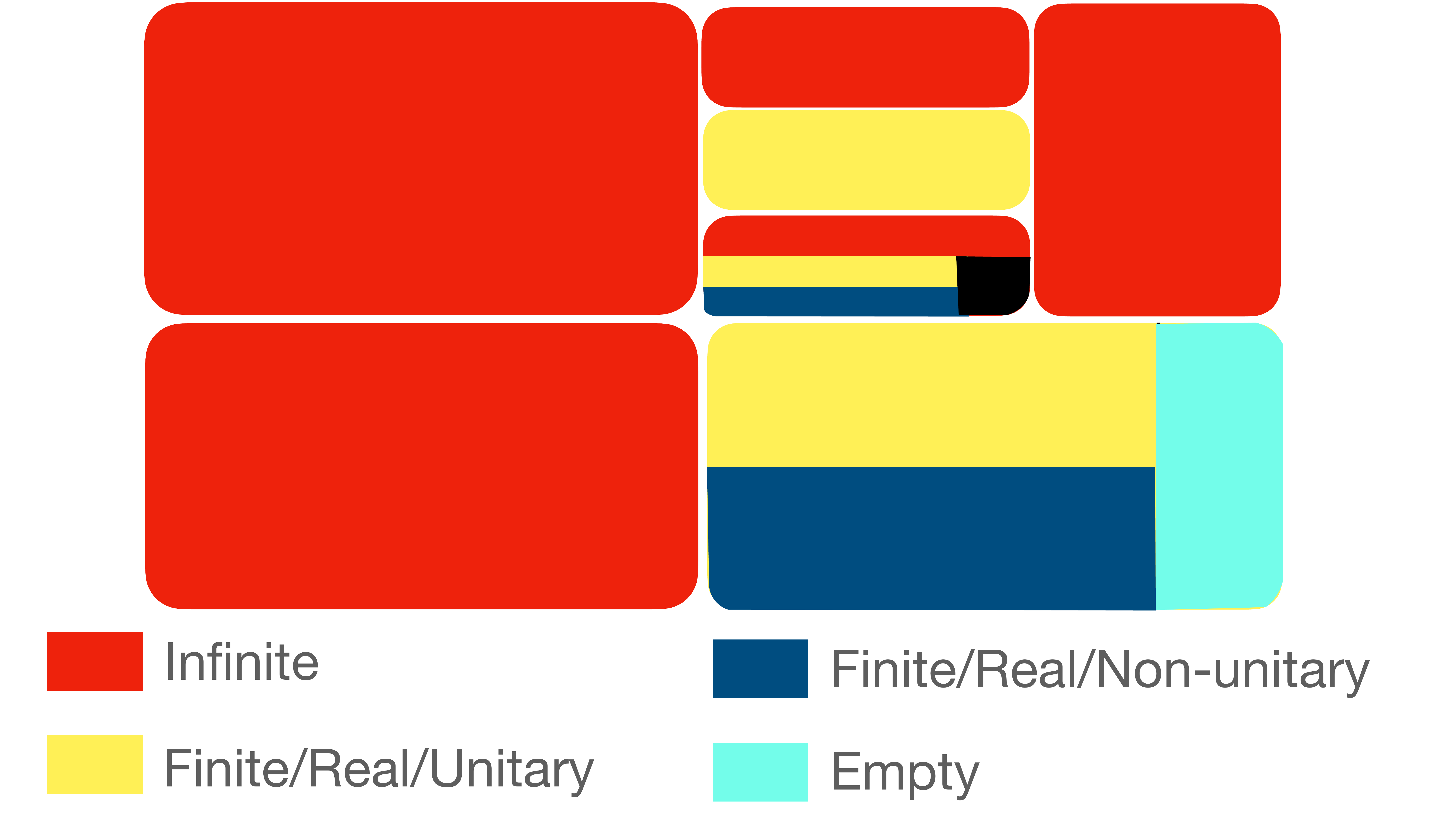}
	\end{center}
	\caption{Two classifications of closed non-hyperbolic 3-manifolds: (Left) according to the existence of essential 2-sphere or 2-torus and (Right) based on the properties of 3D index and  $SL(2,\mathbb{C})$ irreducible flat connections on the 3-manifolds. The relation between two classifications are illustrated in the two diagrams. The 3-manifolds of finite/complex types are always hyperbolic and thus do not appear here. 
	Among prime/torodial 3-manifolds, it is of finite type if all  essential tori are non-separating and of infinite type otherwise.
	The 3-manifolds of infinite type can be further divided into two categories depending on whether it is  real or complex but the refined division  is not drawn in the diagram. Among toroidal SFS, toroidal 3-fibered Seifert-spaces, i.e. $S^2 ((p_1,q_1)(p_2,q_2)(p_3,q_3))$ with $\sum_{i}\frac{q_i}{p_i}=0$, are of finite and real while  $\Sigma_{g>1}\times S^1$  are of infinite type.   We could not fully identify the type of other  toroidal SFS and  leave them in a black color.  SOL manifolds are always finite/real/unitary while $S^2 ((p_1,q_1)(p_2,q_2)(p_3,q_3))$ could be either finite/real/unitary or finite/real/non-unitary as in \eqref{unitary/non-unitary 3-manifolds}. }
	\label{fig: census of non-hyperbolics}
\end{figure}

\subsubsection{Conventional mathematical classification}
The left diagram of Fig.~\ref{fig: census of non-hyperbolics}  summarizes a conventional  classification  of closed non-hyperbolic 3-manifolds according to the existence of embedded essential 2-sphere or embedded essential torus. An embedded 2-sphere  is essential if it does not bound a 3-ball. See  \cite{2018arXiv181211940D} for details on the classification and related mathematical materials. 

Depending on the existence of embedded essential 2-spheres, closed 3-manifolds can be divided into two classes:  reducible (if exists) or irreducible.  When the essential sphere is separating, the 3-manifold can be given as a connected sum of the two separated manifolds. So the ``connected sum''  implies ``reducible''. The only exceptional 3-manifold, which is reducible but not connected sum, is $S^2\times S^1$, which is regarded as a lens space $L(0,1)$ in this paper.  The prime 3-manifold means the 3-manifold which can not be given by a connected sum of two manifolds neither of which is a 3-sphere $S^3$.\footnote{3-sphere can be regarded as an identity under the connect sum operation $\sharp$, i.e. $M \sharp S^3 =M$ for all $M$. } Depending on the existence of the essential torus, 3-manifolds can further be divided into two classes: toroidal (if exists) or atoroidal.

\begin{figure}[htbp]
	\begin{center}
		\includegraphics[width=.3\textwidth]{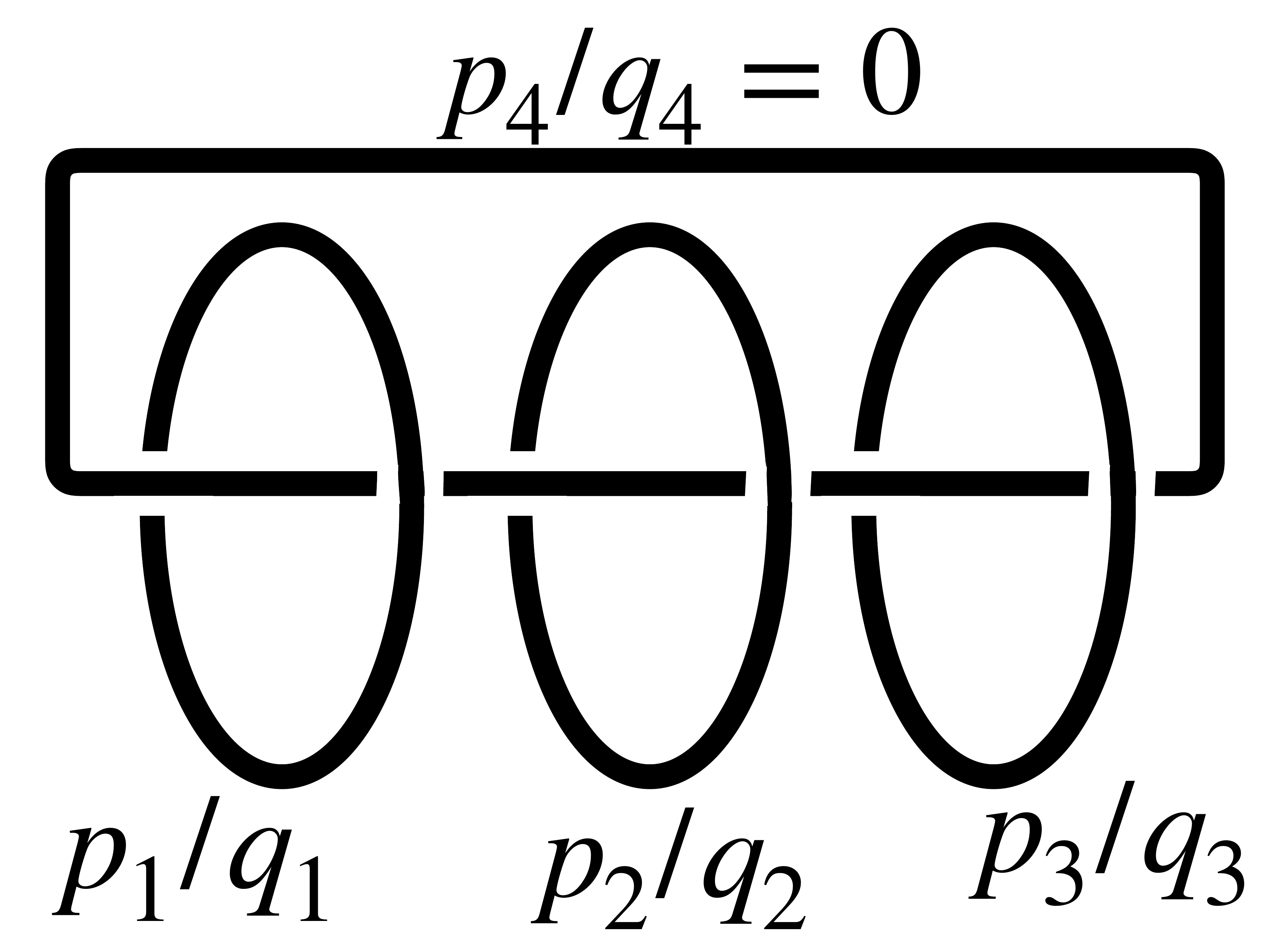}
	\end{center}
	\caption{Dehn surgery description of 3-fibered  Seifert  manifold $S^2\left((p_1, q_1),(p_2,q_2),(p_3,q_3)\right)$. }
	\label{fig: Seifert fibered manifolds}
\end{figure}

All hyperbolic 3-manifolds are prime and atoroidal. Among non-hyperbolic 3-manifolds, the only prime/atoroidal manifolds are  small Seifert fibered spaces (small SFS). There are two kinds of small Seifert fibered manifolds, one is lens space and the other is 3-fibered Seifert spaces $S^2 \left((p_1,q_1)(p_2, q_2),(p_3, q_3)\right)$ over $S^2$ which is not toroidal, i.e. $\sum_{i} \frac{q_i}{p_i} \neq 0$. 
\begin{align}
	\textrm{Prime/atoroidal } \begin{cases}
		\textrm{Hyperbolic }
		\\
		\textrm{Small Seifert fibered } \begin{cases}
			\textrm{Lens space $L(p,q)$  }
			\\
			S^2 \left((p_1,q_1)(p_2, q_2),(p_3, q_3)\right)\big{|}_{\sum_i \frac{q_i}{p_i} \neq 0 }
		\end{cases}
	\end{cases}
	\label{Prime/ator}
\end{align}
When $\sum_{i}\frac{q_i}{p_i} = 0$, the manifold $S^2 \left((p_1,q_1)(p_2, q_2),(p_3, q_3)\right)$  is toroidal and corresponds to a Seifert-toroidal manifold. The manifold $S^2\left((p_1, q_1),(p_2,q_2),(p_3,q_3)\right)$ can be described by a Dehn surgery along a link with 4 components depicted as in Fig. \ref{fig: Seifert fibered manifolds}. Here, $(p_i,q_i)$ are coprime with $p_i>0$.

SOL manifolds refer to  torus bundles $\mathbb{T}^2 \times_\varphi S^1$ with Anosov monodromy $\varphi$. The torus bundles are defined with a  monodromy matrix $\varphi  \in SL(2,\mathbb{Z})$ and an equivalence relation given by
\begin{align}
	\mathbb{T}^2\times_\varphi S^1 = \big{(}\mathbb{T}^2\times [0,1]\big{)}/\big{(}(x,0)\sim (\varphi(x),1)\big{)}\;, 
\end{align}
where $x $ is a point in the two-torus $\mathbb{T}^2$ and $\varphi(x)$ is the image of $x$ under the mapping class group element $\varphi$. $\varphi$ is called a Anosov monodromy if $|\textrm{Tr}(\varphi)|>2$.

Graph manifolds generally refer to prime non-hyperbolic 3-manifolds that do not include hyperbolic pieces in JSJ decomposition \cite{jaco1979seifert,johannson2006homotopy}. In the usual definition, the space of graph manifolds includes all prime Seifert-fibered spaces and SOL manifolds. However, graph manifolds in this paper refer only to graph manifolds which are neither Seifert-fibered spaces nor SOL manifolds. Similarly, the space of SFS's generally includes lens spaces, but our terminology for Seifert-fibered spaces in this paper refers only to Seifert-fibered spaces which are not lens spaces. 

\subsubsection{Classification using 3D index and $SL(2,\mathbb{C})$ flat connections}
We shall propose a new classification scheme for closed 3-manifolds based on the basic properties of  {\it 3D index} \cite{Dimofte:2011py,Gang:2018wek} and irreducible $SL(2,\mathbb{C})$ flat connections as follows
\begin{itemize}
	\item \textbf{Def : Infinite/Finite/Empty} \\
	i. A closed 3-manifold $M$ is of {\it empty type} if the 3D index  vanishes, i.e. $\CI_M(q)=0$.\\
	ii. A non-empty closed 3-manifold $M$  is  of {\it finite (resp. infinite) type} if $\CI_M(q)$ converges (diverges) as a  formal Laurent series. 
\end{itemize}
\begin{itemize}
	\item \textbf{Def : Complex/Real} \\
	i. A $SL(2,\mathbb{C})$ flat connection $\rho$ is 
	$\begin{cases}\textrm{real\,, \ \ if \ Tr}(\rho(a)) \in \mathbb{R} \ \textrm{for all} \ a \in \pi_1 M. \\\textrm{complex\,, \ \ otherwise.}\end{cases}$\\
	ii. A closed 3-manifold $M$ is
	$\begin{cases}\textrm{real\,, \ \ if all irreducible flat}\ SL(2,\mathbb{C})\ \textrm{connections are real.}\\
		\textrm{complex\,, \ \ otherwise.}
	\end{cases}$ \label{def : complex/real}
\end{itemize}
The 3D index is a 3-manifold invariant introduced as an $SL(2,\mathbb{C})$ Chern-Simons partition function with a quantized level $k=0$. The invariant for 3-manifolds $N$ with torus boundaries was firstly studied in \cite{Dimofte:2011py} and then generalized to cover closed 3-manifold $M$ in \cite{Gang:2018wek}. The index, $\CI_{N}(\vec{m},\vec{e};q)$ or $\CI_{M}(q)$, gives a formal Laurent polynomial in $q^{1/2}$ for each $(\vec{m},\vec{e}) \in \mathbb{Z}^{2n}$ where $n$ is the number of boundary tori. 
Using a state-integral model developed in \cite{Dimofte:2011py,Gang:2018wek,Gang:2018gyt},  the index is  schematically given in the following form
\begin{align}
	\mathcal{I}(q) = \sum_{\vec{s} \in \mathbb{Z}^T} \CI (\vec{s};q)\;. \label{3D index : summation}
\end{align}
Here $T$  is a positive integer and we suppress the dependence on $(\vec{m},\vec{e})$ for simplicity. The $\CI(\vec{s};q)$ for each $\vec{s} \in \mathbb{Z}^T$ is $q^{\epsilon(\vec{s})} \times (\textrm{power series in $q^{1/2}$ starting with $I_0 (\vec{s})+O(q^{1/2})$})$ with $
\epsilon(\vec{s}) , I_0 (\vec{s}) \in \mathbb{Z}/2$. The precise definition of the 3D index and some examples will be presented in Section \ref{subsec : 3D-index}. We say the 3D index converges (as a formal series in $q^{1/2}$) when one  needs to sum over only finitely many $\vec{s} \in \mathbb{Z}^T$  to obtain the invariant up to $\CO(q^{\frac{\rm max} 2})$ for arbitrary $\textrm{max} \in \mathbb{Z}_+$.
\\
\\
 We define the (adjoint) irreducible flat connections as follows:\footnote{We remark that our definition of the irreducibility of the $SL(2,\mathbb{C})$ flat connection is  somewhat different from the definition used in many other math literature. In the math literature, irreducible means (fundamental) irreducible and our irreducible flat connections  are called non-Abelian flat connection. Our notion of  irreducibility is relevant in the study of $SL(2,\mathbb{C})$ Chern-Simons theory and class R theories.}
 \begin{itemize}
 	\item \textbf{Def : Irreducibility of  flat connection} \\
 	A $SL(2,\mathbb{C})$ flat connection $\rho \in \textrm{Hom}[\pi_1 M \rightarrow SL(2,\mathbb{C})]/\sim $ is called {\it irreducible} if\\
 	$\textrm{Stab}(\rho) := \{g \in SL(2,\mathbb{C}) \;:\; [g, \rho(a)]=0 \quad \forall a \in\pi_1 M \}$ is a finite subgroup of $SL(2,\mathbb{C})$.
 \end{itemize}
 It turns out that the only possible finite subgroup here is the $\mathbb{Z}_2$ center symmetry. 
 We also regard two homomorphisms $\rho_1, \rho_2 \in \textrm{Hom}[\pi_1 M \rightarrow SL(2,\mathbb{C})]$ as the same $SL(2,\mathbb{C})$ flat connection if their characters are all identical. Namely, we have the following equivalence relation:
 \begin{align}
 	\rho_1 \sim \rho_2 \quad \textrm{if $\textrm{Tr}(\rho_1 (a)) = \textrm{Tr}(\rho_2 (a))$ for all $a \in \pi_1  M$}\;. \label{character equivalence}
 \end{align}
 We define a set of irreducible flat connections as
 \begin{align}
 	\chi^{SL(2,\mathbb{C})}_{\rm irred}(M) := \{ \textrm{a set of inequivalent irreducible $SL(2,\mathbb{C})$ flat connections on $M$}\}\;. \label{chi-irred}
 \end{align}
As we will see in the next section, this set of irreducible flat connections will play a crucial role in the study of 3D class R theories.

We are particularly interested in the closed 3-manifolds of finite and real type since they enjoy the correspondence with topological field theories $\textrm{TFT}[M]$ established in \cite{Cho:2020ljj} (see also \cite{Cui:2021lyi} and \cite{Cui:2021yes}). A closed 3-manifold $M$ of finite and real type is non-hyperbolic, and the class R construction for two M5-branes wrapped around such 3-manifold $M$ provides a 3D field theory, which flows at low energy to either a unitary topological field theory (TQFT) or a (unitary) superconformal field theory (SCFT) containing a sub-sector realizing a non-unitary TQFT structure. The details will be discussed in the next section. We can distinguish the non-hyperbolic 3-manifolds in terms of the TQFT structures of the associated 3D class R theories. We define
\begin{itemize}
	\item \textbf{Def : Unitary/Non-unitary } \\
	$M$ of finite/real type is of 
	$\begin{cases} \text{unitary type, \ if TFT$[M]$ is unitary.} \\
		\text{non-unitary type, \ if TFT$[M]$ is non-unitary.} \end{cases}$
\end{itemize}
In this definition, the TFT$[M]$ denotes the IR topological theory arising from the class R theory when the 3-manifold $M$ is of unitary type, while it means the sub-sector of the class R theory admitting a non-unitary TQFT structure at low energy when $M$ is of non-unitary type. 
We note that unlike other properties above, the unitarity/non-unitarity is  defined only for the 3-manifolds of finite  and real type.

\subsubsection{Relation between two classifications}
Now let us  relate our new classification scheme  to the conventional classification of closed 3-manifolds.  The relation is summarized in Fig.~\ref{fig: census of non-hyperbolics}.

\paragraph{Lens space $\Rightarrow$ Empty type}As we will see in Section \ref{subsec : 3D-index}, the empty type 3-manifolds can alternatively be characterized by the emptiness of  irreducible $SL(2,\mathbb{C})$ flat connections:
\begin{align}
\begin{split}
&\textrm{$M$ is of empty type if there are no irreducible $SL(2,\mathbb{C})$ flat connections on $M$} 
\\
&\qquad \qquad \qquad \qquad \qquad  i.e.\;\; \chi_{\rm irred}^{SL(2,\mathbb{C})} (M)= \emptyset\;. \label{2nd def of emptiness}
\end{split}
\end{align}
Lens spaces $L(p,q)$ are obviously of empty type since its fundamental group is Abelian, $\pi_1 (L(p\neq 0,q)) = \mathbb{Z}_p$   and $\pi_1(L(0,1)) = \mathbb{Z}$, and thus all $SL(2,\mathbb{C})$ flat connections are reducible. 
 \paragraph{No separating essential sphere and torus  $\Rightarrow $ 3D index converges (i.e. empty or finite)} 
 The convergence of the 3D index has been extensively studied in \cite{Garoufalidis:2013axa, Garoufalidis:2016ckn} using an interesting relationship between the 3D index and the normal surface counting.  The main result of these studies is that the 3D indices for cusped 3-manifolds only converge when the 3-manifolds are irreducible and atoroidal. 
 These mathematical studies have focused on 3-manifolds with torus boundaries and hence the conclusion is not directly applicable to 3D indices for closed 3-manifolds. But as we will see in  Section \ref{subsec : 3D-index}, their basic logic can be applicable even to closed 3-manifolds via the 3D-3D correspondence. Motivated by these circumstantial evidences and  confirmation with  explicit examples in Appendix \ref{App : 3D index examples}, we  claim that 
 \begin{align}
 	\begin{split}
 	&\textrm{If a closed 3-manifold $M$ is irreducible/atoroidal} \; 
 	\\
 	&\Rightarrow \textrm{3D index $\CI_M(q)$ converges as a formal series in $q^{1/2}$ }\ .
 	\end{split}
 \end{align}
 In addition to these irreducible/atoroidal 3-manifolds, we found that following reducible or toroidal 3-manifolds have convergent 3D indices.
  \begin{align}
  	\begin{split}
  		&\textrm{$S^2\times S^1$, Torus bundle $\mathbb{T}^2\times_{\varphi}S^1$ and  $S^2((p_1, q_1),(p_2, q_2),(p_3, q_3))|_{\sum_{i=1}^3 q_i/p_i=0}$}
  		\\
  		&:\textrm{Reducible or toroidal closed 3-manifolds having  convergent 3D indices} \;.
  	\end{split}
  \end{align}  
  One characteristic property of these manifolds is that the essential spheres and tori in these manifolds are non-separating. So, we naturally speculate that closed 3-manifolds with no separating essential sphere and torus have convergent 3D indices. For $S^2 \times S^1$ (a lens space), as we saw above, the index vanishes, which is trivially compatible with the above statement. 
 
 \paragraph{$\exists$ Separating essential sphere or torus $\Rightarrow $ Infinity} In the case, we expect that the 3D index diverges. We will give a physical reasoning for this in Section \ref{subsec : 3D-index} and confirm it with many explicit examples in Appendix \ref{App : 3D index examples}. 
 
\paragraph{$\mathbb{T}^2\times_{\varphi} S^1$ or  $S^2 \left((p_1,q_1)(p_2, q_2),(p_3, q_3)\right)$ $\Rightarrow $ Finite and Real} Two basic examples of finite/real 3-manifolds  are  $S^2\left((p_1, q_1),(p_2,q_2),(p_3,q_3)\right)$ and $\mathbb{T}^2 \times_\varphi S^1$. 
The fundamental group of the 3-fibered Seifert 3-manifold is
\begin{align}
	\begin{split}
	&\pi_1 \left(S^2\left((p_1, q_1),(p_2,q_2),(p_3,q_3)\right)\right) 
	\\
	&= \langle x_1, x_2, x_3, h \;:\; x_i^{p_i}\cdot h^{q_i}=1 \; (i=1,2,3),\; x_1  x_2 x_3 =1,\; \textrm{$h$ is central}\rangle\;,
	\end{split}
\end{align}
and that of the torus bundle is
\begin{align}
	\pi_1 (\mathbb{T}^2 \times_\varphi S^1) =  \langle x, y, h\;:\; h^{-1}x h  = x^a y^c,\; h^{-1}y h = x^b y^d \rangle, \; \textrm{when } \varphi = \begin{pmatrix}
a & b \\
c & d 
\end{pmatrix}\;.
\end{align}
Here $(x,y)$ are the two generators of the fundamental group of the fiber $\mathbb{T}^2$ while $h$ represents the loop along the base $S^1$. The relations in the fundamental group represent how the two generators transform under the mapping class group element $\varphi$.
For these 3-manifolds of finite/real, it turns out  to be \cite{to-apper}
\begin{align}
\begin{split}
&\textrm{TFT}[S^2\left((p_1, q_1),(p_2,q_2),(p_3,q_3)\right)]  \textrm{ is } \begin{cases}
\textrm{unitary},\quad q_i \in \pm 1 (\textrm{mod\;} p_i)
\\
\textrm{non-unitary}, \quad \textrm{otherwise}
\end{cases}
\\
&\textrm{and}
\\
&\textrm{TFT}[\mathbb{T}^2 \times_{\varphi }S^1] \textrm{ is always unitary}\;. \label{unitary/non-unitary 3-manifolds}
\end{split}
\end{align}

\subsection{3D index}  \label{subsec : 3D-index}
We now review the 3D index $\CI_M(q)$ in details.  We introduce two  seemingly unrelated interpretations of the invariant: one is the partition function of the complex $SL(2,\mathbb{C})$ Chern-Simons theory  \cite{Dimofte:2012qj,Gang:2017cwq} and the other is weighted sum over the normal surfaces \cite{Garoufalidis:2013axa,Garoufalidis:2016ckn}. As we will see, the 3D-3D correspondence provides a unified picture for these two interpretations.

\paragraph{3D index as a weighted sum over surfaces} In a series of mathematical papers \cite{Garoufalidis:2013axa,Garoufalidis:2016ckn}, they found intricate relation between  the 3D index for cusped 3-manifolds and the normal surfaces. In the 3D-3D correspondence, the 3D index counts BPS  local operators  of the 3D class R theories.  In the M-theoretic set-up of the 3D-3D correspondence, the BPS local operators originate from BPS M2-branes wrapped on surfaces of the internal 3-manifold \cite{Gadde:2013wq}, and it explains the mathematician's observations. In the M-theoretical picture, essential surfaces  play special roles. M2-branes can wrap an essential surface $\Sigma$ arbitrarily many times.   They correspond to the local operators $\CO_\Sigma$, $\CO_\Sigma^2, \CO_\Sigma^3, \cdots$ and their contribution to the index is of the form $\sum_n (\pm q^{k_\Sigma} )^n$ where $k_{\Sigma}$ is a half-integer determined by the topology of $\Sigma$. It means that $\CO_\Sigma$ corresponds to $1/2$ BPS chiral primary operators. From the analysis in \cite{Garoufalidis:2013axa,Garoufalidis:2016ckn}, it turns out that essential surface of Euler characteristic $\chi$ contributes  following terms to the index 
\begin{align}
\sum_n (-q^{1/2})^{-\chi n}\left(1+ O(q^{1/2})\right) \;:\; \textrm{contributions from essential surface to 3D index} \ . \label{essential surface to 3D index}
\end{align}
These terms make the index diverges when the essential surface is 2-sphere ($\chi=2$) or 2-torus ($\chi=0$). Although there is no similar mathematical analysis for the 3D index of closed 3-manifolds, the M-theoretic picture is still valid and one would expect the essential sphere and torus in a closed 3-manifold make its index diverge.  From direct computations, we check that the index indeed diverges  for all reducible or toroidal closed 3-manifolds except for the cases when the essential spheres and tori are non-separating. 

Such exceptional cases include $S^2\times S^1$, $\mathbb{T}^2\times_\varphi S^1$ and toroidal $S^2 ((p_1, q_1),(p_2, q_2),(p_3, q_3))$ whose 3D indices are simply $0,1$  and $1\, (\textrm{or } 2)$ respectively. Note that the 3D indices for these exceptions are $q$-independent. We do not have a clear understanding,  but the observation suggests that the  {\it separating} essential  (instead of merely essential) spheres or tori make the  index divergent for closed 3-manifold cases  (unlike cupsed 3-manifold cases). We will come back to this issue below using another physical interpretation of the 3D index.

\paragraph{3D index as partition function of  $SL(2,\mathbb{C})$ CS theory}  The 3D index for a 3-manifold $M$ computes the partition function of an $SL(2,\mathbb{C})$ Chern-Simons theory on $M$ at Chern-Simons level $k=0$ which takes a path-integral expression
\begin{align}
\begin{split}
&\CI_M (q) = \int \frac{[D\mathcal{A}D \overline{\CA}]}{(\textrm{gauge})}  \exp \bigg{(} \pi i (k+i s) CS[\mathcal{A};M] + \pi i (k-i s) CS[\overline{\mathcal{A}};M]\bigg{)} \bigg{|}_{k=0, s= \frac{4\pi }{\log q }}\;,
\\
&\textrm{where  } CS[\mathcal{A};M] = \frac{1}{8\pi^2} \int_M \textrm{Tr}\left( \mathcal{A}d\mathcal{A}+\frac{2}3 \mathcal{A}^3 \right)\;, \label{3D index : path-integral}
\end{split}
\end{align}
where $\mathcal{A}$ is the $SL(2,\mathbb{C})$ gauge field and $s$ is the imaginary part of the Chern-Simons level, which exists for a complex gauge group.
The path-integral can be evaluated exactly using so-called state-integral model based on a Dehn surgery description $M= N_{pA+qB}$ and an ideal triangulation of $N$ \cite{Dimofte:2012qj,Gang:2017cwq}. 

The index can be factorized into holomorphic/anti-holomorphic blocks in the following way \cite{Beem:2012mb}:
\begin{align}
	\CI_M (q) =  \frac{1}2  \sum_{[\rho] \in \chi^{SL(2,\mathbb{C})}_{\rm irred} (M) }   B^\rho (q^{\frac{1}2}) B^{\bar{\rho}} (q^{-\frac{1}2})\;. \label{3D index from Bethe-sum}
\end{align}
Here $\chi^{SL(2,\mathbb{C})}_{\rm irred}(M)$ is the set of irreducible $SL(2,\mathbb{C})$ flat connections on $M$ as defined in \eqref{chi-irred}. 
 $B^\rho$ is the holomorphic block of the complex CS theory defined as
\begin{align}
B^{\rho} (q) := \int_{\Gamma^{\rho}}\frac{[D\CA]}{(\rm gauge)}  e^{- \frac{4\pi^2}\hbar CS[\CA,M]} \;\; \textrm{with $q=e^\hbar$}\;, \label{holomorphic block}
\end{align}
where $\Gamma^\rho$ is the Lefschetz thimble in the functional space of $\CA$ associated to the flat connection $\rho$. $B^{\bar{\rho}}$ is the holomorphic block for the complex conjugate flat connection $\bar{\rho}$.

The factorization formula makes sense only when the set $\chi^{SL(2,\mathbb{C})}_{\rm irred}$ is finite, i.e. when it contains a finite number of elements, and otherwise we expect the index to diverge. For example, a 3-manifold given by a connected sum, which means $M=M_1 \sharp M_2$ and none of the $M_i$'s is a 3-sphere, always has an infinite number of  inequivalent irreducible flat connections.\footnote{It simply follows from the fact that $\pi_1 (M_1 \sharp M_2)  = \pi_1 (M_1) * \pi_1(M_2)$ where $*$ is the free product of groups. Let $\rho_1$ and $\rho_2$ be (not necessarily irreducible) homomorphisms of $\pi_1 (M_i)\rightarrow SL(2,\mathbb{C})$ with $\textrm{Im}[\rho_i]\neq \{\pm 1\} = (\textrm{Center subgroup $\mathbb{Z}_2$})$. We define $\rho_g\in \textrm{Hom}(\pi_1 M \rightarrow SL(2,\mathbb{C}))$ by $\rho_g(a) =\rho_1 (a)$ for  $a\in \pi_1 (M_1) \subset \pi_1 M$ and $\rho_g(b) =g\rho_2 (b)g^{-1}$ for $b\in \pi_1 (M_2) \subset \pi_1 M$ with $g\in SL(2,\mathbb{C})$. Then the set $\{ \rho_g\}_{g\in SL(2,\mathbb{C})}$ contains  infinitely many inequivalent irreducible flat $SL(2,\mathbb{C})$ connections on $M$.} This is compatible with the fact that 3-manifolds given by a connected sum  are always of infinite type.

Now let us try to understand the behavior of the 3D index using the  partition function representation.
First, the factorization formula of the 3D index implies \eqref{2nd def of emptiness}. For example, the 3-manifold $S^2\times S^1$ does not admit any irreducible $SL(2,\mathbb{C})$ flat connections since its fundamental group is $\mathbb{Z}$ and is Abelian. This implies that $\chi_{\rm irred}^{SL(2,\mathbb{C})}=\emptyset$ and thus the 3D index vanishes.

In the Cardy limit $q\rightarrow 1$ (or $\hbar\rightarrow 0$), the holomorphic blocks are expanded as
\begin{align}
\begin{split}
&B^{\rho}(q^{\frac{1}2}) \xrightarrow{\quad \hbar = \log q \rightarrow 0  \quad } \exp \left(-\frac{4\pi^2}\hbar CS[\rho]+\ldots\right )\;,
\\
&B^{\bar{\rho}}(q^{-\frac{1}2})\xrightarrow{\quad \hbar = \log q \rightarrow 0  \quad } \exp \left(\frac{4\pi^2}\hbar \overline{CS[\rho]}+\ldots\right )\;,
\\
&\Rightarrow  B^{\rho}(q^{\frac{1}2}) B^{\bar{\rho}}(q^{-\frac{1}2}) \xrightarrow{ \quad \hbar =\log q \rightarrow 0 \quad } \exp \left( -\frac{8\pi^2 i }\hbar \textrm{Im}[CS[\rho]]+\ldots\right)\;. \label{Cardy limit}
\end{split}
\end{align}
This shows that the asymptotic behavior of the 3D index essentially depends on whether the 3-manifold is of complex or real type.
A 3-manifold of complex type has a complex flat connection $\rho$ with non-zero $\textrm{Im}[CS[\rho]]$ and this gives an exponentially growing contribution to the index in the Cardy limit.\footnote{In 3-manifolds, there is always a pair of complex flat connections, $\rho$ and $\bar{\rho}$. One of them gives exponentially growing contribution to the index in the $\hbar \in e^{i\delta} \mathbb{R} \rightarrow 0$ limit unless $\delta =0$ or $\pi$.} On the other hand, for 3-manifolds of finite/real type, the 3D index becomes finite and $\hbar$-independent in  the asymptotic limit $\hbar\rightarrow 0$ (or equivalently $q\rightarrow 1$) since $\textrm{Im}[CS[\rho]]=0$. This partly explains why the 3D indices for $\mathbb{T}^2\times_{\varphi} S^1$ and $S^2 ((p_i, q_i))|_{i=1}^3$  are finite and $q$-independent. Indeed these 3-manifolds have only a finite number of irreducible $SL(2,\mathbb{C})$ flat connections which are all real. 

Next, the partition function for a 3-manifold $\Sigma_g\times S^1$ is expected to compute  the ($q$-independent) dimension of Hilbert space of the $SL(2,\mathbb{C})$ Chern-Simons theory on the genus $g$ Riemann surface $\Sigma_g$. The Hilbert space can be obtained by quantizing the phase space $P(\Sigma_g)$ defined as
\begin{align}
P(\Sigma_g) = \chi^{SL(2,\mathbb{C})}_{\rm irred}(\Sigma_g) :=(\textrm{set of inequivalent irreducible $SL(2,\mathbb{C})$ flat connections on $\Sigma_g$})\;.
\end{align}
The phase space is empty for $g=0$ since $\pi_1 (\Sigma_{g=0,1})$ is Abelian, whereas for $g>1$, it is a non-compact K\"ahler manifold of complex dimension $(g-1)$. 
We thus assert that
\begin{align}
\CI_{\Sigma_g\times S^1}(q) = \begin{cases} 0\;, \quad \text{for } g=0\;,
	\\
	\textrm{$q$-independent infinity}\;, \quad  \text{for } g>1\;.
\end{cases}
\end{align}
This behavior is expected since the 3-manifolds $\Sigma_{g=0} \times S^1$ do not contain separating essential 2-sphere and torus, which we will call ${\rm SES}_{g=0}$ and  ${\rm SES}_{g=1}$ respectively, and they have no irreducible $SL(2,\mathbb{C})$ flat connections, while there are separating essential 2-tori and infinitely many irreducible flat connections in $\Sigma_{g>1}\times S^1$.\footnote{The separating essential 2-torus is given by a direct product of the $S^1$ in $S^1\times \Sigma_g$ and a separating non-contractible $S^1$ in $\Sigma_g$. The separating non-contractible 1-cycle exists only when $g>1$.   }
\\

Combining the above discussions with  experimental evidences,  we propose that the 3D index of a 3-manifold $M$
%
\begin{align}
\CI_M(q) = \begin{cases}
	\textrm{$1+O(q^{1/2})$ in $q$ expansion} &\ \Longleftrightarrow \  \textrm{No ${\rm SES}_{g=0,g=1}$ and Complex}
	    \\
	0  &\ \Longleftrightarrow   \chi^{SL(2,\mathbb{C})}_{\rm irred} =\emptyset
	\\
	| \cap_\rho  \textrm{Inv}(\rho)| \in \{1,2\} &\ \Longleftrightarrow  \ \textrm{No ${\rm SES}_{g=0,g=1}$ and Real}
	\\
	\textrm{Divergent} &\ \Longleftrightarrow  \exists\; {\rm SES}_{g=0} \textrm{ or } {\rm SES}_{g=1}
	\end{cases} \label{3D index pattern}
\end{align}
We have tested this proposal for a large number of 3-manifolds by explicit calculations. Some concrete examples will be presented below and we also refer to Appendix \ref{App : 3D index examples} for more detailed computations. 

 For hyperbolic 3-manifolds, we expect the 3D index to converge as a $q^{1/2}$-series since hyperbolic 3-manifolds are irreducible and atoroidal. Also, the 3D index of a hyperbolic 3-manifold should show exponentially growing  behavior in the Cardy limit since it always contains at least one complex flat connection, say $\rho_{\rm hyp}$, associated with the hyperbolic structure.
For lens spaces, as explained above, we expect $\mathcal{I}_M(q)=0$ since they have no irreducible flat connection. This is not manifest at all in the actual computations of the 3D indices using the state-integral model, but after delicate cancellations, one can see that the 3D indices vanish for these cases. For reducible or toroidal cases, the 3D index diverges if some of essential spheres or tori are separating.

The 3-manifolds in the 3rd line of the above proposal are of finite and real type. For these 3-manifolds, it is argued in \cite{Cho:2020ljj} that there is an associated topological field theory ${\rm TFT}[M]$ and the 3D index computes the partition function of ${\rm TFT}[M]$ on $S^2\times S^1$.  Note that for  topological theories, the  $S^2\times S^1$ partition function is always $1$ since there are no local excitations contributing to the partition function. 
However, we observe that the index sometimes becomes $2$ instead. This is puzzling, but we can understand this as follows. Let us first introduce some mathematical definitions. The cohomology group $H^1 (M, \mathbb{Z}_2) = \textrm{Hom}(\pi_1 M \rightarrow \mathbb{Z}_2)$ naturally acts on the elements in $\chi^{SL(2,\mathbb{C})}_{\rm irred}$ as follows
\begin{align}
	\begin{split}
		&\eta  \in H^1 (M, \mathbb{Z}_2) \;:\; \rho \in \chi^{SL(2,\mathbb{C})}_{\rm irred} \quad \longrightarrow \quad \rho \otimes \eta \in \chi^{SL(2,\mathbb{C})}_{\rm irred} 
		\\
		& \textrm{where} \; (\rho\otimes \eta )(a):= \rho(a) \eta (a)\;.
	\end{split}
\end{align}
The invariant subgroup $\textrm{Inv}(\rho) \subset H^1 (M, \mathbb{Z}_2)$ is then defined as
\begin{align}
	\textrm{Inv} (\rho):= \{ \eta \in H^1 (M, \mathbb{Z}_2) \;:\; \rho \otimes \eta \sim \rho \}\;. 
\end{align}
Recall from equivalence \eqref{character equivalence} that two homomorphisms $\rho_1$ and $\rho_2$ are considered identical if their characters are the same.
Among finite/real 3-manifolds, we find that the following 3-fibered Seifert spaces  have non-trivial  $\cap_{\rho}\textrm{Inv}(\rho)$ as
\begin{align}
S^2 ((2,1),(2,1),(p,q))\;:\; \cap_{\rho}\textrm{Inv}(\rho)= \mathbb{Z}_2\;. \label{Inv-rho}
\end{align}
In \cite{Cho:2020ljj}, it is claimed that the ${\rm TFT}[M]$ can be obtained from a progenitor TQFT, denoted as $\widetilde{{\rm TFT}}[M]$, by gauging the 1-form symmetry $H^1 (M, \mathbb{Z}_2)$. But sometimes  a subgroup of the 1-form symmetry acts trivially on all observables and thus decouples in IR. This decoupling happens when the subgroup of  $H^1(M, \mathbb{Z}_2)$ trivially acts  on $\chi^{SL(2,\mathbb{C})}_{\rm irred} $. We note that  $\cap_{\rho}\textrm{Inv}(\rho) $ is a subgroup of $H^1(M, \mathbb{Z}_2)$ trivially acting on the $\chi^{SL(2,\mathbb{C})}_{\rm irred} $. In the case when $\cap_{\rho}\textrm{Inv}(\rho) = \mathbb{Z}_2$, the 3D index  counts $2$ instead of $1$ since the ground state of ${\rm TFT}[M]$ on $S^2$ comes with multiplicity $2$. An example of this type of decoupled 1-form symmetry will be presented in Section \ref{sec : m006p}.

\paragraph{Example : $M=(S^3\backslash \mathbf{4}_1)_{P \mu+Q \lambda}$.} $(S^3\backslash \mathbf{4}_1)_{P \mu+Q \lambda}$ denotes a closed 3-manifold obtained by a Dehn surgery along the  figure-eight knot ($\mathbf{4}_1$) with slope $P/Q$. The 3D index can be written as
 \begin{align}
 	\mathcal{I}_{(S^3\backslash \mathbf{4}_1)_{P\mu +Q \lambda}}\big(q\big)
 	:=\!
 	\sum_{(m,e)\in \mathbb{Z}^2} \!\!\!\!\! K(m,e;P,Q;q)\text{ }
 	\mathcal{I}_{S^3\backslash \mathbf{4_1}}\big( m,e;q\big)\ ,  \label{Index Dehn filling}
 \end{align}
 where $K(m,e;P,Q;q)$ is a kernel of $SO(3)$ type \cite{Gang:2018wek} given by
 \begin{align}
 	K &:= 
 	\frac{1}{2}(-1)^{R m + 2 S e}
 	\Big(
 	\delta_{P m + 2 Q e , 0} (x^{\frac{R m + 2 S e}{2}} + x^{-\frac{R m + 2 S e}{2}})
 	-\delta_{P m + 2 Qe,-2}
 	-\delta_{P m + 2 Q e,2}
 	\Big)\ .
 	\nonumber\\
 \end{align}
 Here the integers $R,S\in\mathbb{Z}$ are determined by a condition $QR-PS=1$.\footnote{This condition cannot fix $(R,S)$ uniquely. However, the index will not depend on this freedom.} The 3D index for the figure-eight knot complement is 
 \begin{align}
 \CI_{S^3\backslash \mathbf{4}_1} (m,e;q) = \sum_{e_2 \in \mathbb{Z}} \CI^c_{\Delta} (m-e_2, m+e-e_2) \CI^c_{\Delta} (e-e_2,-e_2) \;.
 \end{align}
Here $\CI^c_\Delta (m,e)$ is the tetrahedron index in charge basis defined as follows.
\begin{align}
	&\mathcal{I}_{\Delta}(m, u) := \prod_{r=0}^\infty \frac{1-q^{r-\frac{ m}2+1} u^{-1}}{1- q^{r- \frac{m}2 } u } = \sum_{e \in \mathbb{Z}}\CI^c_{\Delta} (m,e) u^e
	 \label{tetraheron index}\;.
\end{align}
The index $\CI_{S^3\backslash \mathbf{4}_1} (m,e)$ gets contributions from the surface in the knot complement along  the 1-cycle $2e \mu + m\lambda$ at the boundary \cite{Garoufalidis:2016ckn}. In the knot complement, there are embedded essential once-punctured  Klein bottles ending on the 1-cycle $\pm 4 \mu  \pm \lambda$ at the boundary. These essential surfaces can be detected by looking at the leading term of the 3D indices in the $q$ expansion as follows.
\begin{align}
\begin{split}
&\CI_{S^3\backslash \mathbf{4}_1}(\pm 1,\pm 2;q) = -q^{1/2}+ q^{5/2}+ 4 q^{7/2}+\cdots \;,
\\
&\CI_{S^3\backslash \mathbf{4}_1}(\pm 2,\pm 4;q) = q-q^4-2 q^5+\cdots \;,
\\
&\CI_{S^3\backslash \mathbf{4}_1}(\pm 3,\pm 6;q) = -q^{3/2}+q^{11/2}+ 2q^{13/2}+\cdots \;.
\end{split}
\end{align}
From these results, one can confirm that $\CI_{S^3\backslash \mathbf{4}_1}(m=\pm k,e=\pm 2k;q) $ starts with $(-q^{1/2})^{k}$, which can be identified with the contribution from a BPS M2-brane winding around the essential surface $k$-times. Note that the once-punctured Klein bottle has Euler characteristic $-1$ which is compatible with \eqref{essential surface to 3D index}. 
For comparison, one can compute the index with $(m,e)=k (\pm 1,\pm1)$ as
\begin{align}
\begin{split}
&\CI_{S^3\backslash \mathbf{4}_1}(\pm 1,\pm 1;q) =-q-q^2+2 q^3+\cdots\;,
\\
&\CI_{S^3\backslash \mathbf{4}_1}(\pm 2,\pm 2;q) = 2q^3 +2 q^4+2 q^5+\cdots\; , \\
&\CI_{S^3\backslash \mathbf{4}_1}(\pm 3,\pm 3;q) = -q^5-3q^6-4q^7+\cdots\;.
\end{split}
\end{align}
One may notice that unlike the above cases, the lowest power in $q^{1/2}$ is not linearly increasing. This implies  the absence of embedded essential surfaces ending on boundary 1-cycle $\pm 2 \mu \pm \lambda$.

After the Dehn filling, the 3D index becomes
\begin{align}
&\CI_{(S^3\backslash \mathbf{4}_1)_{P \mu+Q\lambda}} (q) =
\begin{cases}
         0\;, \quad (P,Q)=(\pm 1 , 0)
         \\
	1\;, \quad P/Q=0, \pm 1, \pm 2 \textrm{ and }\pm 3
	\\
         (\sum_{k=0}^\infty 1) -2q^2-2q^3-4q^4-4q^5-6 q^6+\cdots \;, \quad P/Q= \pm 4
        \\
	\textrm{Power series in $q^{1/2}$ starting with }1+\CO(q^{1/2})\;, \quad  \textrm{otherwise}	\\
\end{cases}
\end{align}
This is indeed consistent with our proposal in \eqref{3D index pattern} for the 3D index and with the following topological facts \cite{2018arXiv181211940D}.
\begin{align}
	(S^3\backslash \mathbf{4}_1)_{P \mu + Q \lambda}   = 
	\begin{cases}
	        S^3\;, \quad (P,Q) = (\pm 1, 0)
	        \\
		\mathbb{T}^2 \times_{\varphi } S^1\textrm{ with } \varphi = \begin{pmatrix}  2 & 1 \\ 1 & 1 \end{pmatrix}\;, \quad P/Q =0
		\\
		\textrm{Atoroidal SFS $S^2 ((2,1),(3,1),(7,-6))$}\;, \quad P/Q=\pm 1
		\\
		\textrm{Atoroidal SFS $S^2 ((2,1),(4,1),(5,-4))$}\;, \quad P/Q=\pm 2
		\\
		\textrm{Atoroidal SFS $S^2 ((3,1),(3,1),(4,-3))$}\;, \quad P/Q=\pm 3
		\\
		\textrm{Graph}\;, \quad P/Q=\pm 4
		\\
		\textrm{Hyperbolic}\;, \quad \textrm{otherwise}
	\end{cases}
\end{align}
The divergent term $(\sum_{k=0}^\infty 1)$ for $P/Q=\pm 4$ comes from $(-q^{1/2})^{k }$  in $\CI_{S^3\backslash \mathbf{4}_1} (\pm k, \pm 2k;q)$ after the Dehn filling in \eqref{Index Dehn filling}. We note that this term can be  regarded as contributions from  BPS M2-branes winding around the essential torus in $(S^3\backslash \mathbf{4}_1)_{\pm 4 \mu + \lambda}$. Here, the essential torus comes from the once-punctured Klein bottle in the knot complement. The puncture is closed after the Dehn filling. Then the puncture-closed essential surface has Euler characteristic $\chi=0$ and it can therefore give the divergent contribution in the 3D index as noticed in \eqref{essential surface to 3D index}.

\section{IR phases of class R theory} \label{sec : IR phases}

In this section, we discuss low-energy phases of 3D class R theories arising from compactifications of the 6d $\mathcal{N}=(2,0)$ theory on M5-branes wrapped around closed 3-manifolds. We particularly focus on the 3D theories from the $A_1$-type 6D (2,0) theory.
When the 3-manifold $M$ is hyperbolic, the low-energy theories become 3D $\mathcal{N}=2$ superconformal field theories, which have been extensively studied in \cite{Dimofte:2011ju,Dimofte:2011py}. See also \cite{Gang:2018wek} for more discussions. On the other hand, the low-energy phases for the 3D theories engineered by non-hyperbolic 3-manifolds have much richer possibilities. For example, as discussed in \cite{Cho:2020ljj}, compactifications on non-hyperbolic 3-manifolds can lead to 3D theories in gapped phases described by TQFTs and also to SCFTs embedding a sub-sector related to a non-unitary TQFT structure. 

More specifically, we study how the IR phases of class R theories change under the exceptional Dehn fillings, and what kinds of IR phases can appear from the class R theories engineered by closed non-hyperbolic 3-manifolds. Interestingly, the 
exceptional Dehn filling corresponds to gauging of $SU(2)$ symmetry with a small Chern-Simons (CS) level. The small CS level means that the theory becomes strongly coupled and thus the IR phase can be drastically changed after the gauging due to wild non-perturbative quantum effects. This is in a nice parallelism with the 3-manifold side story,  drastic change of geometrical structures under the exceptional Dehn filling. We will make the parallelism more precise. We will use basic features of 3-manifolds explained in the previous section and determine the IR phases of class R theories. The main result is summarized in \eqref{proposal for IR phases}. Using the result, one can easily determine the IR phases of class R theories without   analyzing  the non-perturbative quantum effects. The emergent IR phases after the exceptional gaugings turn out to be very rich. One can find SUSY breaking phases, gapped topological phases, gapless phases, IR phases with SUSY enhancement and etc, which is also compatible with the mathematical fact that non-hyperbolic manifolds covers 7 geometries out of Thurston's 8 geometries.  We will provide concrete examples for these phases.

\subsection{3D class R theory from 6D (2,0) theory }
Let $M$ be a closed (i.e. without boundary) 3-manifold. Then the 3D class R theory $\CT[M]$ associated to the $M$ is defined as follows:
\begin{align}
	(\textrm{6D $A_1$ (2,0) theory on $\mathbb{R}^{1,2}\times M$}) \xrightarrow{\qquad \textrm{size}(M)\rightarrow 0 \qquad  } (\textrm{3D $\CT[M]$ theory on $\mathbb{R}^{1,2}$})\;. \label{6D set-up}
\end{align} 
In the compactification, we perform a topological twisting along the internal 3-manifold using a $SO(3)$ subgroup of $SO(5)$ R-symmetry of the 6D theory. The topological twisting preserves 4 supercharges
and thus the low energy theory under the compactification becomes a 3D  $\CN=2$ supersymmetric theory.  The 3D theory has $SO(2)=U(1)$ R-symmetry originated from the $SO(2) \subset SO(3)\times SO(2) \subset SO(5)$  subgroup of the 6D R-symmetry preserved in the topological twisting.  For some cases, the $\CT[M]$ can have accidental Abelian flavor symmetries and the $U(1)$ R-symmetry can be mixed with them in IR. To distinguish it, we denote the R-symmetry originated from the 6D theory by $U(1)_{R_{\rm geo}}$ and its charge by $R_{\rm geo}$:
\begin{align}
	U(1)_{R_{\rm geo}} \subset SO(2)_{R_{\rm geo}} \times SO(3) \subset (\textrm{$SO(5)$ R-symmetry of the 6D theory}) \;.\label{R-geo}
\end{align}
In this paper, we consider the 3D class R theory $\mathcal{T}[M]$ proposed in \cite{Dimofte:2011ju,Gang:2018wek},  which will be briefly reviewed in the next subsection. The theory is sometimes denoted as $\CT_{\rm irred}[M]$ since it only sees an {\it irreducible sub-sector} of the 6D compactification \cite{Chung:2014qpa}. So our $\mathcal{T}[M]$ is different from $\mathcal{T}_{\rm full}[M]$ (which is expected to see the full sector) studied in \cite{Pei:2015jsa,Gukov:2017kmk,Eckhard:2019jgg}. Unlike $\mathcal{T}_{\rm irred}[M]$, we still do not have 3D field theoretic descriptions for general $\mathcal{T}_{\rm full}[M]$ theories.  It is even not sure whether $\mathcal{T}_{\rm full}[M]$ for a general $M$  (especially for a hyperbolic $M$) can be described by a genuine 3D theory instead of the 6D theory compactified on $M$ \cite{Gang:2018wek}. As we will see below,  $\mathcal{T}_{\rm irred}[M]$ for non-hyperbolic 3-manifolds  have richer IR phases than $\mathcal{T}_{\rm full}[M]$. For example, the spontaneously SUSY breaking phase can not be realized in $\mathcal{T}_{\rm full}[M]$ since there always exists a Bethe-vacuum, i.e. SUSY vacuum on $\mathbb{R}^2\times S^1$, in the theory, which corresponds to the trivial $SL(2,\mathbb{C})$ flat connection on $M$ \cite{Dimofte:2010tz}.

\subsection{Field theoretic construction} \label{subsec: UV field theory}
We start with a brief review of the  field theoretic construction of the $\CT[M]$ theory proposed in \cite{Dimofte:2011ju,Gang:2018wek}. For simplicity, we focus on the 3-manifolds $M$ given by
\begin{align}
M= N_{pA+B} \;,
\end{align}
where $N$ is a 1-cusped hyperbolic 3-manifold.
Then the corresponding 3D theory $\CT[M]$ can be obtained by gauging certain global symmetries in the 3D class R theory  $\CT[N,A;B]$ associated to $N$ 
\begin{align}
	\CT[M = N_{pA+B}] = \begin{cases} \frac{\CT[N,A;B]}{``SO(3)"_p} , \quad {\rm odd \; }A
		\\
		\frac{\CT[N,A;B]}{SU(2)_p},   \quad {\rm even \;}A \ . \label{Dehn filing/Gauging}
	\end{cases}
\end{align}
Here, $\CT[N,A;B]$ is the 3D theory constructed by Dimofte-Gaiotto-Gukov \cite{Dimofte:2011ju} based on an ideal triangulation of $N$ (see section 2.2 of \cite{Gang:2018wek} for a review with some improvements).  $``SO(3)"_p$ and $SU(2)_p$ in the denominators mean particular gauging procedures which we will now explain. 

First, the 3D theory $\CT[N,A;B]$ depends on the choice of the boundary 1-cycle $A$ and always has $U(1)$ flavor symmetry, say $U(1)_A$, associated to the boundary 1-cycle.   The theory can be schematically described as follows:
\begin{align}
\begin{split}
\CT[N,A;B]  &=  \left(U(1)^r_K\; \textrm{gauge theory coupled to $S$ chiral multiplets of charge $Q$}\right)
\\
&\quad +(\textrm{deformed by superpotential }\CW = \sum_{I=1}^{S-\tilde{F}-1} \CO_I)  \label{TNAB} \ ,
\end{split}
\end{align}
where $K$ is a symmetric $(r\times r)$ matrix representing the  mixed Chern-Simons levels and $Q$ is an $(S\times r)$ matrix representing the gauge charges of the chirial multiplets. The number of chrial multiplets denoted by $S$ is equal to the number of  tetrahedra in the ideal triangulation.  The $r$, $K$, $Q$ and $\{\CO\}_{I=1}^{S-\tilde{F}-1}$ are determined by the ideal triangulation of $N$ and choice of the boundary 1-cycle $A$.  
Before the superpotential deformation, the Abelian gauge theory has $U(1)^S$ flavor symmetry, which is broken to $U(1)^{\tilde{F}+1}$ after the deformation. The $U(1)^{\tilde{F}+1}$ always contains the $U(1)_A$, i.e. $U(1)_{A}\subset U(1)^{\tilde{F}+1} $.  The $U(1)_A$ symmetry is expected to exist from the 6D construction of $\CT[N, A;B]$ theory\footnote{For a 1-cusped hyperbolic 3-manifold $N$, the 6D set-up includes a codimension-two defect which gives an $SU(2)_A$ symmetry. The $SU(2)_A$ symmetry is  broken to the $U(1)_A$ symmetry for a generic choice of $A$ due to a symmetry breaking superpotential. The superpotential is forbidden when the cycle $A$ is non-closable. We shall refer to \cite{Gang:2018wek} for details.  }, while the other symmetries are accidental symmetry, which are not manifest in the UV 6D set-up. The existence of the additional flavor symmetry is closely related to the existence of so-called {\it hard internal edges} \cite{Dimofte:2011ju} in the ideal triangulation of $N$ as pointed out in \cite{Gang:2018wek}.

 In this paper, we always choose the  1-cycle $A$ to be a {\it  non-closable} cycle defined as
\begin{align}
\textrm{$A\in H_1 (\partial N, \mathbb{Z}) $ is called  {\it non-closable} if  $N_{A}$  is of  empty type}\;.  \label{condition on A-1}
\end{align}
It then follows that the $U(1)_A$ symmetry of  $\CT[N,A;B]$ is enhanced to  $SU(2)$ flavor symmetry which we denote by $SU(2)_A$ \cite{Gang:2018wek}:
\begin{align}
\textrm{$A$ is non-closable } \Rightarrow  \textrm{$\CT[N,A;B]$ has a $SU(2)_A \supset U(1)_A$ flavor symmetry} \;.\label{su(2)A}
\end{align}
Also, we define an oddness/evenness of the 1-cycle $A \in H_1 (\partial N, \mathbb{Z})$ as follows
\begin{align}
\begin{split}
 A \in H_1 (\partial N, \mathbb{Z})\textrm{ is called } 
\begin{cases}
\textrm{even}, \quad A \in \textrm{Ker}[i_*: H_1(\partial N, \mathbb{Z}) \rightarrow H_1 (N,\mathbb{Z}_2)]
\\
\textrm{odd}, \quad \textrm{otherwise} \ .
\end{cases} \label{even/odd}
\end{split}
\end{align}
If the cycle $A$ is odd, the global structure of the flavor symmetry is $SO(3)$ (i.e. all local operators have integer spin under the $SU(2)_A$).

The choice of the other 1-cycle $B$ then affects the background CS level for the $SU(2)$ flavor symmetry.
In order for the  CS level to be properly  quantized, we need to choose  two 1-cycles ($A$ and $B$) to have different oddness/evenness.
For example, when $N=S^3\backslash \CK$  is a knot complement in $S^3$ we can choose
\begin{align}
A = \mu, \quad B= \lambda\;, \quad \textrm{for $N = S^3\backslash \CK$}\;.
\end{align}
In this case, the meridian $\mu$ is always non-closable  and odd while longitude $\lambda$ is always even.

In order to obtain the $\mathcal{T}[M]$ theory, we gauge the $SU(2)_A$ flavor symmetry of the $\CT[N,A;B]$ theory. The gauging should be performed with a Chern-Simons coupling at level $p$ for the $SU(2)_A$ flavor symmetry. For even $A$ cases, the gauging is straightforward and one obtains the $T[M]$ theory as
\begin{align}
	\begin{split}
		&	\CT[N_{pA+B}] = \frac{\CT[N,A;B]}{SU(2)_p} =(\textrm{Gauging $SU(2)_A$ of $\CT[N,A;B]$ with additional CS level $p$})\;.
	\end{split} \label{T[NpAB-for-even-A]}
\end{align} 

For odd $A$ cases, on the other hand, the gauging of the $SU(2)_A$ flavor symmetry is more involved.
 As explained above, when $A$ is odd, the gauge group is $SO(3)$ rather than $SU(2)$. This means that the $\mathbb{Z}_2$ 1-form symmetry coming from the $\mathbb{Z}_2$ center symmetry of the $SU(2)$ symmetry must also be gauged. However, the $\mathbb{Z}_2$ 1-form symmetry has non-trivial `t Hooft anomaly for odd $p$. In this case, we first need to tensor the $\CT[N,A;B]$ theory with a topological theory to cancel the $\mathbb{Z}_2$ anomaly, and then we gauge the $\mathbb{Z}_2$ 1-form symmetry. We thus propose for an odd $A$ that the $\mathcal{T}[M]$ theory is constructed as
\begin{align}
\begin{split}
&\mathcal{T}[N_{pA+B}]	 =  \frac{\CT[N,A;B]}{``SO(3)"_p} 
\\
&\qquad \qquad \;:= \begin{cases}
\frac{(\textrm{Gauging $SU(2)_A$ of $\CT[N,A;B]$ with additional CS level $p$})}{\mathbb{Z}_2}, \quad \textrm{even $p$}
\\
\frac{(\textrm{Gauging $SU(2)_A$ of $\CT[N,A;B]$ with additional CS level $p$})\otimes U(1)_{-2}}{\mathbb{Z}^{\rm diag}_2}, \quad \textrm{odd $p$} \ .
\end{cases}
\end{split} \label{T[NpAB-for-odd-A]}
\end{align} 
Here $/\mathbb{Z}_2$ denotes the gauging of the $\mathbb{Z}_2$ 1-form symmetry. Note that when $p$ is odd, we tensor the theory with  the $U(1)_{-2}$ theory, which is a topological theory with an anomalous $\mathbb{Z}_2$ 1-form symmetry, such that the tensored theory has an anomaly free 1-form symmetry $\mathbb{Z}_2^{\rm diag}$, and this symmetry is finally gauged. For odd $p$, the above relation  implies that
\begin{align}
\frac{\CT[N,A;B]}{SU(2)_p} = \CT[N_{pA+B}] \otimes U(1)_2\;.\label{T[Np-for-odd-p]}
\end{align} 
After the $SU(2)$ or $``SO(3)"$ gauging,  the final $\CT[M]$ has $U(1)^F$ flavor symmetry, which is a subgroup of $U(1)^{\tilde{F}}$ symmetry commuting with the gauged $SU(2)$ symmetry.

\subsection{IR phases (Main Proposal) }

We now discuss low energy physics of the 3D field theory $\mathcal{T}[M]$ for a closed 3-manifold $M$. The IR phase of this theory is entirely determined by geometric data of the 3-manifold $M$. We thus approach the problem of the classification of the IR phases of the class R theories by examining the geometric properties of the corresponding 3-manifolds.

For example, the IR phases associated with 3-manifolds of type $M=N_{pA+B}$ at large $p$ can be rather easily determined. As explained in the previous subsection, the corresponding UV field theory can be constructed by gauging the $SU(2)$ flavor symmetry of the SCFT $\mathcal{T}[N,A;B]$ with a Chern-Simons coupling at level $p$. When the level $p$ is large, the quantum effect of the gauging is suppressed at $1/|p|\ll 1$, and this implies that the theory $\CT[M]$ after the gauging is expected to flow to a SCFT at low energy. Indeed, the stress-energy central charge of the  $\CT[M]$ theory approaches to  that of the original theory $\mathcal{T}[N,A;B]$ as $|p|$ increases. 
This is compatible with the Thurston's hyperbolic Dehn surgery theorem saying that $M= N_{pA+B}$ is hyperbolic for a large enough $|p|$. Note that $\CT[M]$ for a hyperbolic 3-manifold $M$ always flows to a non-trivial SCFT.

However, on the other hand when $|p|$ is small, the $SU(2)$ gauging leads to strongly interacting theories showing various interesting non-perturbative phenomena at low energy. Thus, the analysis of the IR phases becomes much more complicated for these cases. It turns out that $\CT[M]$ theories for generic closed 3-manifolds, which involve such strong interactions, can flow to non-trivial IR phases other than SCFT phases.

Based on the characteristics of BPS partition functions related to a closed 3-manifold $M$, which we will explain in details in the next subsection, we propose that
\begin{framed}
\noindent {\bf Main proposal}
\begin{align}
	\CT[M]= \begin{cases}
		\textrm{$\mathcal{N}=2$ SCFT},  & \textrm{for a finite/complex type $M$}
		\\
		\textrm{Spontaneous SUSY breaking phase},  & \textrm{for an empty type $M$}
		\\
		\textrm{Unitary TQFT},  &\textrm{for a finite/real/unitary type $M$}
		\\
		\textrm{$\mathcal{N}=4$ (rank-0) SCFT},
		 &\textrm{for a finite/real/non-unitary type $M$}
		\\
		\textrm{$\CN=2$   SCFT $+$ accidental sym/CPO},    &\textrm{for an infinite type $M$} \label{proposal for IR phases}
	\end{cases}
\end{align}
\end{framed}
More precise statements in this proposal are as follows:
\begin{itemize}
	\item When the closed 3-manifold $M$ is of finite and complex type, $M$ is hyperbolic and the $\mathcal{T}[M]$ theory flows to a $\mathcal{N}=2$ superconformal field theory.

	\item When $M$ is of empty type, supersymmetry (SUSY) of the $\mathcal{T}[M]$ theory is spontaneously broken at low energy.

	\item When $M$ is of finite, real, and unitary type, the low energy theory of $\mathcal{T}[M]$ is in a gapped phase described by a (unitary) topological field theory.

	\item When $M$ is of finite, real, and non-unitary type, the $\mathcal{T}[M]$ theory flows to a superconformal field theory that includes sub-sectors realizing a non-unitary TQFT structure. The IR theory also enjoys an $\mathcal{N}=4$ SUSY enhancement, but has no Coulomb and Higgs branches, which means that it is a rank-0 theory.

	\item When $M$ is of infinite type, the low energy phase of $\mathcal{T}[M]$ is rather subtle. The 3D index in this case diverges, which is an unexpected behavior for a consistent IR phase. A natural mechanism to remedy this is the emergence of an accidental flavor symmetry at low energy.   R-symmetry  can mix with   this accidental flavor symmetry, and the 3D superconformal index takes an appropriate index expression, capturing a finite degeneracy of BPS states for a given charge, only if it is computed with  the correct IR R-symmetry. The divergence of the 3D index is due to a wrong choice of the IR superconformal R-symmetry. Geometrically, this divergence comes from the contributions of separating essential surface $\Sigma$  with  a genus $g\leq 1$. The essential surface corresponds to a $1/2$ BPS chiral primary operator (CPO) in $\mathcal{T}[M]$ theory with $R_{\rm geo} =(g-1) $.  To  satisfy the unitarity bound $R_{\rm IR}\, (\textrm{the correct IR R-charge})\geq \frac{1}2$, the CPO should be charged under the accidental flavor symmetry. Since there is no such a CPO in $\mathcal{T}[N,A;B]$\footnote{Otherwise,  the 3D index for a 1-cupsed hyperbolic 3-manifold $N$ should diverge. But we know that the 3D index always converges \cite{Garoufalidis:2013axa}.}, the CPO should correspond to a gauge-invariant monopole operator of the $SU(2)$ gauge symmetry of $\mathcal{T}[M]$ in \eqref{Dehn filing/Gauging}. This operator is generically $1/4$ BPS operator but we expect that it becomes  $1/2$ BPS operator when  $M$ is of infinite type.    In summary, we  propose  IR phase of this type is  a $\mathcal{N}=2$ SCFT with accidental flavor symmetry, and the $SU(2)$ monopole operators become 1/2 BPS CPOs,  which are charged under the accidental symmetry.  
\end{itemize}

We will now explain how to read off the IR phases by inspecting BPS partition functions, and then utilize it to concrete examples. This will illustrate the physics behind our proposal and also provide supporting evidences for it.

\subsection{Strategy for  reading off IR phases} \label{sec : strategy}
To verify our proposal \eqref{proposal for IR phases}  from the UV description of $\CT[M]$, reviewed in Section \ref{subsec: UV field theory}, we provide a general method to determine the IR phases using the properties of various supersymmetric partition functions.

\paragraph{BPS partition functions}  We consider the following three types of BPS partition functions
\begin{align}
	\begin{split}
&\CI_q (u_i,\nu_i) \;:\; \textrm{Superconformal index (or 3D index)}\;,
\\
&\CZ^{(s=\pm 1)}_{\CM_{g,p}} (m_i, \nu_i)  \;:\; \textrm{Twisted partition function on $\CM_{g,p}$}\;,
\\
&\CZ_b (m_i , \nu_i)\;:\; \textrm{Squashed 3-sphere partition function}\;. 
\end{split}
\label{eq:BPS-partition-functions}
\end{align}
Here, $m_i$ and $u_i$ with the index $i=1,\cdots, F$ are respectively real masses and fugacities for the flavor symmetry $U(1)^F \subset G_F $ of $\CT[M]$. The UV $U(1)_R$ R-symmetry can mix, at low energy, with the Abelian components of the flavor symmetry, and this mixing is captured by the mixing parameters $\nu_i$ as
 \begin{align}
 	R_{\vec{\nu}} = R_{\nu_i =0} +\sum_{i=1}^F J_i \nu_i\;,
\end{align}
where $R_{\nu_i=0}$ stands for a reference R-charge and $J_i$'s denote the Abelian flavor charges. One linear combination,  $\vec{\nu}_{\rm geo}$, gives the $U(1)_{R_{\rm geo}}$ symmetry in \eqref{R-geo}, i.e. $R_{\vec{\nu}_{\rm geo}} = R_{\rm geo}$.

The superconformal index is a Witten index counting supersymmetric (or BPS) local operators in a 3D SCFT that can be defined as a trace over the Hilbert space in radial quantization 
\begin{equation}
	\CI_q(u_i,\nu_i) = \textrm{Tr}_{\mathcal{H}_{\rm rad} (S^2)} \bigg[(-1)^{R_{\vec{\nu}}} q^{\frac{R_{\vec{\nu}}}{2} + j_3} \prod_i u_i^{J_i}\bigg]\ .
\end{equation}
Here, $q$ is the fugacity for a combination of R-charge $R_{\vec\nu}$ and the angular momentum $j_3$ commuting with a pair of supercharges preserved by the local operators.
Via SUSY localization, the index can be computed and the result can be expressed in the following form \cite{Kim:2009wb,Imamura:2011su}
\begin{align}
	\begin{split}
		\CI_q^{N_p} (u_i, \nu_i) &= \sum_{m_x }\oint_{|u_x|=1}\frac{du_x}{2\pi i u_x}  \Delta (m_x, u_x)\CI^{N}_q(m_x, u_x, u_i, \nu_i) u_x^{2 p m_x } \ .
		\label{Index for Np}
	\end{split}
\end{align}
Here  $\CI^N_q (m_x, u_x, u_i ,\nu_i)$ is the superconformal index for $\CT[N,A;B]$ theory. $(m_x,u_x)$ are (monopole flux, fugacity)  coupled to the $SU(2)$ gauge symmetry, and $(u_i, \nu_i)|_{i=1}^F$ are the (fugacities, R-symmetry mixing parameters) associated with $U(1)^F$ flavor symmetry.  
$\Delta(m_x, u_x)$ is the measure factor coming from the $SU(2)$ vector multiplet given by
\begin{align}
	\Delta (m_x, u_x) := (-1)^{2m_x}(q^{m_x/2}u_x -q^{-m_x/2}u_x^{-1})  (q^{m_x/2}u^{-1}_x -q^{-m_x/2}u_x) \times \begin{cases}\frac{1}2,\; m_x =0 \\ 1,\; m_x \neq 0 \end{cases} \ . \label{SCI : SU(2) measure}
\end{align}
The summation range of the $m_x$ will be discussed below. 

One notices that the index at general $\nu_i$ can be obtained by shifting $u_i \to  (-q^{1/2})^{\nu_i} u_i$ in the index at $\vec{\nu}=0$ as
\begin{align}
\CI_q (u_i, \nu_i)= \CI_q (u_i, \nu_i=0)|_{u_i \rightarrow (-q^{1/2})^{\nu_i}u_i}\;. \label{SCI under mixing}
\end{align}
The BPS operators of the IR SCFT can be counted correctly only if we choose the right mixing parameters $\nu_i$, which can be determined by the F-extremization in \cite{Jafferis:2010un}.

We can also tune $\nu_i$'s to compute the spectrum of a sub-sector in which we are particularly interested. For example, certain IR phases exhibit $\mathcal{N}=4$ SUSY enhancement and contain interesting sub-sectors consisting of holomorphic functions on the Coulomb or Higgs branch. In this case, the parameter $\nu$ for the axial R-charge $J_{\rm Axial}$ of the $SU(2)_L\times SU(2)_R$ R-symmetry, which is a flavor charge from the point of view of the $\mathcal{N}=2$ superconformal algebra, can be tuned to compute the Hilbert series of the Coulomb or Higgs branch operators as discussed in \cite{Razamat:2014pta}. We will see some examples below.

Next, $\CZ^{(s=\pm 1)}_{\CM_{g,p}}$ is the twisted partition function on $\CM_{g,p}$, the $S^1$ bundle of degree $p$  over genus $g$ Riemann surface $\Sigma_g$ \cite{Closset:2017zgf,Closset:2018ghr}. The $s$ in the superscript represents the choice of spin-structure along the $S^1$. We choose $s=+1$ ($s=-1$) as periodic (anti-periodic) boundary condition for fermionic fields. For odd $p$, only $s=1$ choice is allowed while both choices are allowed for even $p$.  Especially for $p=0$, the partition function corresponds to the twisted index $I^{(s)}_g (u_i = e^{m_i}, \nu_i):=\CZ^{(s)}_{\CM_{g,p=0}}(m_i, u_i)$ on a genus-$g$ Riemann surface $\Sigma_g$ \cite{Gukov:2015sna,Benini:2015noa,Benini:2016hjo,Closset:2016arn}. 

The twisted index counts supersymmetric ground states with signs on a topologically twisted surface $\Sigma_g$, i.e. 
\begin{align}
	\begin{split}
	&I^{(s=1)}_g (u_i,\nu_i) = \textrm{Tr}_{\mathcal{H}(\Sigma_g;R_{\vec{\nu}})} \bigg{[}(-1)^{2j_3} \prod_i u_i^{J_i}\bigg{]} \;,
	\\
	&I^{(s=-1)}_g (u_i,\nu_i) = \textrm{Tr}_{\mathcal{H}(\Sigma_g;R_{\vec{\nu}})} \bigg{[}(-1)^{R_{\vec{\nu}}} \prod_i u_i^{J_i}\bigg{]} \;.
	\end{split}
\end{align}
Here $\CH(\Sigma_g;R_{\vec{\nu}})$ is Hilbert space on $\Sigma_g$ with a topological twisting using the $U(1)_{R_{\vec{\nu}}}$ symmetry. 
Unlike other BPS partition functions, the mixing parameters $\nu_i$ in the twisted indices (with $g\neq 1$) can take only discrete values satisfying the following Dirac quantization condition:
\begin{align}
	R_{\vec{\nu}}  (g-1)\in \mathbb{Z}\;.
\end{align}
The condition is always satisfied at $\vec{\nu} = \vec{\nu}_{\rm geo}$ since the $U(1)_{R_{\rm geo}}$ is a compact subgroup of a $SO(5)$ symmetry as in \eqref{R-geo}. At $g=1$, the twisted index is a $(u_i, \nu_i)$-independent number, which is equal to the Witten index on 3-torus \cite{Kim:2010mr,Intriligator:2013lca}. When $g=0$ and $p=1$, the twisted partition function is identical to the round 3-sphere partition function, 
\begin{align}
\CZ^{(s=1)}_{\CM_{g=0,p=1}} = \CZ_{b=1}\;.
\end{align}

The twisted partition function can also be written as follows \cite{Closset:2017zgf}
\begin{align}
\CZ^{(s)}_{\CM_{g,p}} = \sum_{a \in \CS_{\rm BE}} (\CH^{(s)}_\alpha (m_i,\nu_i))^{g-1} (\CF^{(s)}_{\a} (m_i, \nu_i))^p\;. \label{twisted ptns from SBE}
\end{align}
Here $\alpha$ labels so-called Bethe-vacua of the theory and $\CH^{(s)}_\a$, $\CF^{(s)}_\a$ are handle-gluing/fibering operators respectively. Note that the $\CF^{(s)}$ is only well-defined up to an overall sign factor, i.e. only $(\CF^{(s)})^2$ is well-defined, for $s=-1$ since the twisted partition function is  defined only for $p\in 2\mathbb{Z}$.  Especially for $p=0$, the twisted indices are given as
\begin{align}
	I_g^{(s=\pm 1)} (u_i, \nu_i)\big{|}_{u_i=e^{m_i}} = \sum_{a \in \CS_{\rm BE}} (\CH^{(s)}_\alpha (m_i,\nu_i))^{g-1}\;. \label{Bethe-sum for Ig}
\end{align}

Lastly, $\mathcal{Z}_b$ is the partition function of a 3D $\mathcal{N}=2$ gauge theory on a squashed three-sphere $S^3_b$ with a squashing parameter $b$.
 Using the Coulomb branch localization in \cite{Hama:2011ea}, the squashed $S^3_b$ partition function for the $\CT[N,A;B]/SU(2)_p$ can be written as 
\begin{equation}
\begin{aligned}
 \mathcal{Z}^{N_p}_b (m_i,\nu_i) &:= (\textrm{Squashed 3-sphere partition function of $\CT[N,A;B]/SU(2)_p$})
\\
&=   \frac{1}{2}\int_{\mathbb{R}^{r+1}} \frac{dX d^{r} Z}{(2\pi \hbar)^\frac{r+1}{2}} \CI^{N_p}_\hbar (X,\vec{Z};W_i)|_{W_i = m_i + (i \pi +\frac{\hbar}2) \nu_i, \;\hbar = 2\pi i b^2}\;,
\end{aligned}
\end{equation}
where $X$ and $\vec{Z}$ parametrize the Coulomb branches of the $SU(2)$ and $U(1)^r$ vector multiplets, see \eqref{TNAB},  in the theory respectively. The integrand $\mathcal{I}_\hbar$ is the collection of 1-loop contributions from the vector and the chiral multiplets on the Coulomb branch. 
The contribution from a chiral multiplet is given by a special function called {\it quantum dilogarithm} \cite{faddeev1994quantum}. The function will be denoted by $\psi_{\hbar} (X)$. We follow the convention used in \cite{Gang:2019jut} and refer to the paper for the definition and some basic properties for the function. 
\begin{align}
\psi_\hbar (Z)  \;:\; \textrm{quantum dilogarithm function}. \label{QDL}
\end{align}
 Interestingly, the $S^3_b$ partition function depends only on a holomorphic combination, $\vec{m}+(i \pi +\frac{\hbar}2)\vec{\nu}$, of $\vec{m}$ and $\vec{\nu}$. As a consequence of this holomorphic property, one finds
 \begin{align}
 	\CZ_{b} (\vec{m}, \vec{\nu}) = \CZ_b \left(\vec{m} +( i \pi +\frac{\hbar}2)(\vec{\nu} - \vec{\nu}_0) , \vec{\nu}_0\right)\;. \label{Zb-at-different-nu}
 \end{align} 
This partition function has  overall factor ambiguity of the following form due to the local counter-terms,
\begin{align}
	\exp \left( \frac{k^{ij}_{FF}W_i W_j+ 2 k^{i}_{FR}W_i (i \pi +\frac{\hbar}2)+ k_{RR} (i \pi +\frac{\hbar}2)^2}{2\hbar }\right)\bigg{|}_{W_i = m_i +(i\pi +\frac{\hbar }2)\nu_i}\;, \label{ambiguity of Zb}
\end{align}
with properly normalized mixed CS levels $k_{FF},k_{FR}, k_{RR}$ between $U(1)^F$ flavor symmetry and $U(1)_R$ symmetry.

The above BPS partition functions are not all independent but related to each other in a sophisticated way. For later use, we will review the relationship between the squashed 3-sphere partition function and  the twisted indices discussed in \cite{Gang:2019jut}.
To relate the two partition functions, we consider the following asymptotic expansion of the integrand $\CI_\hbar$ in the limit $\hbar \rightarrow 0$
\begin{align}
\log \CI^{N_p}_\hbar \xrightarrow{\quad \hbar \rightarrow 0 \quad } \frac{1}{\hbar } \CW_0(X,\vec{Z};m_i,\nu_i) + \CW_1 (X,\vec{Z};m_i,\nu_i) +\cdots\; . 
\label{W0 and W1 from integrand}
\end{align}
The leading part $\CW_0$ corresponds to the twisted superpotential. In order to compute the perturbative series $\CW_n$, one needs to use following asymptotic expansion of  the quantum dilogarithm,
\begin{align}
\log \psi_\hbar (X)  \xrightarrow{\quad \hbar \rightarrow 0 \quad }  \sum_{n=0}^\infty\frac{   B_n }{n !} \textrm{Li}_{2-n} (e^{-Z})  \hbar^{n-1}\;. \label{asymptotic of QDL}
\end{align}
Here $B_n$ are Bernoulli numbers  with $B_1 = + \frac{1}2$. 

The Bethe vacua are the solutions to the Bethe equations that extremize the twisted superpotential as 
\begin{equation}
\mathcal{S}_{\textrm{BE}} = \left\{(x,\vec z)_\alpha : \exp \left.\left( \partial_{\vec Y} \mathcal{W}_0 (\vec{Y};m_i,\nu_i)\right)\right|_{\vec Y = \log \vec y_\alpha} = \vec 1,\;\textrm{triv. isotr.}\right\}/\mathbb{Z}_2^{\rm Weyl}\ , \label{Bethe-vacua set}
\end{equation}
where $\vec{Y}:=(X,\vec{Z})$ and $\vec{y}:=(x,\vec{z})$ and $\alpha$ labels a Bethe vacuum. `triv. isotr.' means that the Bethe solutions invariant under the Weyl group $\mathbb{Z}_2$ should be discarded. The Weyl symmetry acts on $(x,\vec{z})$ as
\begin{align}
\mathbb{Z}_2^{\rm Weyl}\;:\; (x, \vec{z}) \leftrightarrow (1/x, \vec{\tilde{z}})\;.
\end{align}
In general, $\vec{\tilde{z}}$ is  a non-trivial function of $\vec{z}$. 
However, when the $SU(2)_A$ symmetry is manifest in the field theory description of $\CT[N,A;B]$, $\vec{\tilde{z}}$ is just $\vec{z}$. In any case, we should always discard Bethe-vacua with $x=1/x$, i.e. $x=\pm 1$.
The handle-gluing/fibering operators $\mathcal{H}^{(s)}_\alpha$/$\CF^{(s)}_\alpha$  in \eqref{twisted ptns from SBE} with $s=-1$ for each Bethe vacuum $\vec z_\alpha$ can be computed as  
\begin{equation}
\begin{aligned}
&\CF_\alpha (m_i, \nu_i):=\mathcal{F}^{(s=-1)}_\alpha (m_i,\nu_i) = \exp \left( i\frac{ S_0^\alpha(m_i,\nu_i)-m_i\partial_{m_i} S_0^\alpha(m_i,\nu_i)}{2\pi} \right)\ , 
\\
&\CH_\alpha (m_i, \nu_i):=  \mathcal{H}^{(s=-1)}_\alpha (m_i,\nu_i) =\exp \left( -2S_1^\alpha(m_i,\nu_i)\right)\;.
\end{aligned} \label{Handle gluing operator}
\end{equation}
To avoid clutter, we omit the superscript $(s)$ in $\CH_\a$ and $\CF_\a $ when $s=-1$.
We define 
\begin{equation}
\begin{aligned}
&S_0^\alpha (m_i,\nu_i) = (\mathcal{W}_0(\vec Y;m_i,\nu_i) -2\pi i \vec n_\alpha \cdot \vec Y)|_{\vec Y=\log \vec y_\alpha}\ , \\
&S_1^\alpha (m_i,\nu_i) = \left.\left( \mathcal{W}_1(\vec Y;m_i,\nu_i)-\frac{1}{2} \log \det \left(\frac{\partial^2\mathcal{W}_0}{\partial \vec Y \partial \vec Y} \right)  \right)\right|_{\vec Y=\log \vec y_\alpha}\ , \label{S0 and S1}
\end{aligned}
\end{equation}
with
\begin{equation}
	\partial_{\vec Y} \mathcal{W}_0(\vec Y;m_i,\nu_i)|_{\vec Z =\log \vec y_\alpha} = 2\pi i \vec n_\alpha\ . \label{vecn}
\end{equation}
Due to the overall factor ambiguity of $\mathcal{Z}_b$ in \eqref{ambiguity of Zb}, the fibering/handle-gluing operators have following overall factor ambiguities
\begin{align}
	\begin{split}
		&\exp \left(-\frac{i}{4\pi} (k_{FF}^{ij}m_i m_j-  \nu_i \nu_j) - \frac{i \pi}4 (2 k_{FR}^i \nu_i +k_{RR})\right)  \textrm{ for } \mathcal{F}_\a\;,
		\\
		&\exp \left( -k_{FF}^{ij} \nu_i  (m_j+i \pi  \nu_j )-k^i_{FR} (m_i+2 i \pi  \nu_i )-i \pi k_{RR}\right)   \textrm{ for } \mathcal{H}_\a\;. \label{ambiguity in H and F}
	\end{split}
\end{align}
 As commented around \eqref{twisted ptns from SBE}, only $(\CF_\alpha^{(s=-1)})^2$ is well-defined and the above relation for $\CF_\alpha$ should be understood as $(\CF_\alpha^{(s=-1)})^2  = \exp \left(i \frac{S_0- m\partial_m S_0}{\pi}\right)$ for general cases. But for  some cases, the twisted partition function $\mathcal{Z}^{(s)}_{\CM_{g,p \in 2\mathbb{Z}}}$ does not depend on the spin-structure choices. In the case, the above relation should be understood as  $\CF^{(s=+1)}=\CF^{(s=-1)} = \exp \left(i \frac{S_0 - m\partial_m S_0}{2\pi} \right)$.

The partition function $\mathcal{Z}_{b}$ at $b=1$, which corresponds to a round 3-sphere,  i.e. $\CZ^{(s=1)}_{\CM_{(g,p)=(1,1)}}$, can also be written as a sum over the Bethe-vacua, 
\begin{equation}
	\mathcal{Z}_{b=1} (\vec m=0,\vec \nu) = \mathcal{Z}_{b=1} \left(\vec m=2\pi i  (\vec \nu-\vec \nu_0),\vec \nu_0\right) = \sum_{\alpha} \frac{\mathcal{F}_\alpha(\vec m=2\pi i  (\vec \nu-\vec \nu_0),\vec \nu_0)}{\mathcal{H}_\alpha(\vec m=2\pi i  (\vec \nu-\vec \nu_0),\vec \nu_0)}\;.
	\label{round 3-sphere from Bethe-sum}
\end{equation} 
The first equality follows from the relation in \eqref{Zb-at-different-nu} with $\hbar = 2\pi i $.
Here, $\vec \nu_0$ denotes a special choice of $\nu$ satisfying the condition 
\begin{align}
R_{\vec \nu_0}+2j_3 \in 2\mathbb{Z}\;, \label{condition for nu0}
\end{align}
for all local BPS operators \cite{Gang:2019jut}. Namely, the superconformal index at $\vec \nu=\vec \nu_0$ will only contain the terms with integer power of $q$, which is the fugacity conjugate to $\frac{R_{\vec \nu}}{2}+j_3$. It is argued in \cite{Gang:2019jut} that  $(\CH^{(s=1)},\CF^{(s=1)}) =(\CH^{(s=-1)},\CF^{(s=-1)})$ at $\vec \nu= \vec \nu_0$ for non-trivial SCFTs, and we thus have the formula given in \eqref{round 3-sphere from Bethe-sum} in this case. The Bethe sum formula makes it much easier to compute the round 3-sphere partition function $\CZ_{b=1}$ than the Coulomb branch integral formula.

The $S^3$ partition function $\mathcal{Z}_{b=1}$ can be used to obtain the exact superconformal R-symmetry $R_{\rm IR} = R_{\vec{\nu}_{\rm IR}}$ of IR CFT. This is the F-extremization principle in \cite{Jafferis:2010un} which can be summarized as
\begin{align}
	\textrm{F-extremization : } F(\vec{\nu}) := -\log |\CZ_{b=1} (m_i=0, \vec{\nu})|  \textrm{ is extremized at $\vec \nu = \vec \nu_{\textrm{IR}}$.} \label{F-maximization}
\end{align}
\paragraph{Effect of $\mathbb{Z}_2$ 1-form symmetry gauging} So far we have studied the BPS partition functions for $\CT[N,A;B]/SU(2)_p$ theory. The theory is identical to $\CT[N_{pA+B}]$ for even $A$, but the two theories are different for odd $A$. To study the BPS partition functions of $\CT[N_{pA+B}]$ theory for odd $A$, one needs to understand the effect of gauging 1-form $\mathbb{Z}_2$ symmetry on the BPS partition functions.  For this,  we shall use the general formula given by
\begin{align}
	\mathcal{Z} [\CT/{\mathbb{Z}_2} \textrm{ on }\mathcal{B}] =\frac{|H^0(\mathcal{B},\mathbb{Z}_2)|}{|H^1 (\mathcal{B},\mathbb{Z}_2)|} \sum_{[\beta_2] \in H^2 (\mathcal{B}, \mathbb{Z}_2)} \CZ[\mathcal{T} \textrm{ on }\CB ; [\beta_2]]\;. \label{Z under 1-form gauging}
\end{align}
Here $\CT$ is a 3D field theory with non-anomalous $\mathbb{Z}_2$ one-form symmetry and $\CT/\mathbb{Z}_2$ is the theory after gauging the symmetry. $\CB$ is a general curved background  and $\CZ[\mathcal{T} \textrm{ on }\CB ; [\beta_2]]$ is the partition function of $\CT$ on $\CB$ with background flat $\mathbb{Z}_2$ 2-form $[\beta_2]$ coupled to the 1-form symmetry. 

By applying the formula to $\CT = \frac{\CT[N,A;B]}{SU(2)_p}$ and $\CB= S^3_b$ (squashed 3-sphere), one has
\begin{align}
	(\mathcal{Z}_b(m_i, \nu_i) \textrm{ for } \CT[N_{pA+ B}]  \textrm{ with odd $A$}) = \begin{cases}
		2 \CZ_b^{N_p}(m_i, \nu_i)\;, \quad p \in 2\mathbb{Z}
		\\
		\sqrt{2} \CZ_b^{N_p}(m_i, \nu_i)\;, \quad p \in 2\mathbb{Z}+1
	\end{cases}
\end{align}
Here we use the fact that $|H^1(S^3, \mathbb{Z}_2)| =| H^2 (S^3, \mathbb{Z}_2)|=1$, $|H^0 (S^3,\mathbb{Z}_2)|=2$ and \\$\mathcal{Z}[U(1)_{-2} \textrm{ on }S^3_b] = \frac{1}{\sqrt{2}}$. Since the two partition functions are related to each other by an overall numerical factor, the gauging does not affect the F-extremization. 

Gauging the 1-form $\mathbb{Z}_2$ symmetry affects the summation range of $m_x$ (when $A$ is odd) in the superconformal index in \eqref{Index for Np} as
\begin{align}
	& \begin{cases}
		m_x \in  \mathbb{Z}_{\geq 0 }\;, \quad \textrm{for }\frac{\CT[N, A ;B]}{SU(2)_p}
		\\
		m_x \in  \frac{1}2\mathbb{Z}_{\geq 0 }\;, \quad  \textrm{for }\CT[N_{pA+B}] =  \frac{\CT[N, A;B]}{``SO(3)"_p} 
		\label{Index for Np-2}
	\end{cases}
\end{align}
For odd $A$, all local operators have integer spins under the $SU(2)$ gauge symmetry and the Dirac quantization requires that $m_x \in \mathbb{Z}/2$ instead of $m_x \in \mathbb{Z}$. 
The above summation range  reflects the fact that  $SO(3)$ principle bundle on $S^2$ admits half-integer monopole fluxes, $m_x \in \mathbb{Z}/2$, while $SU(2)$ bundle only admits integer monopole fluxes, $m_x \in \mathbb{Z}$.

The superconformal index is a BPS partition function on $\CB =S^2\times S^1$ and $H^2(\mathcal{B},\mathbb{Z}_2) = \mathbb{Z}_2 = \{1,-1\}$. Therefore, unlike in higher dimensions, 3D superconformal index is sensitive to the gauging of 1-form symmetry. The index obtained by summing over  all  $m_x \in \mathbb{Z}_{\geq 0 }$ can be identified with $\CZ[\CT \textrm{ on } S^2\times S^1 ; [\beta_2]=1]$  in \eqref{Z under 1-form gauging} while the index from all $m_x \in \mathbb{Z}_{\geq 0} +\frac{1}2 $ is $\CZ[\CT \textrm{ on } S^2\times S^1 ; [\beta_2]=-1]$.  For odd $p$, the $\mathbb{Z}_2$ gauging does not affect the superconformal index since there is no contribution from $m_x \in \mathbb{Z}+\frac{1}2$. This is indeed expect from the relation in \eqref{T[Np-for-odd-p]}.  For some cases, the 1-form $\mathbb{Z}_2$ symmetry decouples at IR. In that case, the theory $\CT[N,A;B]/SU(2)_p$ is the same as $\CT[N,A;B]/``SO(3)"_p$, and the index gets the same contributions from $m_x \in \mathbb{Z}$ and $m_x \in \mathbb{Z}+\frac{1}2$. Then, we need to sum over only $m_x \in \mathbb{Z}$ for $\CT[N_{pA+B}]$ theory, otherwise the index will start with $2+O(q^{1/2})$. 

The twisted partition function is also affected after gauging the $\mathbb{Z}_2$ 1-form symmetry. For odd $p$, using the relation in \eqref{T[Np-for-odd-p]}, one has 
\begin{align}
	\begin{split}
		&\left(\mathcal{S}_{\rm BE} \textrm{ of } \frac{\CT[N,A;B]}{SU(2)_p}\right)   = \left(\mathcal{S}_{\rm BE} \textrm{ of }  \frac{\CT[N,A;B]}{``SO(3)_p"} \right)^{\otimes 2} \;,
		\\
		&\bigg{\{} \CH_\a \textrm{ of } \frac{\CT[N,A;B]}{SU(2)_{p\in 2\mathbb{Z}+1}}  \bigg{\}}= \bigg{\{} 2\CH_\a \textrm{ of } \frac{\CT[N,A;B]}{``SO(3)"_p}   \bigg{\}} \times \bigg{\{} 1,1 \bigg{\}}\;,
		\\
		&\bigg{\{} \CF_\a \textrm{ of } \frac{\CT[N,A;B]}{SU(2)_{p\in 2\mathbb{Z}+1}}  \bigg{\}}= \bigg{\{} \CF_\a \textrm{ of } \CT[(N)_{p A +B}] \bigg{\}} \times \bigg{\{} 1, e^{\frac{\pi i}2} \bigg{\}}\;.
	\end{split} \label{SBE for odd p}
\end{align}
This follows from that the topological theory $U(1)_2$ theory has two simple objects (or two Bethe-vacua) $\alpha=0$ and $\alpha=1$ with $S_{00} =S_{01}=1/\sqrt{2}$ (i.e. $\CH_{\alpha=0,1} = 2$) and topological spins $h_{\alpha=0}=0$ and $h_{\alpha=1}=\frac{1}4$ (i.e $\CF_{\a=1}/\CF_{\a=0} = \exp (\frac{1}4 \times 2\pi i )$). 

For even $p$, the effect of the gauging  is more subtle. The Bethe-vacua  in \eqref{Bethe-vacua set} can be divided into two classes, $\CS_{\rm BE} = S^{(+1)}_{\rm BE}\bigsqcup  S^{(-1)}_{\rm BE}$,
\begin{align}
	\left(\mathcal{S}^{Q = \pm 1}_{\rm BE} \textrm{ of } \frac{\CT[N,A;B]}{SU(2)_p}\right)  =  \left\{ (x, z)  \in \CS_{\rm BE}\;:\; \exp \left(\frac{1}2 \partial_X \CW_0 \right) = Q  \right\}\;. \label{SBEQ}
\end{align}
Then, the Bethe-vacua for the gauged theory are (when $p\in 2\mathbb{Z}$)
\begin{align}
	\begin{split}
		&\left(\mathcal{S}_{\rm BE} \textrm{ of } \CT[N_{pA+B}] = \big{(}\frac{\CT[N, A;B]}{SU(2)_p}\big{)}/\mathbb{Z}_2 \right)  = \CS^{\rm untwisted}_{\rm BE} \bigsqcup \CS^{\rm twisted}_{\rm BE} \quad \textrm{where}
		\\
		&\CS^{\rm untwisted}_{\rm BE} = \bigg{\{}(x,\vec{z}) \in \CS_{\rm BE}^{Q=1} \;:\; x +\frac{1}x\neq -(x+\frac{1}x)\bigg{\}}/\mathbb{Z}_2^{\rm 1-form}\textrm{ and }
		\\
		&\CS^{\rm twisted}_{\rm BE} = \bigg{\{}(x,\vec{z}) \in \CS_{\rm BE}^{Q=1} \;:\; x +\frac{1}x= -(x+\frac{1}x)\bigg{\}}^{\otimes 2}\;.
	\end{split} \label{SBEQ-2}
\end{align}
The $\mathbb{Z}_2$ 1-form symmetry acts as\footnote{The Bethe-vacua represents the supersymmetric vacua on two torus $\mathbb{T}^2$. Since the $H_1 (\mathbb{T}^2,\mathbb{Z}) = \mathbb{Z} \oplus \mathbb{Z}$, one can consider two $\mathbb{Z}_2$ 1-form symmetry charge operators , $Q_{(1,0)}$ and  $Q_{(0,1)}$,  supported on the two generators of the $H_1 (\mathbb{T}^2, \mathbb{Z})$. The set $\CS_{\rm BE}^{Q=1}$ selects the Bethe-vacua with $Q_{(0,1)}=1$ while the $Q_{(1,0)}$ transforms a Bethe-vacuum $(x,\vec{z})$ to another Bethe-vacuum $(-x, \vec{\tilde{z}})$.}
\begin{align}
\mathbb{Z}_2^{\rm 1-form} \;:\; (x,\vec{z})\leftrightarrow (-x,\vec{\tilde{z}})\;.
\end{align}
Like the $\mathbb{Z}^{\rm Weyl}_2$ case, the $\vec{\tilde{z}}$ is in general a nontrivial function of $\vec{z}$.  For the Bethe-vacua invariant under the $\mathbb{Z}_2^{\rm 1-form}$ modulo the $\mathbb{Z}_2^{\rm Weyl}$, i.e $x= \pm i$, we count them with multiplicity 2. For some cases,  all the Bethe-vacua in $\CS_{\rm BE}$ are invariant under the $\mathbb{Z}_2^{\rm 1-form}$. This implies that 1-form symmetry decouples at IR. In this case,  the two theories $\CT[N_{pA+B}]$ and $\CT[\frac{\CT[N,A;B]}{SU(2)_p}]$ are identical and thus they have the same Bethe vacua.
The handle-gluing and fibering operators of the Bethe-vacua are
\begin{align}
	\begin{split}
		&\left(\CH_\a \textrm{ of } \CT[N_{pA +B}]_{p \in 2\mathbb{Z}} \right) = \begin{cases} \frac{1}4 \times \left(\CH_\a \textrm{ of } \frac{\CT[N,A;B]}{SU(2)_p} \right), \;\; \a \in \CS^{\rm untwisted}_{\rm BE} 
			\\
			\left(\CH_\a \textrm{ of } \frac{\CT[N,A;B]}{SU(2)_p} \right), \;\; \a \in \CS^{\rm twisted}_{\rm BE} 
		\end{cases}
		\\
		&\left(\CF_\a \textrm{ of } \CT[N_{pA +B}]_{p \in 2\mathbb{Z}} \right)  =  \left(\CF_\a \textrm{ of } \frac{\CT[N,A;B]}{SU(2)_p}\right) \;.  \label{SBEQ-3}
	\end{split}
\end{align}
For some cases, only  $(\CF_\alpha)^2$ (instead of $\CF_\a$) is  well-defined in the $\CT[N_{pA+B}]$ theory since the fibering operator for a Bethe-vacuum with $Q=+1$ can change its sign under the action of $\mathbb{Z}^{\rm 1-form}_2$. It implies that $\CF_\alpha^{(s=-1)} \neq \CF_\alpha^{(s=1)}$ and the twisted partition functions $\mathcal{Z}_{\CM_{g,p\in 2\mathbb{Z}}}^{(s)}$ depends on the spin-structure choice. 

\paragraph{3D-3D relations for 3D index and $\CS_{\rm BE}$} The 3D-3D relation tells us that the supersymmetric partition functions of the $\CT[M]$ theory are related to the topological invariants of the $SL(2,\mathbb{C})$ Chern-Simons theory on the associated 3-manifold $M$. For example, 
\begin{align}
	\begin{split}
	& \big{(}\CI_q (u_i=1, \vec{\nu}= \vec{\nu}_{\rm geo}) \textrm{ of $\CT[M]$ theory}\big{)} = \begin{cases} \CI_M (q)\;, \quad \cap_{\rho} \textrm{Inv}(\rho)=1
		\\
	    \frac{1}2 \CI_M (q)\;, \quad \cap_{\rho} \textrm{Inv}(\rho)=\mathbb{Z}_2
		\end{cases}
	\\
	&
	\big{(}\CS_{\rm BE}  \textrm{ of $\CT[M]$ theory}\big{)} 
	\\
	&= \begin{cases} \big{\{} [\rho] \in \chi^{SL(2,\mathbb{C})}_{\rm irred} \; \textrm{counted with multiplicity }|\textrm{Inv}(\rho)|\big{\}}\;, \quad \cap_{\rho} \textrm{Inv}(\rho)=1
		\\
		\chi^{SL(2,\mathbb{C})}_{\rm irred} \;, \quad \cap_{\rho} \textrm{Inv}(\rho)=\mathbb{Z}_2\;.
	 	\end{cases}
	  \label{3D-3D relations}
	 \end{split}
\end{align}
For the case when $\cap_{\rho} \textrm{Inv}(\rho)=\mathbb{Z}_2$, the corresponding $\CT[N,A;B]/SU(2)_p$ theory has a decoupled $\mathbb{Z}_2^{\rm 1-form}$ symmetry and the 3D-3D relation should be treated separately as discussed above.

Since the BPS partition functions are protected along RG-flows, they can be used to scrutinize the physics of the low energy phases. We now explain a number of possible scenarios for the IR phases of $\mathcal{T}[M]$ captured by the BPS partition functions. 

  \paragraph{Spontaneous SUSY Breaking} We start with the IR phases with spontaneously broken supersymmetries.
 As explained above, the superconformal index counts local BPS operators in radial quantization. In particular, the identity operator corresponding to a unique supersymmetric ground state in the radially quantized theory is the leading contribution to the index, which cannot be cancelled by other operator contributions due to the unitarity bounds. The absence of this SUSY ground state and its index contribution therefore implies spontaneous SUSY breaking. As a result, all the BPS states in this case are paired up to form long multiplets and their contributions to the index are all cancelled.
 This means the superconformal index becomes trivial, i.e. $\mathcal{I}_q=0$.   Twisted indices $I_g$ also vanish for all $g$ since there are fermionic zero modes associated to the broken SUSYs, which implies that $\CS_{\rm BE} =\emptyset$ from \eqref{Bethe-sum for Ig}. 
 Therefore, the supersymmetry of $\mathcal{T}[M]$ is spontaneously broken at low energy if and only if 
 \begin{align}
 (\textrm{Spontaneous SUSY breaking}) \Leftrightarrow (\mathcal{I}_q =0) \Leftrightarrow  (\mathcal{S}_{\rm BE} = \emptyset)\;.
 \end{align}

The 3D-3D relation for the superconformal index in (\ref{3D-3D relations}) implies that the geometric invariant $\mathcal{I}_M(q)$ (or 3D index) becomes trivial in this case. This means that the 3-manifold is of empty type. On the other hand, the relation for $\CS_{\rm BE}$ implies the 3-manifold does not have any irreducible $SL(2,\mathbb{C})$ flat connection, which is another implication that the 3-manifold is of empty type.

\paragraph{Gapped Phase}
The gapped phase is another interesting IR phase of $\mathcal{T}[M]$. One distinguished property of the gapped phase is that it has a unique ground state on $S^2$ and all other local states are gapped out at low energy. From this, we assert that a 3D $\CN=2$ gauge theory has a mass gap only if
  \begin{align}
  	\CI_q (u_i,\nu_i) =1\;.
  	\label{eq:gapped}
  \end{align}
The IR phase in this case is described by a unitary topological field theory (TFT).
This condition can be satisfied only when $M$ is of finite/real type. Otherwise, the index at $u_i=1$ and $\vec{\nu} = \vec{\nu}_{\rm geo}$ is divergent or it shows exponentially growing behavior in the Cardy limit $q\rightarrow 1$ according to the 3D-3D relation \eqref{3D-3D relations} and the analysis below \eqref{Cardy limit}. For a 3-manifold $M$ of  finite and real type, the superconformal index in the specialization always becomes  just $1$. But it does not guarantee that the superconformal index becomes 1 at general value of $(u_i, \nu_i)$. For these finite/real 3-manifolds, one can assign a  topological field theory ${\rm TFT}[M]$  to the 3-manifold as studied in \cite{Cho:2020ljj}. Furthermore, it was claim that the $\CT[M]$ has  a mass gap only when the topological theory is  unitary.  In that case, the superconformal index is $(m_i, \nu_i)$-independent $1$, and the IR physics is described by the unitary topological field theory ${\rm TFT}[M]$. For the case when the ${\rm TFT}[M]$ is non-unitary, the corresponding IR phase will be discussed below. 

The physics of  a bosonic unitary TQFT can be described by a mathematical framework called {\it modular tensor category} (MTC). Let us provide a simple dictionary between MTC data and BPS partition functions that we can use to identify the topological theory appearing at  IR. Firstly, simple objects (anyons) in TQFT are in one-to-one with Bethe-vacua.
\begin{align}
	\alpha \;(\textrm{simple objects in MTC}) \quad \leftrightarrow \quad (x_\a,\vec{z}_\alpha) \in \CS_{\rm BE}\;.
\end{align}
 The modular  S- and T-matrices in MTC can be computed using the handle-gluing/fibering operators of $\mathcal{T}[M]$. We compute them as
 \begin{align}
 S_{0\alpha}^2 =( \mathcal{H}_\alpha (\vec{\nu}_{\rm geo}))^{-1}\;, \quad T_{\a\a}/T_{00} =  \mathcal{F}_\a (\vec{\nu}_{\rm geo})  /F_{\a=0} (\vec{\nu}_{\rm geo}) \;.  \label{S,T from H,F}
 \end{align}
Here $\alpha=0$ corresponds to the trivial simple object in TQFT, and the corresponding Bethe-vacuum $(x_\alpha,\vec{z}_\alpha)_{\alpha=0}$ is chosen to satisfy 
\begin{align}
	|\CZ_b|  = S_{00}= \frac{1}{\sqrt{\CH_{\a=0} (\vec{\nu}_{\rm geo})} } \;. \label{S3 ptn in TQFT}
\end{align}
For a gapped phase, the squashed 3-sphere partition function is  $b$-independent, modulo the local counter-terms in \eqref{ambiguity of Zb}, and its absolute value should equal to the $S_{00}$ \cite{witten1989quantum}.  The above relations make sense since  the handle-gluing and fibering operators for a gapped phase do not depend on the continuous real mass parameters $m_i$, and they satisfy
\begin{align}
	|\CF_\a|=1 \textrm{ and } \CH_\a \in \mathbb{R}_+ \textrm{  for all Bethe-vacua $\a$}\; \label{F, H in TQFT}
\end{align}
in a proper choice of  the local counter-terms $(k_{FF}^{ij}, K_{FR}^i, K_{RR})$  in \eqref{ambiguity in H and F}.
The topological spin $h_\alpha \in \mathbb{R}/\mathbb{Z}$ of an anyon can be read off from the T-matrix as
\begin{align}
e^{2\pi i h_\alpha} = T_{\a\a}/T_{00}\;. \label{topological spin from T-matrix}
\end{align}
For some cases, as seen below \eqref{SBEQ-3}, only $(\CF_\alpha)^2$ (instead of $\CF_\alpha$) is well-defined. In that case, the partition functions on $\CM_{g,p \in 2\mathbb{Z}}$ depend on the spin-structure choices, and thus the topological theory is actually spin (fermionic) TQFT. For spin TQFTs,  the topological spin $h_\alpha$ is well defined only up to mod $1/2$ (instead of $1$).

We can extract some constraints on the handle-gluing operators for a unitary TQFTs from the BPS partition functions.
Since there is a unique ground state in a TQFT on $S^2$, the twisted index at $g=0$ and the handle-gluing operators must satisfy
\begin{align}
I_{g=0} = \sum_{\a} (\CH_\a)^{-1}= 1\;. \label{sum of 1/H in TQFT}
\end{align}
In addition, the modular S-matrix for a unitary TQFT possess a unitarity condition providing an inequality among the handle-gluing operators as
\begin{align}
(\textrm{Unitarity})\;:\; |S_{00}| \leq |S_{0\a}| \; \Rightarrow \;  \CH_{\alpha=0} \geq \CH_{\a}\;. \label{unitarity of TQFT}
\end{align}

\paragraph{Rank-0 $\CN=4$ SCFT Phase}
Interestingly, the ${\rm TFT}[M]$ associated with finite/real 3-manifold $M$ can be  non-unitary as found in \cite{Cho:2020ljj}. In that case, we call the 3-manifold is of finite/real/non-unitary type. For that case, it is claimed that the  $\CT[M]$ does  not have a mass gap. Instead, the theory flows to an $\CN=2$ SCFT with a $U(1)$ flavor symmetry, which contains the non-unitary TQFT structure in a sub-sector. Modular S- and T-matrices for the TQFT can be obtained using the expressions in  \eqref{S,T from H,F} and \eqref{S3 ptn in TQFT}. One crucial difference with the gapped case is that i) the $\CH$, $\CF$ as well as $\CZ_{b}$ depend on $(m, \nu)$ associated with the $U(1)$ symmetry, and the modular matrices appear only at $m=0$ and $\nu=\nu_{\rm geo}$, and ii) the modular S-matrix violates the unitarity condition in \eqref{unitarity of TQFT}. Recently, it is found that non-unitary TQFTs  appear in degenerate limits of 3D $\CN=4$ rank-0 SCFTs \cite{Gang:2021hrd}. The degenerate limit is the Coulomb/Higgs branch limits studied in  \cite{Razamat:2014pta} to compute the Hilbert-series.

Motivated by these curious observations, we claim that the 3D $\mathcal{T}[M]$ theory associated with the finite/real/non-unitary 3-manifolds $M$ flows at low energy to a rank-0 SCFT with an enhanced $\mathcal{N}=4$ superconformal algebra. 

In this case, as we will see with explicit examples below, the UV field theory constructed by the method in Section \ref{subsec: UV field theory} always has a $U(1)$ flavor symmetry, which is an accidental flavor symmetry from the point of view of the class R theory construction. Hence, we can turn on the real mass/fugacity $m/u$ and the mixing parameter $\nu$ for the flavor symmetry in the BPS partition function computations using the UV field theory description.
For convenience, we shall set the mixing parameter $\nu=1$ as the geometrical $R$-charge choice, i.e. 	$\nu_{\rm geo} = 1$.
As we will explicitly check from the F-extremizations on various examples,  it turns out that the superconformal R-charge $R_{\nu_{\rm IR}}$ in the low energy theory is always at $\nu_{\rm IR} =1 \pm 1$. Then, by a proper redefinition of the mixing parameter as $\nu \leftarrow 2-\nu$ or $\nu \leftarrow \nu$, one can always set
\begin{align}
\nu_{\rm IR} = 0 \;, \quad \nu_{\rm geo}=1\;.
\end{align}

Our claim is that the accidental $U(1)$ flavor symmetry becomes the $U(1)_{\rm Axial}$ subgroup of the $SO(4)\simeq SU(2)_L \times SU(2)_R$ R-symmetry of the enhanced $\mathcal{N}=4$ superconformal algebra in the low energy theory. The charge $J_{\rm Axial}$ for this accidental symmetry is identified with the axial combination of the enhanced R-charges as
\begin{align}
	J_{\rm Axial} = (J^3_L-J^3_R) \in \mathbb{Z}\;,
\end{align}
where $J^3_{L/R}$ are the Cartan charges of $SU(2)_{L/R}$ normalized as $J^3_{L/R}\in \frac{\mathbb{Z}}2$. The mixing parameter $\nu$ is then
\begin{align}
	R_\nu = (J^3_L+J^3_R) + \nu (J_L^3 -J_R^3)\;.
\end{align}

To confirm the SUSY enhancement in the low energy theory, we first compute the superconformal index using the R-charge of the IR $\mathcal{N}=2$ superconformal algebra. The additional SUSY current multiplets for the enhanced $\CN=4$ superconformal algebra contains a $1/4$ BPS operator with $j_3=R = 1$ and $\Delta = 2$ (see, for example, \cite{Cordova:2016emh}), and this operator contributes to the superconformal index at order $q^{3/2}$. Therefore, the existence of the enhanced SUSY ensures the following pattern of the index:  
\begin{align}
	\CI_q (u,\nu=0) \ni -\left(u+\frac{1}{u}\right)q^{3/2}\;. \label{SUSY enhancement from index}
\end{align}
The appearance of the above term is a strong evidence\footnote{But it can not be a proof since the term $-q^{3/2}$ could also come from a chiral primary operator with $R =\Delta =3$.} that the the theory has a SUSY enhancement to $\CN=4$ and the accidental $U(1)$ symmetry becomes the axial $U(1)_{\rm Axial}$ subgroup of the enhanced R-symmetry.

We further conjecture that the resulting IR $\CN=4$ SCFT is a rank-0 theory having no Higgs/Coulomb branches. The 3D $\CN=4$ rank-0 SCFTs are extensively studied in \cite{Gang:2021hrd} and  they propose several non-trivial relations  among SUSY partition functions of the rank-0 SCFTs.  We list some of the relations which we will use in the next section,
\begin{align}
	&\bullet \CI_q(u=1,\nu=\pm1) = 1\;,       \label{properties of ran0 SCFT-0}
	\\
	&\bullet |\mathcal{F}_\alpha(m,\nu=\pm1)|=1, \;  \mathcal{H}_\alpha(m,\nu=\pm1) \in \mathbb{R}_+ \textrm{ and }\; \sum_\alpha \big(\mathcal{H}_\alpha(m=0,\nu=\pm1)\big)^{-1}=1\;,  \label{properties of ran0 SCFT-1}
	\\
	&\bullet \exists  \textrm{ a Bethe-vacuum } \alpha=0 \textrm{ such that } |\mathcal{Z}_{b=1} (m=0,\nu = \pm 1)| = \mathcal{H}^{-1/2}_{\alpha=0}(m=0,\nu=\pm1)\;,  \label{properties of ran0 SCFT-2}
	\\
	&\bullet |\mathcal{Z}_{b=1} (m=0,\nu = 0)| = \textrm{min}_\alpha[\mathcal{H}_{\alpha}(m=0,\nu=\pm1)^{-1/2}]\;.
     \label{properties of ran0 SCFT-3}
\end{align}
The 1st equation implies that the theory has no Coulomb/Higgs branches since the index in the degenerate limit, $u=1$ and  $\nu \rightarrow \pm 1$, counts Coulomb/Higgs branch operators and their descendants \cite{Razamat:2014pta}. 
The 2nd equation implies the emergence of a pair of TQFT structures in the degenerate limits where $m=0$ and $\nu \rightarrow \pm 1$. This can be compared with the relations  \eqref{F, H in TQFT} and \eqref{sum of 1/H in TQFT} for unitary TQFTs. The 3rd one implies that the emergent TQFT structure  in the degenerate limit is of non-unitary type unlike the gapped cases in \eqref{S3 ptn in TQFT}.
The emergence of non-unitary TQFT structures in the $\CN=4$ rank-0 SCFTs was already observed in \cite{Gang:2021hrd} and dubbed as (rank-0  SCFT)/(non-unitary TQFT) correspondence. The last equation is one of (and the most surprising) dictionary of the correspondence proposed in that work. Thus, the confirmation of the above equations not only provides a strong evidence for the emergence of 3D $\CN=4$ rank-0 theory at IR but also non-trivial test for the dictionary.

\section{Examples}\label{sec:examples}
In this section, we check the our proposal in \eqref{proposal for IR phases} using non-hyperbolic 3-manifolds $M$ obtained by exceptional Dehn fillings on five simplest 1-cusped hyperbolic manifolds ($m003, m004= S^3\backslash \mathbf{4}_1, m006,m007,m009$) and $m015=S^3\backslash \mathbf{5}_2$.\footnote{Dehn fillings on $m005$ and $m008$ always yield non-hyperbolic manifolds.} 

\begin{table}[h]
	\centering
	\begin{tabular}{ |c|c|c|c| } 
		\hline
		 $M$ & Topological type & IR phase & UV  theory
		 \\ 
		\hline
		$(S^3 \backslash \mathbf{4}_1)_{p\mu+\lambda : 0\leq |p|\leq 3}$  & Finite/Real/Unitary & Gapped/TQFT &  Sec. \ref{subsec : Field for 41p}
		\\
		$(S^3 \backslash \mathbf{4}_1)_{p\mu+\lambda : |p|= 4}$  & Infinite & $\CN=2$ SCFT with CPOs&  
		\\ 
		\hline
		$(S^3 \backslash \mathbf{5}_2)_{p\mu+\lambda : 1\leq p\leq 3}$  & Finite/Real/Non-unitary & $\CN=4$ rank-0 SCFT &  Sec. \ref{subsec : field theory for 52p}
		\\
		$(S^3 \backslash \mathbf{5}_2)_{p\mu+\lambda : p=0,4}$  & Infinite &  $\CN=2$ SCFT with CPOs &  
		\\ 
		\hline
		$(m003)_{pA_{\rm sp}+B_{\rm sp} : p=1,-2}$  & Finite/Real/Unitary & Gapped/TQFT &  Sec. \ref{subsec : field theory for m003p}
		\\
		$(m003)_{pA_{\rm sp}+B_{\rm sp} : p=0,-1}$  & Empty & SUSY broken &  
		\\
		\hline
		$(m006)_{pA_{\rm sp}+B_{\rm sp} : p=0}$  & Empty & SUSY broken &  Sec. \ref{subsec : field theory for m006p}
		\\
		$(m006)_{pA_{\rm sp}+B_{\rm sp} : p=\pm 1}$  & Finte/Real/Unitary & Gapped/TQFT &  
		\\
		$(m006)_{pA_{\rm sp}+B_{\rm sp} : p=-2}$  & Finte/Real/Non-unitary & $\CN=4$ rank-0 SCFT &  
		\\
        $(m006)_{pA_{\rm sp}+B_{\rm sp} : p=-3}$  & Infinite & $\CN=2$ SCFT with CPOs &  
        \\
        \hline
        $(m007)_{pA_{\rm sp}+B_{\rm sp} : p=0,2}$  & Finte/Real/Unitary & Gapped/TQFT &    Sec. \ref{subsec : field theory for m007p}
        \\
        $(m007)_{pA_{\rm sp}+B_{\rm sp} : p=-2,-1,1}$  & Finte/Real/Non-unitary & $\CN=4$ rank-0 SCFT &  
        \\
        $(m007)_{pA_{\rm sp}+B_{\rm sp} : p=-3}$  & Infinite & $\CN=2$ SCFT with CPOs & 
        \\
        \hline
        $(m009)_{pA_{\rm sp}+B_{\rm sp} : -2\leq p\leq 2}$  & Finte/Real/Unitary & Gapped/TQFT &    Sec. \ref{subsec : field theory for m009p}
        \\
        $(m009)_{pA_{\rm sp}+B_{\rm sp} : |p|=3}$  & Infinite & $\CN=2$ SCFT with CPOs & 
        \\
        \hline
	\end{tabular}
	\caption{IR phases of class R theories associated to non-hyperbolic 3-manifolds obtained by exceptional Dehn fillings on $S^3\backslash \mathbf{4}_1, S^3\backslash \mathbf{5}_2, m003,m006,m007$ and  $m009$. }
\end{table}

\subsection{Example : $M=(S^3\backslash \mathbf{4}_1)_{p\mu+\lambda} = (m004)_{pA_{\rm SP}+B_{\rm SP}}$} 
\subsubsection{Field theory} \label{subsec : Field for 41p}
$\CK=\mathbf{4}_1$ is the figure-eight knot, the twist knot with two half-twists. We denote its complement in 3-sphere by $S^3\backslash \mathbf{4}_1$, which is a hyperbolic cusped 3-manifold. The corresponding 3D class R theory  is \cite{Dimofte:2011ju}
\begin{align}
\begin{split}
&\CT[(S^3\backslash \mathbf{4}_1)_{p\mu+\lambda}] = \frac{\CT[S^3\backslash  \mathbf{4}_1, \mu;\lambda]}{``SO(3)"_p}\;, \textrm{ with }
\\
&\CT[S^3\backslash \mathbf{4}_1,\mu;\lambda] = (U(1)_{0}  \;\CN=2 \textrm{ gauge theory  coupled to  two chirals of charge +1})\;.  \label{Field theory  for 41}
\end{split}
\end{align}
The $\CT[S^3\backslash \mathbf{4}_1, \mu;\lambda]$ has $SU(2)_\Phi \times U(1)_{\rm top}$ which is enhanced to $SU(3)$ at IR \cite{Benini:2018umh,Gaiotto:2018yjh,Gang:2018wek}.  The $SU(2)_A$ subgroup in \eqref{su(2)A} of $SU(3)$ we are gauging is chosen such that
\begin{align}
(\mathbf{3} \textrm{ of }SU(3)) = (\textrm{adjoint of } SU(2)_A)\;. \label{su(2)A in 41}
\end{align}
After gauging the $``SO(3)"$ symmetry with CS level $p$, the resulting $\CT[(S^3\backslash \mathbf{4}_1)_{p\mu+\lambda}] $ is expected not to have any continuous flavor symmetry for generic $p$. 

\subsubsection{SUSY partition functions}
The UV field theory $\CT[(S^3\backslash \mathbf{4}_1)_{p \mu+\lambda}]$ above does not have any flavor symmetry, i.e. $\textrm{rank}(G_F)=0$, and the supersymmetric partition functions do not depend on  real masses/fugacities or R-symmetry mixing parameters.
\paragraph{Squashed 3-sphere partition function} Using localization, the partition function for the theory $\frac{\CT[S^3\backslash \mathbf{4}_1,\mu;\lambda]}{SU(2)_p}$ is given as
\begin{align}
	\begin{split}
\mathcal{Z}^{(4_1)_p}_b =&\frac{1}2 \int \frac{dXdZ}{(2\pi \hbar)} \CI^{(4_1)_p}_\hbar (X,Z) \quad \textrm{with}
\\
\CI^{(4_1)_p}_\hbar (X,Z)   = &  (4 \sinh X \sinh \frac{2\pi i X}\hbar ) \psi_\hbar (Z+2X) \psi_\hbar  (Z)
\\
 & \times  \exp \left( \frac{(2p+8)X^2+Z^2+8 X Z - (2\pi i +\hbar)(2X+Z) }{2\hbar}\right)\;.  \label{squashed 3-sphere for 41}
 \end{split}
\end{align}
Here, $\psi_{\hbar} $ is the quantum dilogarithm function in \eqref{QDL}. 

\paragraph{Superconformal index} The index is
\begin{align}
	\begin{split}
		\CI_q^{(4_1)_p} &= \sum_{(m_x, m_z)}\oint_{|u_x|=1}\frac{du_x}{2\pi i u_x} \oint_{|u_z|=1}\frac{du_z}{2\pi i u_z} \Delta (m_x, u_x) \CI_{\Delta}(m_z+2m_x, u_z u^2_x)
		\\
		& \quad \quad  \times   \CI_{\Delta}(m_z, u_z )  u_x^{(2p+8) m_x} u_z^{m_z} u_x^{4m_z}u_z^{4m_x} (-q^{1/2})^{-2m_x-m_z}\;.
		\label{Index for 41}
	\end{split}
\end{align}
Here $(m_z,u_z)$ and $(m_x,u_x)$ are (monopole flux, fugacity)  coupled to the $U(1)$ and $SU(2)$ gauge symmetry respectively.  The $\CI_\Delta (m,u)$ is the  tetrahedron index in \eqref{tetraheron index} and 
$\Delta(m_x, u_x)$ is the measure factor, given in \eqref{SCI : SU(2) measure}, coming from the $SU(2)$ vector multiplet. 
The summation range of the monopole fluxes is, see \eqref{Index for Np-2},
\begin{align}
	& \begin{cases}
		m_x \in  \mathbb{Z}_{\geq 0 },\;m_z \in \mathbb{Z}\;, \quad \textrm{for }\frac{\CT[S^3\backslash \mathbf{4}_1, \mu \;\lambda]}{SU(2)_p}
		\\
		m_x \in  \frac{1}2\mathbb{Z}_{\geq 0 },\;m_z \in \mathbb{Z}\;, \quad \textrm{for }\CT[(S^3\backslash \mathbf{4}_1)_{p\mu +\lambda }]=\frac{\CT[S^3\backslash \mathbf{4}_1, \mu \;\lambda]}{``SO(3)"_p} \ .
		\label{Index for 41-2}
	\end{cases}
\end{align}

\paragraph{Bethe-vacua and handle-gluing/fibering operator} Expanding the integrand $\CI^{(4_1)_p}_{\hbar} (X,Z)$ in the limit $\hbar \rightarrow 0$, see \eqref{W0 and W1 from integrand} and \eqref{asymptotic of QDL}, we have 
\begin{align}
	\begin{split}
&\CW^{(4_1)_p}_0 (X,Z)= \textrm{Li}_2 (e^{-Z-2X}) +  \textrm{Li}_2 (e^{-Z}) + (p+4) X^2 +\frac{1}2 Z^2 +  4XZ  - i \pi (2X+Z) \pm 2 \pi i X\;,
\\
&\CW^{(4_1)_p}_1 (X,Z)= - \frac{1}2 (2X+Z +\log(1-e^{-2X-Z}) +\log (1-e^{-Z}))+ \log \left( 2 \sinh X \right)\;.
\end{split}
\end{align}
Here we used the following asymptotic expansion (assuming $\textrm{Im}(\hbar) \neq 0$)
\begin{align}
	\log (2 \sinh(\frac{2\pi i X}\hbar )) \xrightarrow{\quad \hbar \rightarrow 0 \quad } \pm \frac{2\pi i  X}\hbar +(\textrm{exponentially suppressed})\;.
\end{align}
The overall sign depends on the sign of $\textrm{Im}(\hbar)$ but this choice is not important for our purpose. Extremizing the twisted superpotential $\CW^{(4_1)_p}_0$, we have
\begin{align}
	\begin{split}
&\left(\mathcal{S}_{\rm BE} \textrm{ of } \frac{\CT[S^3\backslash \mathbf{4}_1,\mu;\lambda]}{SU(2)_p}\right)  
\\
&= \left\{ (x, z) \;:\; (x^{p+2}z(x^2 z-1))^2=1,\;\frac{x^2(1-z)(x^2 z-1)}{z}=1\;, x^2 \neq 1\right\}/\mathbb{Z}^{\rm Weyl}_2 \ . \label{Bethe-vacua for 41p}
\end{split}
\end{align}
The Weyl $\mathbb{Z}_2$ acts as
\begin{align}
\mathbb{Z}^{\rm Weyl}_2 \;:\;(x,z) \; \leftrightarrow \;(1/x,\tilde{z})\;.
\end{align}
The handle-gluing and fibering operators are 
\begin{align}
	\begin{split}
&\left(\CH_\a \textrm{ of } \frac{\CT[S^3\backslash \mathbf{4}_1,\mu;\lambda]}{SU(2)_p}\right)   = \frac{2 x^2 \left((p-4) x^2 z^2-p+2 \left(4 x^2+1\right) z-4\right)}{\left(x^2-1\right)^2 z} \bigg{|}_{(x,z)=(x^{(\a)},z^{(\a)}) \in \mathcal{S}_{\rm BE}} \;,
\\
&\left(\CF_\a \textrm{ of } \frac{\CT[S^3\backslash \mathbf{4}_1,\mu;\lambda]}{SU(2)_p}\right)  =  \exp \left(\frac{i (\CW_0-2\pi i n^{(\a)}_x X -2\pi i n^{(\a)}_z Z)}{2\pi }\right) \bigg{|}_{(X,Z) = (\log x^{(\a)}, \log z^{(\a)})}\;. \label{handle/fiber for 41p}
\end{split}
\end{align}
Two integers $(n_x^{(\a)}, n_z^{(\a)})$ for each Bethe-vacuum $\a$ are chosen such that, see \eqref{vecn},
\begin{align}
\begin{split}
&\partial_X (\CW_0-2\pi i n^{(\a)}_x X -2\pi i n^{(\a)}_z Z)\big{|}_{(X,Z) = (\log x^{(\a)}, \log z^{(\a)})} =0\;,
\\
&\partial_Z (\CW_0-2\pi i n^{(\a)}_x X -2\pi i n^{(\a)}_z Z)\big{|}_{(X,Z) = (\log x^{(\a)}, \log z^{(\a)})} =0\;.
\end{split}
\end{align}
The theory $\frac{\CT[S^3\backslash \mathbf{4}_1,\mu;\lambda]}{SU(2)_p}$ has 1-form $\mathbb{Z}_2$ symmetry originated from the $\mathbb{Z}_2$ center symmetry of the gauged $SU(2)$ symmetry.

\subsubsection{IR phases}
According to \cite{2018arXiv181211940D}, the Dehn filled manifolds are
\begin{align}
	(S^3\backslash \mathbf{4}_1)_{p \mu + \lambda}   = 
	\begin{cases}
		\textrm{Torus bundle  with } \varphi = \begin{pmatrix}  2 & 1 \\ 1 & 1 \end{pmatrix}\;, \quad p=0
		\\
		\textrm{Atoroidal SFS $S^2 ((2,1),(3,1),(7,-6))$}\;, \quad p=\pm 1
		\\
		\textrm{Atoroidal SFS $S^2 ((2,1),(4,1),(5,-4))$}\;, \quad p=\pm 2
		\\
		\textrm{Atoroidal SFS $S^2 ((3,1),(3,1),(4,-3))$}\;, \quad p=\pm 3
		\\
		\textrm{Graph}\;, \quad p=\pm 4
		\\
		\textrm{Hyperbolic}\;, \quad |p|>4 \ .
	\end{cases}
\end{align}
According to \eqref{unitary/non-unitary 3-manifolds}, the  topological field theory $\textrm{TFT}[M]$ for the above Seifert fibered manifolds  are all  unitary. 
Combined with the mathematical facts, our proposal \eqref{proposal for IR phases} predicts that
\begin{align}
\CT[(S^3\backslash \mathbf{4}_1)_{p\mu+\lambda}]  =
	\begin{cases}
		\textrm{Unitary  TQFT}\;, \quad p=0, \pm 1, \pm 2, \pm 3
		\\
		\textrm{$\CN=2$ SCFT with $U(1)$ flavor symmetry and CPOs}\;, \quad p= \pm 4
		\\
		\textrm{$\mathcal{N}=2$ SCFT}\;, \quad  |p|>4 \ .
	\end{cases} \label{IR phase of 41-p}
\end{align}
Now let us carefully check the expected IR phases using the general methods outlined in Section \ref{sec : strategy}.

Using the formulas in \eqref{Index for 41} and \eqref{Index for 41-2}, one can compute the superconformal index and check that
\begin{align}
\begin{split}
&\left(\CI_q \textrm{ of } \frac{\CT[S^3\backslash  \mathbf{4}_1, \mu;\lambda]}{SU(2)_p}\right) =  \left(\CI_q \textrm{ of } \frac{\CT[S^3\backslash  \mathbf{4}_1, \mu;\lambda]}{``SO(3)"_p}\right) 
\\
&= 
\begin{cases}
	1\;, \quad p=0, \pm 1, \pm 2, \pm 3
	\\
	\left(\sum_{n=0}^\infty 1 \right) \times q^0 -q^2-q^3-2q^4-2q^5-3q^6+O(q^7)\;, \quad p= \pm 4
	\\
	\textrm{Non-trivial power series in $q^{1/2}$ starting with }1+\CO(q^{1/2})\;, \quad  |p|>4 \ .
\end{cases} \\
\end{split}
\end{align} 
For $p=\pm 4$, the $\mathbb{Z}_2$ 1-form symmetry decouples at IR and we need to sum over $m_x \in \mathbb{Z}$ (not $m_x \in \mathbb{Z}/2$) even for the $``SO(3)"$ gauged theory. See the discussion below Eq. \eqref{Index for Np-2}. 
The result is compatible with the expectation in \eqref{IR phase of 41-p}. For the $p=\pm 4$ cases, the index gets contribution of the form $\left( 1-q^{ 2m_x +1 }+(\textrm{higher order}) \right)$ from each $m_x \in \frac{1}2 \mathbb{Z}_{\geq 0}$ and the coefficient of $q^0$ goes to infinity as we sum over all values of  $m_x$. The contributions are from 1/2 BPS $SO(3)$ monopole operators with $R_{\rm geo}=0$. It implies that there should be an emergent Abelian symmetry  and the correct IR superconformal R-charge $R_{\rm IR}$ is given as a linear combination of $R_{\rm geo}$ and the  charge of the accidental $U(1)$ and the monopole operators satisfy the unitarity bound $R_{\rm IR} \geq \frac{1}2$.

\paragraph{Unitary topological field theory when $p=0,\pm 1,\pm 2,\pm 3$} To probe the topological field theories in  $p=0,\pm 1, \pm 2$ and $\pm3$ cases, we compute the handle-gluing/fibering operators given in \eqref{handle/fiber for 41p} 
\begin{align}
\begin{split}
& \textrm{ For } \frac{\CT[S^3\backslash \mathbf{4}_1,\mu;\lambda]}{SU(2)_{p=0}},\quad \{ \CH_\a^{-1}  \}  =\{ 20^{-1},20^{-1},5^{-1},5^{-1},4^{-1},4^{-1}\}  \textrm{ and } 
\\
& \qquad \qquad \qquad \qquad \qquad  \{ \CF_\a  \} = e^{\frac{\pi i}{6}} \{ 1 ,  1 , e^{2\pi i /5} ,  e^{-2\pi i /5}, 1,  -1\}\;,
\\
& \textrm{ For } \frac{\CT[S^3\backslash \mathbf{4}_1,\mu;\lambda]}{SU(2)_{p=\pm 1}},\quad \{ \CH_\a^{-1}  \}  =\{ (\zeta_5^1)^2,   (\zeta_5^2)^2 ,  (\zeta_5^3)^2\} \times \{1,1\}\textrm{ and } 
\\
& \qquad \qquad \qquad \qquad \qquad    \{ \CF_\a  \} = e^{\frac{5 \pi i}{28}} \{ 1 ,  e^{2\pi i /7} ,    e^{-4\pi i /7} \} \times \{1,e^{\frac{\pi i}2}\}\;,
\\
& \textrm{ For } \frac{\CT[S^3\backslash \mathbf{4}_1,\mu;\lambda]}{SU(2)_{p=\pm 2}}, \quad \{ \CH^{-1}_\a  \}   =\{ \frac{(\zeta_3^1)^2}{2}, \frac{(\zeta_3^1)^2}{2} ,  \frac{(\zeta_3^2)^2}{2}, \frac{(\zeta_3^2)^2}{2},  (\zeta_3^1)^2,   (\zeta_3^2)^2\}  \textrm{ and }  
\\
& \qquad \qquad \qquad \qquad \qquad  \{ \CF_\a  \} = e^{\frac{23 \pi i}{120}} \{ 1 ,  -1 , e^{-4\pi i /5} ,  e^{\pi i /5},   e^{5\pi i /8},   e^{-7\pi i /40} \}\;,
\\
& \textrm{ For } \frac{\CT[S^3\backslash \mathbf{4}_1,\mu;\lambda]}{SU(2)_{p=\pm 3}}, \quad \{ \CH^{-1}_\a  \}   =\{ 8^{-1},8^{-1},4^{-1} \} \times \{ 1,1\}\textrm{ and }  
\\
& \qquad \qquad \qquad \qquad \qquad \{ \CF_\a  \}= e^{\frac{17 \pi i}{24}} \{ 1 ,  e^{-3\pi  i/2},  e^{-7\pi i /8}\}\times \{1, e^{\frac{\pi i}2}\}\;.
\end{split}
\end{align}
Here we defined
\begin{align}
\zeta_k^m = \left(S_{0,\a =m-1} \textrm{ of } SU(2)_k \right) = \sqrt{\frac{2}{k+2}}\sin \left( \frac{\pi m}{k+2}\right)\;. \label{zetakm}
\end{align}
Notice that $|\CF_\alpha|=1$  and $\CH_\alpha \in \mathbb{R}_+$  for all $\alpha$, which are another evidences for emergence of TQFT at IR, see \eqref{F, H in TQFT}. Furthermore, one can check that
\begin{align}
\sum_\alpha \frac{1}{\CH_\alpha}=1\;,	
\end{align}
which is also expected from \eqref{sum of 1/H in TQFT}. Using the above computations combined with the relation in \eqref{S,T from H,F}, one can determine the $S^2_{0\a}$ and $T_{\a \a}/T_{00}$ of the IR topological field theories. 

Now let us compute the twisted partition functions for $\CT[(S^3\backslash \mathbf{4}_1)_{p\mu+\lambda}] =\frac{\CT[S^3\backslash \mathbf{4}_1,\mu;\lambda]}{``SO(3)"_{p}}$ using the relation in \eqref{SBEQ}, \eqref{SBEQ-2} and  \eqref{SBEQ-3}.
For odd $p$, the above computation is compatible with the general expectation in the Eqn.\eqref{SBE for odd p} and one can see that 
\begin{align}
	\begin{split}
		& \textrm{ For } \frac{\CT[S^3\backslash \mathbf{4}_1,\mu;\lambda]}{``SO(3)"_{p=\pm 1}}, \; \{ \CH_\a^{-1}  \} =\{ 2(\zeta_5^1)^2,   2(\zeta_5^2)^2 , 2 (\zeta_5^3)^2\} \textrm{ and }  \{ \CF_\a  \} = e^{\frac{5 \pi i}{28}} \{ 1 ,  e^{2\pi i /7} ,    e^{-4\pi i /7} \} \;,
        \\
		& \textrm{ For } \frac{\CT[S^3\backslash \mathbf{4}_1,\mu;\lambda]}{SU(2)_{p=\pm 3}},\; \{ \CH^{-1}_\a  \}   =\{ 4^{-1},4^{-1},2^{-1} \} \textrm{ and }  \{ \CF_\a  \}= e^{\frac{17 \pi i}{24}} \{ 1 ,  e^{-3\pi  i/2},  e^{-7\pi i /8}\}\;.
	\end{split}
\end{align}
For even $p$, on the other hand,  the Bethe-vacua  in \eqref{Bethe-vacua for 41p} can be divided into two classes, $\CS_{\rm BE} = S^{(+1)}_{\rm BE}\bigsqcup  S^{(-1)}_{\rm BE}$,
\begin{align}
	\left(\mathcal{S}^{Q = \pm 1}_{\rm BE} \textrm{ of } \frac{\CT[S^3\backslash \mathbf{4}_1,\mu;\lambda]}{SU(2)_p}\right)  =  \left\{ (x, z)\in \CS_{\rm BE} \;:\; x^{p+2} z(x^2 z-1) =Q\right\}\;.
\end{align}
Then, using the Eqns. in \eqref{SBEQ}, \eqref{SBEQ-2}, \eqref{SBEQ-3}, one has
\begin{align}
	\begin{split}
		& \textrm{ For } \frac{\CT[S^3\backslash \mathbf{4}_1,\mu;\lambda]}{``SO(3)"_{p=0}},
		\\
		&\{ \CH_\a^{-1}  \}   =\{ 5^{-1},5^{-1},5^{-1},5^{-1},5^{-1}\} \textrm{ and } \{ \CF_\a  \} = e^{\frac{\pi i}{6}} \{ 1  , e^{2\pi i /5} , e^{2\pi i /5} , e^{-2\pi i /5}, e^{-2\pi i /5}\}\;,
		\\
		& \textrm{ For } \frac{\CT[S^3\backslash \mathbf{4}_1,\mu;\lambda]}{SU(2)_{p=\pm 2}},
		\\
		&\{ \CH^{-1}_\a  \}  =\{ 2(\zeta_3^1)^2, 2(\zeta_3^2)^2 \} \textrm{ and }  \{ (\CF_\a)^2  \}= e^{\frac{23 \pi i}{60}} \{ 1 , e^{2\pi i /5} \}\;.
	\end{split}
\end{align}
For  $p=0$, the last 4 Beth-vacua correspond to the $\CS_{\rm}^{\rm twisted}$. For $p=2$, the $\CF_\alpha$ for Bethe-vacua with $Q=1$ changes its sign under the $\mathbb{Z}_2^{\rm 1-form}$ action, and only $(\CF_\a)^2$ is well-defined after gauging the 1-form symmetry. It implies that the IR TQFT is a fermionic (spin) TQFT.

\subsection{Example : $M=(S^3\backslash \mathbf{5}_2)_{p\mu+\lambda}= (m015)_{(p-2)A_{\rm SP}+B_{\rm SP}}$ }

\subsubsection{Field theory} \label{subsec : field theory for 52p}
$\CK=\mathbf{5}_2$ is the twist knot with three half-twists.   The field theory is \cite{Gang:2018wek}
\begin{align}
	\begin{split}
		&\CT[(S^3\backslash \mathbf{5}_2)_{p\mu+\lambda}]  =  \frac{\CT[S^3\backslash \mathbf{5}_2, \mu ;\lambda]}{``SO(3)_p"} 
		\\
		&=\begin{cases} 
			\frac{(U(1)_{-\frac{1}2}\times SU(2)_{p-4} \; \CN=2\; \textrm{gauge theory  coupled to a chiral in } \textrm{(Adj)}_1)}{\mathbb{Z}_2}\;, \quad \textrm{even $p$}
			\\
			\frac{(U(1)_{-\frac{1}2}\times SU(2)_{p-4} \; \CN=2\; \textrm{gauge theory  coupled to a chiral in } \textrm{(Adj)}_1)\otimes U(1)_{-2}}{\mathbb{Z}^{\rm diag}_2}, \quad \textrm{odd $p$} \ .
		\end{cases}
	\end{split}
	\label{Field theory  for 52}
\end{align}
$\textrm{Adj}_q$ means that the matter field is in the adjoint representation of $SU(2)$ and has charge $q$ under $U(1)$. The theory has a $U(1)_{\rm top}$ symmetry (which is an unexpected accidental symmetry in 6D set-up \eqref{6D set-up}).
\subsubsection{SUSY partition functions}
\paragraph{Squashed 3-sphere partition function} The $S^3_b$ partition function of the theory, \\$\frac{\CT[S^3\backslash \mathbf{5}_2, \mu ;\lambda]}{SU(2)_p} = \big{(}U(1)_{-1/2}\times SU(2)_{p-4} +\Phi$ (in Adj$_{+1}$)$\big{)}$, is given by ($\hbar:=2\pi i b^2$)
\begin{align}
	\begin{split}
		&\mathcal{Z}_b^{(5_2)_p} \left(W= m + (i \pi +\frac{\hbar}2)\nu \right)
		\\
		&= \frac{1}2 \int \frac{dXdZ}{2\pi \hbar} \left( 4 \sinh (X) \sinh(\frac{2\pi i X}{\hbar})\right) \psi_\hbar (Z+2X) \psi_\hbar (Z-2X)  \psi_\hbar (Z) 
		\\
		& \quad \quad  \times \exp \left(\frac{Z^2}{2\hbar} +(p-2)\frac{X^2}\hbar + \frac{Z(W-2\pi i -\hbar)}{\hbar}\right)\;.
	\end{split}
\end{align}
Here $m$ is the (rescaled) real mass, $m=b\times (\textrm{real mass})$,  for the $U(1)_{\rm top}$ symmetry, topological symmetry for the $U(1)$ gauge symmetry. The real mass parameter can be identified with the  FI parameter.  $\nu$ parametrizes the mixing between $U(1)_R$ and $U(1)_{\rm top}$ and
the R-symmetry at $\nu=1$ corresponds to the geometrical R-symmetry in \eqref{R-geo}, i.e. $R_{\nu=1} = R_{\rm geo}$.\footnote{One can check that there is a Bethe-vacuum,  $\alpha = (\textrm{hyp})$, satisfying that  $\textrm{Im}[S^{(\textrm{hyp})}_0 (m=0, \nu=1)]$, defined in \eqref{S0 and S1},  of $\CT[\frac{\CT[S^3\backslash \mathbf{5}_2, \mu ;\lambda]}{SU(2)_p} ]$ is equal to the hyperbolic volume of $(S^3\backslash \mathbf{5}_2)_{p\mu +\lambda}$ when the Dehn filled manifold is hyperbolic.}

\paragraph{Superconformal index} The index is 
\begin{align}
\begin{split}
\CI_q^{(5_2)_p} (u,\nu) &= \sum_{(m_x, m_z)}\oint_{|u_x|=1}\frac{du_x}{2\pi i u_x} \oint_{|u_z|=1}\frac{du_z}{2\pi i u_z} \Delta (m_x, u_x) \CI_{\Delta}(m_z+2m_x, u_z u_x^2)
\\
& \quad \quad  \times   \CI_{\Delta}(m_z-2m_x, u_z /u_x^2)  \CI_{\Delta}(m_z, u_z)   u_z^{m_z} u_x^{(2p-4) m_x} (u (-q^{1/2})^{\nu-2} )^{m_z}\;.
\label{Index for 52}
\end{split}
\end{align}
Here $(m_z,u_z)$ and $(m_x,u_x)$ are (monopole flux, fugacity)  coupled to the $U(1)$ and $SU(2)$ gauge symmetry respectively. $u$ is the fugacity for the $U(1)_{\rm top}$ symmetry.  The summation range of the monopole fluxes is
\begin{align}
& \begin{cases}
m_x \in  \mathbb{Z}_{\geq 0 },\;m_z \in \mathbb{Z}\;, \quad \textrm{for }\frac{\CT[S^3\backslash \mathbf{5}_2, \mu ;\lambda]}{SU(2)_p}
\\
m_x \in  \frac{1}2\mathbb{Z}_{\geq 0 },\;m_z \in \mathbb{Z}\;, \quad \textrm{for }\CT[(S^3\backslash \mathbf{5}_2)_{p\mu +\lambda }]=\frac{\CT[S^3\backslash \mathbf{5}_2, \mu ; \lambda]}{``SO(3)_p"} \ .
\label{Index for 52-2}
\end{cases}
\end{align}

\paragraph{Bethe-vacua and handle-gluing/fibering operators} Expanding the integrand of $\CZ_b^{(5_2)_p}$ in the limit $\hbar \rightarrow 0$, see \eqref{W0 and W1 from integrand} and \eqref{asymptotic of QDL}, we have 
\begin{align}
	\begin{split}
		\CW^{(5_2)_p}_0 (X,Z;m, \nu)&= \textrm{Li}_2 (e^{-Z-2X}) +  \textrm{Li}_2 (e^{-Z+2X}) +  \textrm{Li}_2 (e^{-Z}) + (p-2) X^2 +\frac{1}2 Z^2 
		\\
		&\quad +Z (m+i \pi \nu) \pm 2\pi i X\;,
		\\
		\CW^{(5_2)_p}_1 (X,Z;m,\nu)&=- \frac{1}2 \left((2-\nu)Z+\log(1-e^{-2X-Z}) +\log (1-e^{2X-Z}) +\log (1-e^{-Z})\right)
		\\
		& \quad + \log \left( 2 \sinh X \right)\;. \nonumber
	\end{split}
\end{align}
Extremizing the twisted superpotential $\CW^{(5_2)_p}_0$, we have
\begin{align}
	\begin{split}
		&\left(\mathcal{S}_{\rm BE} \textrm{ of } \frac{\CT[S^3\backslash \mathbf{5}_2,\mu;\lambda]}{SU(2)_p}\right)  
		\\
		&= \left\{ (x, z) \;:\; \frac{x^{2p-8}(1-x^2 z)^2}{(x^2-z)^2}=1,\;\frac{e^{m+ i\pi \nu}(z-1)(1-x^2 z)(1-x^{-2}z)}{ z^2}=1\;, x^2 \neq 1\right\}/\mathbb{Z}^{\rm Weyl}_2 \ . \label{SBE for 52p}
	\end{split}
\end{align}
The Weyl $\mathbb{Z}^{\rm Weyl}_2$ acts as
\begin{align}
	\mathbb{Z}^{\rm Weyl}_2 \;:\;(x,z) \; \leftrightarrow \;(1/x,z)\;.
\end{align}
The handle-gluing and fibering operator are
\begin{align}\hspace{-1cm}
	\begin{split}
		&\left(\CH_\a \textrm{ of } \frac{\CT[S^3\backslash \mathbf{5}_2,\mu;\lambda]}{SU(2)_p}\right)   
		\\
		&=\frac{2 z^{-\nu -1} \left(x^4 z (-p+2 z+2)+x^2 \left((p-2) z^3-(p-6) z+2 (p-6)\right)+z (-p+2 z+2)\right)}{\left(x^2-1\right)^2} \bigg{|}_{(x,z)=(x^{(\a)},z^{(\a)})} \;,
		\\
		&\left(\CF_\a \textrm{ of } \frac{\CT[S^3\backslash \mathbf{5}_2,\mu;\lambda]}{SU(2)_p}\right)  =  \exp \left(\frac{i (\CW^{(5_2)_p}_0-2\pi i n^{(\a)}_x X -2\pi i n^{(\a)}_z Z-mZ)}{2\pi }\right) \bigg{|}_{(X,Z) = (\log x^{(\a)}, \log z^{(\a)})}\;. \label{handle/fiber for 52p}
	\end{split}
\end{align}
Two integers $(n_x^{(\a)}, n_z^{(\a)})$ for each Bethe-vacuum $\a$ are chosen as in \eqref{vecn}.

\subsubsection{IR phases}
According to \cite{2018arXiv181211940D}, the Dehn filled manifolds are
\begin{align}
	(S^3\backslash \mathbf{5}_2)_{p \mu + \lambda}   = 
	\begin{cases}
		\textrm{Atoroidal SFS $S^2 ((2,1),(3,1),(11,-9))$}\;, \quad p=1
		\\
		\textrm{Atoroidal SFS $S^2 ((2,1),(4,1),(7,-5))$}\;, \quad p=2
		\\
		\textrm{Atoroidal SFS $S^2 ((3,1),(1,3),(5,-3))$}\;, \quad p=3
		\\
		\textrm{Graph}\;, \quad p=0, 4
		\\
		\textrm{Hyperbolic}\;, \quad |p-2|>2 \ .
	\end{cases} \label{topological types of 52p}
\end{align}
According to \eqref{unitary/non-unitary 3-manifolds}, the  topological field theory $\textrm{TFT}[M]$ for the above Seifert fibered manifolds  are all  non-unitary. 
Combined with the mathematical facts, our proposal \eqref{proposal for IR phases} predicts that
\begin{align}
	\begin{split}
	&\CT[(S^3\backslash \mathbf{5}_2)_{p\mu+\lambda}]  
	\\
	&=
	\begin{cases}
		\textrm{$\mathcal{N}=4$ rank-0 SCFT}\;, \quad p=1,2,3
		\\
		\textrm{$\CN=2$ SCFT with $U(1)$ flavor symmetry and CPOs}\;, \quad p= 0, 4
		\\
		\textrm{$\mathcal{N}=2$ SCFT}\;, \quad |p-2|>2 \ .
	\end{cases}
\end{split}
\end{align}
Interestingly, the $\CN=2$ gauge theories  in \eqref{Field theory  for 52} for $p=1,2,3$ are expected to have SUSY enhancement to $\CN=4$. Below, we will  confirm the expected IR phases using the general methods outlined in Section \ref{sec : strategy}.

\begin{figure}[htbp]
	\begin{center}
		\includegraphics[width=.3\textwidth]{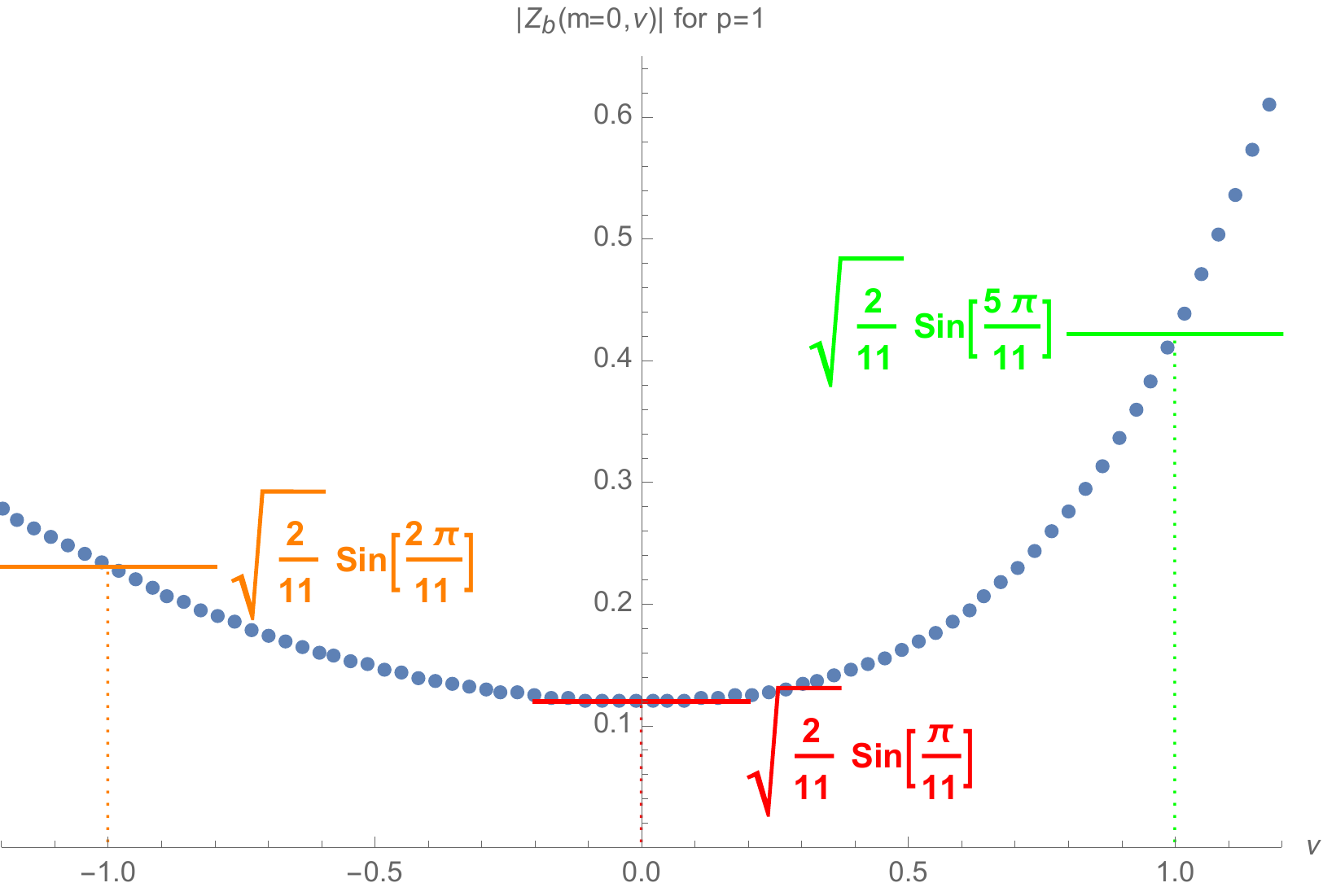}
		\includegraphics[width=.3\textwidth]{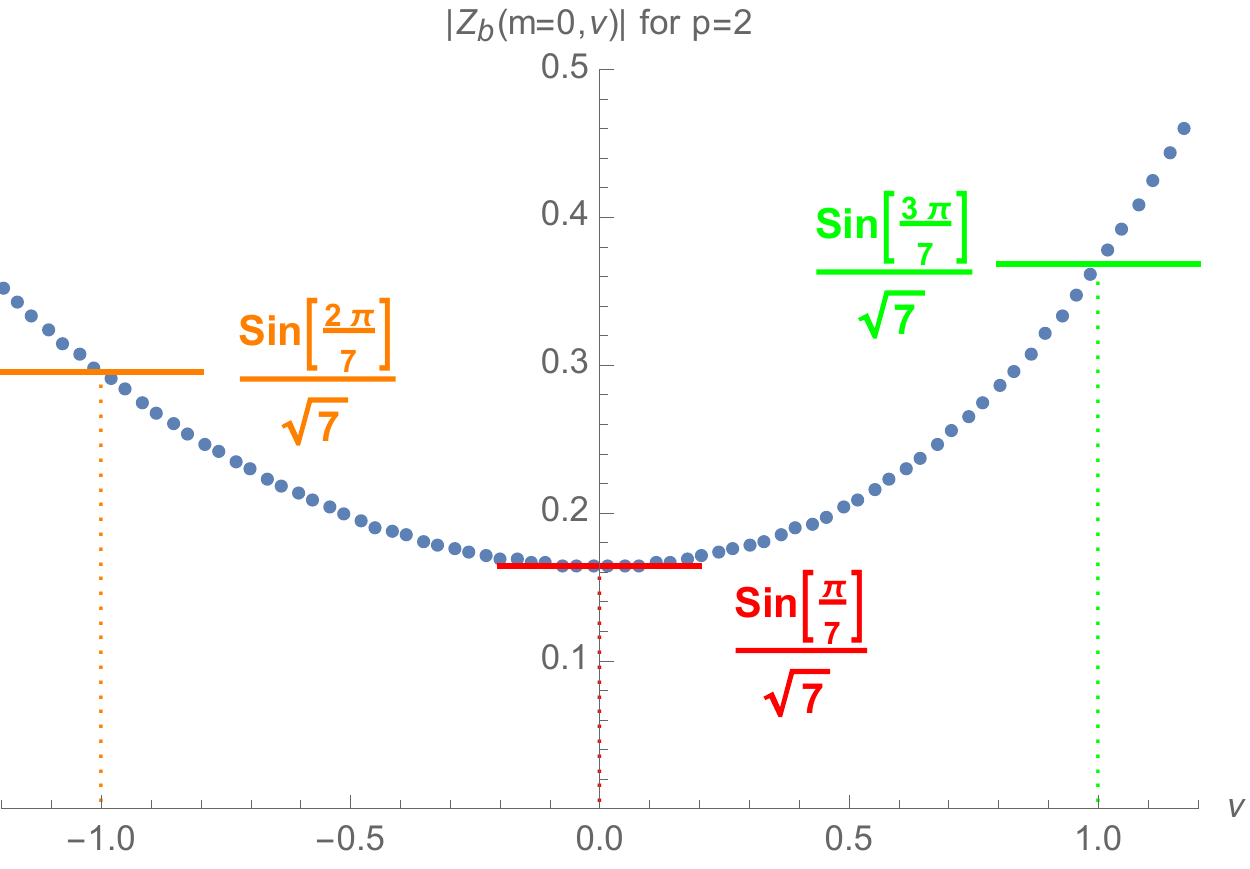}
		\includegraphics[width=.3\textwidth]{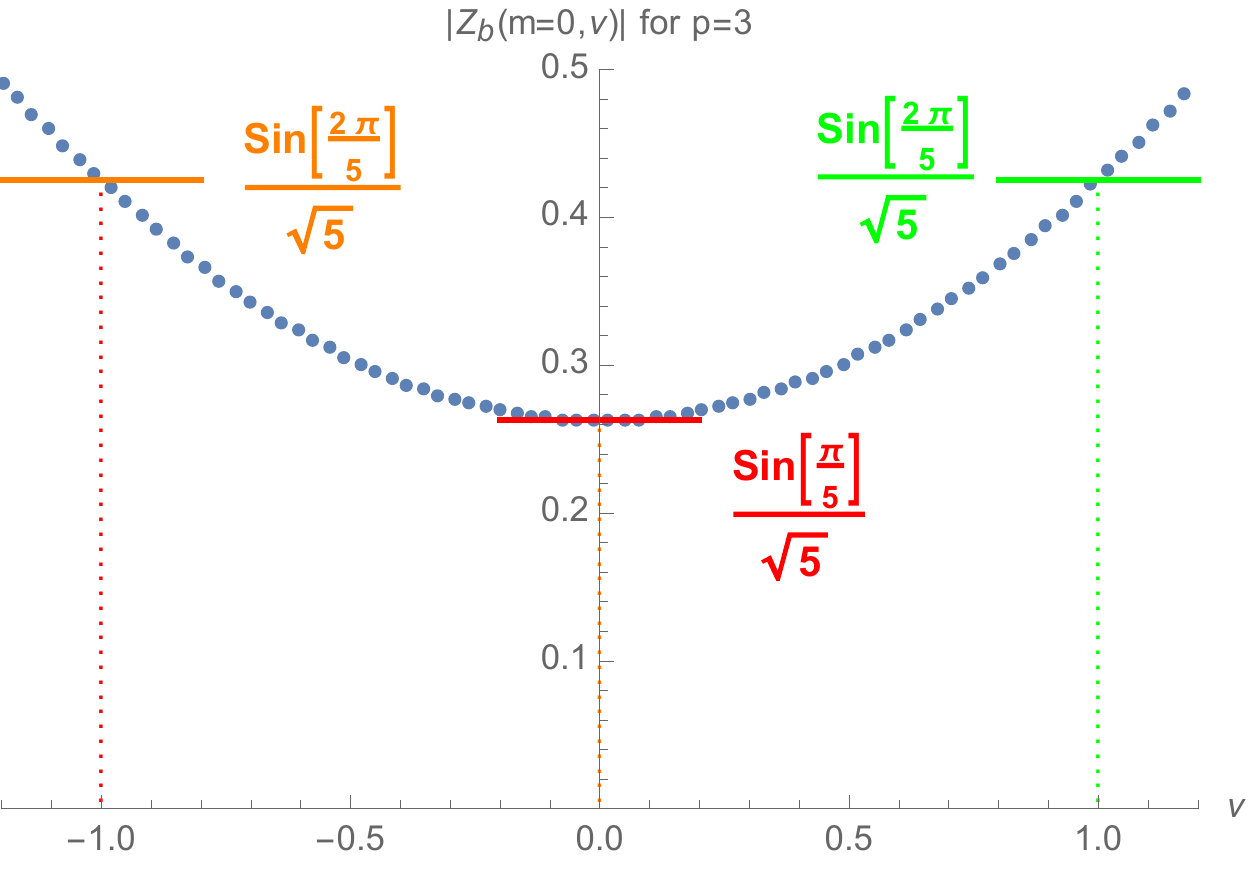}
	\end{center}
	\caption{$|\mathcal{Z}_{b=1}^{(5_2)_{p}} (m=0, \nu)|$ as a function on $-1.2 < \nu < 1.2$ for $p=1,2,3$. The minimums are located at $\nu=0$ and its values are $\sqrt{\frac{2}{11}} \sin(\frac{\pi}{11}), \sqrt{\frac{1}{7}} \sin(\frac{\pi}{7})$ and $\sqrt{\frac{1}{5}} \sin(\frac{\pi}{5})$ for $p=1,2,3$ respectively.}
	\label{fig: Z for 52p}
\end{figure}

\paragraph{SUSY enhancement when $p=1,2,3$} The superconformal indices, computed using the formulae in  \eqref{Index for 52} and \eqref{Index for 52-2}, for $p=1,2,3$ are
\begin{align}
	\begin{split}
		&\left(\CI_q (u;\nu=0)\textrm{ of } \frac{\CT[S^3\backslash \mathbf{5}_2, \mu ;\lambda ]}{SU(2)_{p=1}}\right) = \left(\CI_q (u;\nu=0)\textrm{ of } \frac{\CT[S^3\backslash \mathbf{5}_2, \mu ;\lambda ]}{``SO(3)"_{p=1}}\right)  
		\\
		& =  1-q-\left(u+\frac{1}{u}\right) q^{3/2}-2 q^2-u q^{5/2}+\left(-1+\frac{1}{u^2}\right)q^3+ \left(\frac{1}u-u \right)q^{7/2}+O\left(q^{9/2}\right)\ ,
		\\
		&\left(\CI_q (u;\nu=0)\textrm{ of } \frac{\CT[S^3\backslash \mathbf{5}_2, \mu ;\lambda ]}{SU(2)_{p=2}}\right) = \left(\CI_q (u;\nu=0)\textrm{ of } \frac{\CT[S^3\backslash \mathbf{5}_2, \mu ;\lambda ]}{``SO(3)"_{p=2}}\right)  
		\\
		& =  1-q-\left(u+\frac{1}{u}\right) q^{3/2}-2 q^2-u q^{5/2}+\left(-1+\frac{1}{u^2}\right)q^3+ \left(\frac{1}u-u \right)q^{7/2} + \frac{q^4}{u^2}+O\left(q^{9/2}\right)\ ,
		\\
		&\left(\CI_q (u;\nu)\textrm{ of } \frac{\CT[S^3\backslash \mathbf{5}_2, \mu ;\lambda ]}{SU(2)_{p=3}}\right) = \left(\CI_q (u;\nu)\textrm{ of } \frac{\CT[S^3\backslash \mathbf{5}_2, \mu ;\lambda ]}{``SO(3)"_{p=3}}\right)  
		\\
		& = 1-q-\left(\frac{1}{u}+u\right) q^{3/2}-2 q^2-\left(\frac{1}{u}+u\right) q^{5/2}-2 q^3-\left(\frac{1}{u}+u\right) q^{7/2}-2q^4 +O(q^5) \ .
	\end{split}
\end{align}
The index for general R-charge mixing $\nu$ can be obtained by the relation in \eqref{SCI under mixing}. To compute the correct IR superconformal index, one needs to determine the  IR superconformal R-symmetry, $\nu_{\rm IR}$, using the F-maximization in \eqref{F-maximization}. Using the Bethe-sum formula of round 3-sphere partition function in \eqref{round 3-sphere from Bethe-sum} along with \eqref{SBE for 52p} and \eqref{handle/fiber for 52p}, one can confirm  that (see Fig.~\ref{fig: Z for 52p})
\begin{align}
	|\mathcal{Z}_{b=1}^{(5_2)_{p}} (m=0, \nu=0)| <|\mathcal{Z}_{b=1}^{(5_2)_{p}} (m=0, \nu\neq 0)|\;, \quad \textrm{for $p=1,2,3$} \ .
\end{align}
So the $\nu= 0$ corresponds to the correct IR superconformal R-charge according to the F-maximization, i.e. $\nu_{\rm IR}=0$.
Applying the formula in  \eqref{round 3-sphere from Bethe-sum}, one can choose arbitrary $\nu_0\in 2\mathbb{Z}+1$ at which the superconformal index contains only  $q^{\textrm{integer}}$ terms. Putting back the IR superconformal R-charge choice to the superconformal index, one can confirm that the   index satisfies the necessary condition in \eqref{SUSY enhancement from index} for the SUSY enhancement. In the degenerate limits $u=1$ and $\nu \rightarrow \pm 1$, the superconformal indices become trivial, i.e.
\begin{align}
	\begin{split}
	&\textrm{For $p=1,2,3$},
	\\
	&\left(\CI_q (u=1;\nu=\pm 1)\textrm{ of } \frac{\CT[S^3\backslash \mathbf{5}_2, \mu ;\lambda ]}{SU(2)_{p}} \textrm{ and } \frac{\CT[S^3\backslash \mathbf{5}_2, \mu ;\lambda ]}{``SO(3)"_{p}}\right)=1\;.
	\end{split}
\end{align}
It is a strong evidence for emergence of non-unitary TQFTs in the degenerate limits.  
To probe  the emergent non-unitary TQFTs, one can compute the handle-gluing and fibering operators. For $p=1$ and $3$, we have
\begin{align}
	\begin{split}
		&\left(\{\CH_\alpha^{-1} (m=0, \nu =+  1)\} \textrm{ of } \frac{\CT[S^3\backslash \mathbf{5}_2, \mu ;\lambda ]}{SU(2)_{p=1}}\right) =\bigg{ \{} (\zeta_9^5)^2, (\zeta_9^1)^2,(\zeta_9^2)^2,(\zeta_9^3)^2, (\zeta_9^4)^2\bigg{ \}}\times \bigg{\{}1, 1 \bigg{\}}\;.
		\\
		&\left(\{\CF_\alpha (m=0, \nu = + 1)\}\textrm{ of } \frac{\CT[S^3\backslash \mathbf{5}_2, \mu ;\lambda ]}{SU(2)_{p=1}}\right) = e^{\frac{19 \pi i}{132}}\bigg{ \{}  1,   e^{\frac{2\pi i }{11}} , e^{\frac{10\pi i }{11}},e^{\frac{16\pi i }{11}}  ,e^{\frac{20\pi i }{11}} \bigg{ \}} \times \bigg{\{}1, e^{\frac{\pi i}2} \bigg{\}}\;.
		\\
		&\left(\{\CH^{-1}_\alpha (m=0, \nu =-  1)\} \textrm{ of } \frac{\CT[S^3\backslash \mathbf{5}_2, \mu ;\lambda ]}{SU(2)_{p=1}}\right) =\bigg{ \{} (\zeta_9^2)^2, (\zeta_9^1)^2, (\zeta_9^3)^2, (\zeta_9^4)^2, (\zeta_9^5)^2\bigg{ \}}\times \bigg{\{}1, 1 \bigg{\}}\;.
		\\
		&\left(\{\CF_\alpha (m=0, \nu = - 1)\} \textrm{ of } \frac{\CT[S^3\backslash \mathbf{5}_2, \mu ;\lambda ]}{SU(2)_{p=1}}\right) = e^{\frac{43 \pi i}{132}} \bigg{ \{}   1,  e^{\frac{20\pi i }{11}} , e^{\frac{18\pi i }{11}},e^{\frac{8\pi i }{11}},e^{\frac{14\pi i }{11}}, \bigg{ \}} \times \bigg{\{}1, e^{\frac{\pi i}2} \bigg{\}}\;.
		\\
		&\left(\{\CH^{-1}_\alpha (m=0, \nu =\pm  1)\} \textrm{ of } \frac{\CT[S^3\backslash \mathbf{5}_2, \mu ;\lambda ]}{SU(2)_{p=3}}\right) = \bigg{ \{}  \frac{ (\zeta_3^2)^2}2 , \frac{ (\zeta_3^2)^2}2,\frac{ (\zeta_3^1)^2}2 , \frac{ (\zeta_3^1)^2}2\bigg{ \}}\times \bigg{\{}1, 1 \bigg{\}}\;.
		\\
		&\left(\{\CF_\alpha (m=0, \nu = \pm 1)\} \textrm{ of } \frac{\CT[S^3\backslash \mathbf{5}_2, \mu ;\lambda ]}{SU(2)_{p=3}}\right) =e^{-\frac{3\pi i }{10}}  \bigg{ \{}   1,e^{\frac{\pi i}2} ,e^{\frac{\pi i}{10}}, e^{\frac{8\pi i}5}\bigg{ \}}  \times \bigg{\{}1, e^{\frac{\pi i}2} \bigg{\}} \;.
			\end{split}
	\end{align}
It  is compatible with the general expectation in \eqref{SBE for odd p}.  For $p=2$ case, 
\begin{align}
\begin{split}
		\\
		&\left(\{\CH^{-1}_\alpha (m=0, \nu =+  1)\} \textrm{ of } \frac{\CT[S^3\backslash \mathbf{5}_2, \mu ;\lambda ]}{SU(2)_{p=2}}\right) 
		\\
		&= \bigg{ \{}  \frac{(\zeta_5^3)^2}2 ,\frac{ (\zeta_5^3)^2}2, \frac{(\zeta_5^1)^2}2,\frac{ (\zeta_5^1)^2}2, \frac{(\zeta_5^5)^2}2, \frac{(\zeta_5^5)^2}2, (\zeta_5^2)^2, (\zeta_5^4)^2,(\zeta_5^6)^2\bigg{ \}} \;.
		\\
		&\left(\{\CF_\alpha (m=0, \nu = + 1)\} \textrm{ of } \frac{\CT[S^3\backslash \mathbf{5}_2, \mu ;\lambda ]}{SU(2)_{p=2}}\right) 
		\\
		&= e^{\frac{197 \pi i}{168}}\bigg{ \{}   1,e^{\pi i }, e^{\frac{\pi i }7},e^{\frac{8\pi i}{7}}, e^{\frac{5\pi i }7}, e^{\frac{12\pi i }7}, e^{\frac{47\pi i }{56}},e^{\frac{9\pi i }{8}},e^{\frac{15\pi i }{56}} \bigg{ \}} \;,
		\\
		&\left(\{\CH^{-1}_\alpha (m=0, \nu =-  1)\} \textrm{ of } \frac{\CT[S^3\backslash \mathbf{5}_2, \mu ;\lambda ]}{SU(2)_{p=2}}\right) 
		\\
		&= \bigg{ \{}  \frac{ (\zeta_5^5)^2}2 , \frac{(\zeta_5^5)^2}2, \frac{(\zeta_5^1)^2}2,\frac{ (\zeta_5^1)^2}2, \frac{(\zeta_5^3)^2}2, \frac{(\zeta_5^3)^2}2, (\zeta_5^2)^2, (\zeta_5^4)^2,(\zeta_5^6)^2\bigg{ \}} \;.
		\\
		&\left(\{\CF_\alpha (m=0, \nu = - 1)\} \textrm{ of } \frac{\CT[S^3\backslash \mathbf{5}_2, \mu ;\lambda ]}{SU(2)_{p=2}}\right) 
		\\
		&= e^{\frac{53 \pi i}{168}} \bigg{ \{}   1,e^{\pi i }, e^{\frac{6\pi i }7},e^{\frac{13\pi i}{7}}, e^{\frac{4\pi i }7}, e^{\frac{11\pi i }7}, e^{\frac{9\pi i }{8}},e^{\frac{95\pi i }{56}} ,e^{\frac{111\pi i }{56}} \bigg{ \}}  \;.
	\end{split}
\label{evaluation of H/F for 52p}
\end{align}
%
One can check that the above $\CH_\alpha$ and $\CF_\alpha$ satisfy the condition in \eqref{properties of ran0 SCFT-1}.
The Bethe-vacuum $\alpha=0$ is chosen  to satisfy  the following relation 
\begin{align}
|\CZ_{b}(m=0, \nu=\pm 1)| = \bigg{|}\sum_\a \frac{\CF_\a (m=0, \nu=\pm 1) }{\CH_\a (m=0, \nu=\pm 1)} \bigg{|} = \frac{1}{\sqrt{\CH_{\a=0} (m=0,\nu=\pm 1)}}\;.
\end{align}
Using the relation in  \eqref{S,T from H,F}, one can compute $|S_{0\alpha}|$ for the topological field theories in the degenerate limits and confirm that the unitarity condition \eqref{unitarity of TQFT} is violated. It implies that the emergent TQFTs are indeed non-unitary. Furthermore, from Fig.~\ref{fig: Z for 52p} and \eqref{evaluation of H/F for 52p}, one can confirm that
\begin{align}
	|\CZ_{b=1}(m=0, \nu=0)| =  \textrm{min}_\alpha[\mathcal{H}_{\alpha}(m=0,\nu=\pm1)^{-1/2}]  = \begin{cases}
		\sqrt{\frac{2}{11}} \sin \left( \frac{\pi}{11}\right), \quad p=1
		\\
		\sqrt{\frac{1}{7}} \sin \left( \frac{\pi}{7}\right), \quad p=2
		\\
		\sqrt{\frac{1}{5}} \sin \left( \frac{\pi}{5}\right), \quad p=3
	\end{cases} \ ,
\end{align}
which is expected from \eqref{properties of ran0 SCFT-3}.
For the case with $p=3$, one can directly check the SUSY enhancement using known IR dualities. Using a chain of dualities, one can prove that\footnote{Duality we are using here is IR equivalence between $(SU(2)_{-1}+ \textrm{adjoint } \Phi) = (\textrm{free chiral} \;X)\otimes U(1)_2$. In the duality, the $U(1)_\Phi$ flavor symmetry acting on $\Phi$ with charge $+1$ in $SU(2)$ gauge theory becomes $U(1)_X$ flavor symmetry acting on $X$ with charge $+2$ in the free chiral theory. The gauge invariant $\textrm{Tr}(\Phi^2)$ in the $SU(2)$ theory is mapped to $X$ in the free chiral theory. The background CS term of CS level $k$ for $U(1)_\Phi$ in $SU(2)$ gauge theory is mapped to background CS level $k+1/2$ for $U(1)_X$.}
\begin{align}
	\begin{split}
		&(\CT[M=(S^3\backslash \mathbf{5}_2)_{p\mu+\lambda:p=3}] \textrm{ in } \eqref{Field theory  for 52})
		\\
		&= \frac{[\left( U(1)_{0}+X (\textrm{of charge +2}) \right) \otimes U(1)_{2} ]\otimes U(1)_{-2}}{\mathbb{Z}_2^{\rm diag}}  \quad (\textrm{using duality appetizer \cite{Jafferis:2011ns}})
		\\
		&= \left( U(1)_{0}+X (\textrm{of charge +2}) \right)\otimes  \left(\frac{U(1)_2 \otimes  U(1)_{-2}}{\mathbb{Z}_2^{\rm diag}}  \right) \;.
	\end{split}
\end{align}
The theory in the 2nd factor is just a (trivial) topological theory and the  theory in 1st factor is actually known to have $\CN=4$ symmetry enhancement and flows to the $\CN=4$ SCFT of rank-0 at IR  \cite{Gang:2018huc}.  

\begin{figure}[htbp]
	\begin{center}
		\includegraphics[width=.5\textwidth]{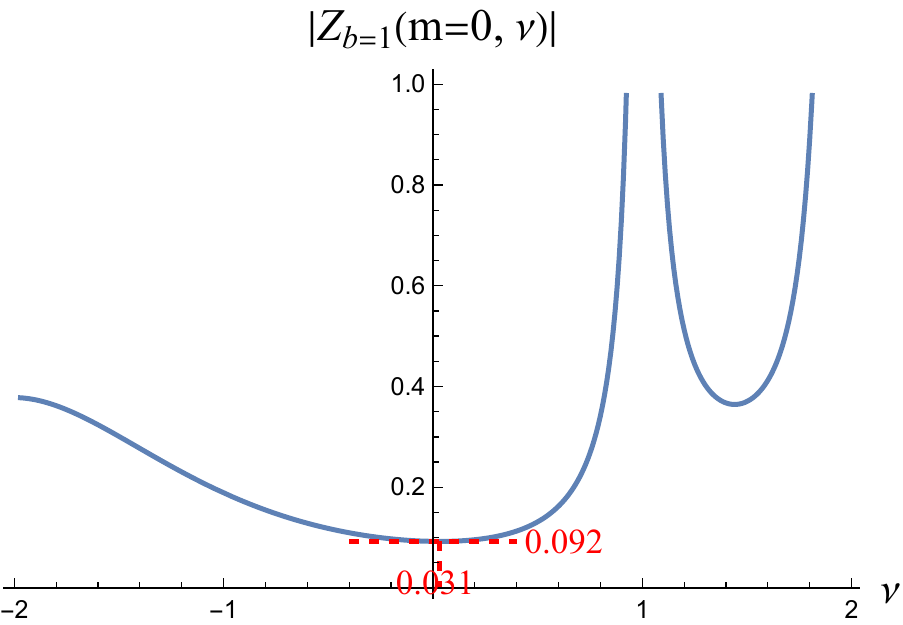}
	\end{center}
	\caption{$|\mathcal{Z}_{b=1}^{(5_2)_{p=0}} (m=0, \nu)|$ as a function of $\nu \in (-2,2)$. The minimum is located at $\nu\simeq 0.031$ and its value is approximately $0.092$.}
\label{fig: m015-0}
\end{figure}

\paragraph{$U(1)$ flavor symmetry and chiral primary operators when $p=0,4$} The superconformal indices, computed using the formulae in  \eqref{Index for 52} and \eqref{Index for 52-2}, for $p=0$ and $4$ are\footnote{When $p=4$, the UV $\mathbb{Z}_2$ 1-form symmetry decouples at IR and we need to sum over only $m_x \in \mathbb{Z}$ even for the $``SO(3)"$ gauging as explained below  \eqref{Index for Np-2}. }
\begin{align}\hspace{0cm}
	\begin{split}
		&\left(\CI_q (u;\nu=0)\textrm{ of } \frac{\CT[S^3\backslash \mathbf{5}_2, \mu ;\lambda ]}{SU(2)_{p=0}}\right) 
		\\
		& = 1+\left(\frac{1}{u^2}-1\right)q-u q^{3/2}+\left( \frac{1}{u^4}-2\right)q^2 +\left(\frac{1}{u}-u \right)q^{5/2} + \frac{q^3}{u^6}
		\\
		&\quad +\left( \frac{2}u-u\right)q^{7/2}+\left(1+\frac{1}{u^8}-\frac{1}{u^2}\right) q^4+O(q^{9/2})\;,
		\\
		&\left(\CI_q (u;\nu=0)\textrm{ of } \frac{\CT[S^3\backslash \mathbf{5}_2, \mu ;\lambda ]}{``SO(3)"_{p=0}}\right)  
		\\
		& =  1-\frac{q^{1/2}}u + \left( \frac{1}{u^2}-2\right)q-\left(u+\frac{1}{u^3}\right) q^{3/2}+\left( \frac{1}{u^4}-3\right)q^2
		\\
		&\quad +\left( \frac{2}u - 2u -\frac{1}{u^5}\right)q^{5/2}+\frac{q^3}{u^6}-\left(\frac{1}{u^7}-\frac{4}{u}+u\right) q^{7/2}+O\left(q^{4}\right)\;,
		\\
		& \left(\CI_q (u;\nu=0)\textrm{ of } \frac{\CT[S^3\backslash \mathbf{5}_2, \mu ;\lambda ]}{SU(2)_{p=4}}\right) = \left(\CI_q (u;\nu=0)\textrm{ of } \frac{\CT[S^3\backslash \mathbf{5}_2, \mu ;\lambda ]}{``SO(3)"_{p=4}} \right)  
		\\
&=\left(\sum_{n=0}^\infty 1\right)\times q^{0}-q-u q^{3/2}-2 q^2-u q^{5/2}+O\left(q^{3}\right)\ .
	\end{split}
\end{align}
When $p=0$, the index at $\nu=1$ (geometrical R-charge) becomes
\begin{align}
\begin{split}
&\left(\CI_q (u;\nu=1)\textrm{ of } \frac{\CT[S^3\backslash \mathbf{5}_2, \mu ;\lambda ]}{SU(2)_{p=0}}\right)  = \left(\CI_q (u;\nu=0)\textrm{ of } \frac{\CT[S^3\backslash \mathbf{5}_2, \mu ;\lambda ]}{SU(2)_{p=0}}\right) \bigg{|}_{u\rightarrow u(-q^{1/2})}
\\
& =\left( \sum_{n=0}^\infty u^{-2n}\right) \times q^0  -q+\left(u- \frac{1}u-2 \right)q^2 +\left(u-\frac{2}u-\frac{1}{u^2}\right)q^3+O(q^4)\;,
\\
&\left(\CI_q (u;\nu=1)\textrm{ of } \CT[(S^3\backslash \mathbf{5}_2)_{0\mu+\lambda}]  = \frac{\CT[S^3\backslash \mathbf{5}_2, \mu ;\lambda ] }{``SO(3)"_{p=0}}\right) 
\\
& =\left( \sum_{n=0}^\infty u^{-n}\right) \times q^0  -2q+\left(u- \frac{2}u-3 \right)q^2 +\left(2u-\frac{4}u-\frac{2}{u^2}\right)q^3+O(q^4)\;.
\end{split}
\end{align}
As expected from \eqref{3D index pattern}, \eqref{3D-3D relations} and  \eqref{topological types of 52p}, the index diverges at $\nu=1$ and $u=1$. The infinity is regularized by the fugacity $u$. The infinity   is just  due to the bad choice of R-charge, $\nu=1$, which is different from the correct IR superconformal R-charge.  The correct R-charge can be determined from the F-maximization principle \eqref{F-maximization}, and we numerically find that (See Fig.~\ref{fig: m015-0}.)
\begin{align}
\nu_{\rm IR}\simeq 0.031\;\;, \quad |\CZ_{b=1}(m=0,\nu=\nu_{\rm IR})| \simeq 0.0924563\ .
\end{align}
The term $\sum_{n=0}^\infty u^{-n}$ comes from 1/2 BPS bare monopole operators, $(V_{(\frac{1}2,-1)})^n$, with $SU(2)$ monopole flux $m_x=\frac{n}2$ and $U(1)$ monopole flux $m_z =-n$. The $V_{(\frac{1}2,-1)}$ is a chiral primary operator whose R-charge is 
\begin{align}
	R_{\nu_{\rm IR}}(V_{(\frac{1}2,-1)}) = 1-\nu_{\rm IR} \simeq 0.969\;.
\end{align}
So, we can conclude that the  theory $\CT[(S^3\backslash \mathbf{5}_2)_{0\mu+\lambda}]$ flows to a 3D $\CN=2$ SCFT with chiral primary operators, which is compatible with our proposal in \eqref{proposal for IR phases}.

When $p=4$, the analysis is more subtle. Like $p=0$ case, the superconformal index diverges at $\nu=1$ and $u=1$ as expected from \eqref{3D index pattern}, \eqref{3D-3D relations} and  \eqref{topological types of 52p}. In this case, however,  the infinity can not be regularized by the fugacity $u$ since the $(\sum_{n=0}^\infty 1) \times q^0$ comes from 1/2 BPS bare monopole operators $V_{(\frac{n}2,0)} (n\geq 0)$ neutral under the $U(1)_{\rm top}$ symmetry. Note that the bare monopole operator is gauge-invariant when $p=4$ since the bare Chern-Simons level for $SU(2)$ gauge symmetry is zero, see  \eqref{Field theory  for 52}, and matter fields are in a real representation of the $SU(2)$, which means that there is no 1-loop CS level shift.
The only possible explanation for the infinity is that there is an  additional $U(1)$ flavor symmetry emerging at IR under which the monopole operators are charged. So we expect that the IR theory will be described by a $\CN=2$ SCFT, which has at least two $U(1)$s as flavor symmetry. The monopole operator becomes a chiral primary operator in the IR SCFT. This is compatible with our proposal in \eqref{proposal for IR phases}.

\subsection{Example : $M=(m003)_{p A_{\rm SP}+B_{\rm SP}}$}

\subsubsection{Field theory} \label{subsec : field theory for m003p}
The $A_{\rm SP}$ is non-closable  since $(m003)_{A_{\rm SP}} =L(10,3)$ \cite{2018arXiv181211940D}. $A_{\rm SP}$ is an even cycle while $B_{\rm SP }$ is an odd cycle. The field theory $\CT[m003,A_{\rm SP}; B_{\rm SP}]$  is \cite{Gang:2018wek}
\begin{align}
\CT[m003,A_{\rm SP}; B_{\rm SP}] = (U(1)_{0}  \;\CN=2 \textrm{ gauge theory  coupled to  two chirals of charge +1})\;.
\end{align}
The theory is actually identical to $\CT[S^3\backslash \mathbf{4}_1,\mu ]$ in \eqref{Field theory  for 41}. The theory has manifest $SU(2)$ flavor symmetry rotating the two chirals, which can be identified with the $SU(2)_A$ in  \eqref{su(2)A}.  The $SU(2)_A$ is different from the $SU(2)_A$  in  \eqref{su(2)A in 41} for  $\CT[S^3\backslash \mathbf{4}_1, \mu]$ case. Background CS level for the $SU(2)$ is $+1/2$.  Then, the field theories associated to the Dehn filled manifolds are (see \eqref{T[NpAB-for-even-A]})
\begin{align}
	\begin{split}
		&\CT[(m003)_{p A_{\rm SP} + B_{\rm SP}}] = \frac{\CT[m003, A_{\rm SP};B_{\rm SP}]}{SU(2)_p}
		\\
		& = (U(1)_0 \times SU(2)_{p+1/2} \textrm{ coupled to a chiral $\Phi$ in $\mathbf{2}_{1}$})\;.\label{Field theory  for m003}
	\end{split}
\end{align}
The theory has $U(1)_{\rm top}$ symmetry associated to the $U(1)_0$ gauge field. 
\subsubsection{SUSY partition functions}
\paragraph{Squashed 3-sphere partition function} The squashed 3-sphere partition function for the theory $\mathcal{T}[(m003)_{pA_{\rm SP}+B_{\rm SP}}]$ is 
\begin{align}
	\begin{split}
		\mathcal{Z}_b^{(\textrm{m003})_p} \left(W = m +(i \pi +\frac{\hbar}2)\nu \right)=&\frac{1}2 \int \frac{dXdZ}{(2\pi \hbar)} (4 \sinh X \sinh \frac{2\pi i X}\hbar ) \psi_\hbar (Z+X) \psi_\hbar  (Z-X)
		\\
		& \times  \exp \left( \frac{2(p+1)X^2+Z^2+ 2 (W-(i\pi +\frac{\hbar}2)) Z}{2\hbar}\right)\;.
	\end{split}
\end{align}
Here $m$ is the real mass parameter for the $U(1)_{\rm top}$ symmetry, i.e. FI parameter, and the $\nu$ parametrizes the mixing between the $U(1)_R$ symmetry and the $U(1)_{\rm top}$ symmetry.  The choice $\nu=1$ corresponds to the geometrical $U(1)$ R-symmetry. 

\paragraph{Superconformal index} The superconformal index for the theory $\mathcal{T}[(m003)_{pA_{\rm SP}+B_{\rm SP}}]$ is
\begin{align}
	\begin{split}
		\CI^{(m003)_p}_q (u, \nu) &= \sum_{m_x \in \mathbb{Z}_{\geq 0}, m_z \in \mathbb{Z}} \oint_{|u_x|=1}\frac{du_x}{2\pi i u_x} \oint_{|u_z|=1}\frac{du_z}{2\pi i u_z}\Delta (m_x, u_x) 
		\\
		& \times  \CI_{\Delta}(m_z+m_x, u_z u_x) \CI_{\Delta}(m_z-m_x, u_z /u_x) u_x^{2(p+1) m_x} u_z^{m_z} (u(-q^{1/2})^{(\nu-1)})^{m_z}\;. \label{Index for m003}
	\end{split}
\end{align}
Here $(m_z,u_z)$ and $(m_x,u_x)$ are (monopole flux, fugacity)  coupled to the $U(1)$ and $SU(2)$ gauge symmetry respectively.  $u$ is the fugacity for the  flavor $U(1)_{\rm top}$ symmetry.

\paragraph{Bethe-vacua and handle-gluing/fibering operators}
Expanding the integrand of $\mathcal{Z}_b^{(\textrm{m003})_p}$ in the limit in which $\hbar \to 0$, we get
\begin{equation}
\begin{aligned}
&\mathcal{W}_0^{(\textrm{m003})_p} (X,Z;m,\nu) \\
&=  \textrm{Li}_2 (e^{-Z-X})+\textrm{Li}_2 (e^{-Z+X}) + (p+1)X^2 + \frac{1}{2}Z^2 + (m+i\pi (\nu-1))Z \pm 2\pi i X  \ ,\\
&\mathcal{W}_1^{(\textrm{m003})_p} (X,Z;m,\nu) \\
&=-\frac{1}{2} \left( \log (1-e^{-Z-X})+\log (1-e^{-Z+X}) \right) + \frac{\nu - 3}{2}Z + \pi i+\log (2\sinh X)\ .
\end{aligned}
\end{equation}
Extremizing the twisted superpotential $\mathcal{W}_0^{(\textrm{m003})_p}$, we obtain	
\begin{equation}
\begin{aligned}
&\left(\mathcal{S}_{\textrm{BE}} \textrm{ of } \frac{\CT[m003, A_{\rm SP};B_{\rm SP}]}{SU(2)_p}\right) \\
&= \left\{ (x,z) \;\; : \;\; \frac{x^{2 p+1} (1-x z)}{x-z}=1\ , \;  \frac{e^{m+i \pi  \nu} (x-z) (x z-1)}{x z}=1\ , \; x^2 \neq 1 \right\} / \mathbb{Z}^{\rm Weyl}_2\ ,
\end{aligned}
\end{equation}
where $x=e^X, z=e^Z$ and the Weyl group $\mathbb{Z}^{\rm Weyl}_2$ acts as
\begin{equation}
\mathbb{Z}^{\rm Weyl}_2 \;\; : \;\; (x,z) \; \leftrightarrow \; (1/x,z)\ .
\end{equation}
Then, the handle-gluing and fibering operators are given by
\begin{equation}\label{f-m003}
\begin{aligned}
&\left(\mathcal{H}_\alpha \textrm{ of } \frac{\CT[m003, A_{\rm SP};B_{\rm SP}]}{SU(2)_p}\right) =\left. -\frac{x z^{1-\nu} \left(x \left(2 (p+1) z^2-2 p+x z\right)+z\right)}{\left(x^2-1\right)^2} \right|_{(x,z) = (x^{(\alpha)},z^{(\alpha)})}\ , \\
&\left(\mathcal{F}_\alpha \textrm{ of } \frac{\CT[m003, A_{\rm SP};B_{\rm SP}]}{SU(2)_p}\right)
\\
& = \left. \exp \left(\frac{i (\CW^{(\textrm{m003})_{p}}_0-2\pi i n^{(\a)}_x X -2\pi i n^{(\a)}_z Z - m Z)}{2\pi }\right) \right|_{(X,Z) = (\log x^{(\a)}, \log z^{(\a)})}\ . \\
\end{aligned}
\end{equation}
Two integers $(n_x^{(\a)}, n_z^{(\a)})$ for each Bethe-vacuum $\a$   are chosen as in \eqref{vecn}.

\subsubsection{IR phases}
According to \cite{2018arXiv181211940D}, the Dehn filled manifolds are
\begin{align}
	(m003)_{p A_{\rm SP}+ B_{\rm SP}}   = 
	\begin{cases}
		\textrm{Atoroidal SFS $S^2 ((2,1),(3,2),(-3,1))$}\;, \quad p= -2, 1
		\\
		\textrm{Lens space $L(5,1)$}\;, \quad p= -1, 0
		\\
		\textrm{Hyperbolic}\;, \quad p<-2 \textrm{ or } p>1 \ .
	\end{cases}
\end{align}
According to \eqref{unitary/non-unitary 3-manifolds}, the $\textrm{TFT}[M]$ for the above Seifert fibered manifolds are all unitary. From our proposal in \eqref{proposal for IR phases}, we expect that
\begin{align}
	\mathcal{T}[(m003)_{p A_{\rm SP}+ B_{\rm SP}} ] =
	\begin{cases}
		\textrm{Unitary  TQFT}\;, \quad p=-2, 1
		\\
		\textrm{SUSY is spontaneously broken}\;, \quad p= -1, 0
		\\
		\textrm{$\mathcal{N}=2$ SCFT}\;, \quad p<-2 \textrm{ or } p>1 \ .
	\end{cases}
\end{align}
When $p=0$ or $p=-1$, one can check that
\begin{align}
	(\CI^{(m003)_p}_q (u,\nu)   \textrm{ in \eqref{Index for m003}})= 0  \textrm{ and } (\mathcal{S}_{\rm BE} \textrm{ in \eqref{Bethe-vacua set}}) \textrm{ is an empty set}\;.
\end{align}
This implies that the theories at $p=0$ or $p=-1$ enjoy dynamical spontaneous supersymmetry breaking.
When $p=1$ or $p=-2$, on the other hand,
\begin{align}
	(\CI^{(m003)_p}_q (u,\nu)   \textrm{ in \eqref{Index for m003}})= 1 \;. 
\end{align}
Interestingly, the dependence on $(u,\nu)$ in $\CI_q$  drops off in the case. It implies that the theory at $p=1$ or $p=-2$ flows to a unitary topological field theory. To probe the topological field theories in these cases, we can compute the handle-gluing/fibering operators given in \eqref{f-m003}
\begin{align}
	\begin{split}
		&\left(\CH^{-1}_{\alpha=0} (m=0, \nu =  1)\textrm{ of } \frac{\CT[m003,A;B]}{SU(2)_{p=1}}\right) = 1 \;,
		\\
		&\left(\CF_{\alpha=0} (m=0, \nu =  1) \textrm{ of } \frac{\CT[m003,A;B ]}{SU(2)_{p=1}}\right) =e^{\frac{\pi i }{4}}  \;,
        \\
        		&\left(\CH^{-1}_{\alpha=0} (m=0, \nu =  1)\textrm{ of } \frac{\CT[m003,A;B]}{SU(2)_{p=-2}}\right) = 1 \;,
		\\
		&\left(\CF_{\alpha=0} (m=0, \nu =  1) \textrm{ of } \frac{\CT[m003,A;B ]}{SU(2)_{p=-2}}\right) =e^{\frac{\pi i }{12}}  \;.
	\end{split}
\end{align}
The $(\CH_\a, \CF_\a)$ satisfy the conditions in \eqref{F, H in TQFT} and \eqref{sum of 1/H in TQFT}. 
It is  another evidence for the emergence of TQFT at IR. 
The handle-gluing operators and fibering operators at arbitrary real $m$ (with $\nu =\nu_{\rm geo} = 1$) are 
\begin{align}
	\begin{split}
		&\left(\CH^{-1}_{\alpha=0} (m, \nu=1)\textrm{ of } \frac{\CT[m003,A;B]}{SU(2)_{p=1}}\right) =e^{2m} \;,
                                \\
        		&\left(\CF_{\alpha=0} (m, \nu =1)\textrm{ of } \frac{\CT[m003,A;B]}{SU(2)_{p=1}}\right) =\exp \left[i \left( \frac{\pi}{4} +\frac{m^2}{4\pi}  \right)\right]\;,
        \\
        		&\left(\CH^{-1}_{\alpha=0} (m, \nu=1)\textrm{ of } \frac{\CT[m003,A;B]}{SU(2)_{p=-2}}\right) =1\;,
                        \\
        		&\left(\CF_{\alpha=0} (m, \nu =1)\textrm{ of } \frac{\CT[m003,A;B]}{SU(2)_{p=-2}}\right) =\exp \left[i \left( \frac{\pi}{12} -\frac{m^2}{4\pi}  \right)\right]\;.
	\end{split}
\end{align}
As expected, they are independent of the continuous real mass parameter $m$ modulo the ambiguities \eqref{ambiguity in H and F} due to local counter-terms.

\subsection{Example : $M=(m006)_{pA_{\rm SP}+B_{\rm SP}}$}  \label{sec : m006p}
 
 \subsubsection{Field theory} \label{subsec : field theory for m006p}
The $A_{\rm SP}$ is non-closable since $(m006)_{A_{\rm SP}} = L(5,2)$. Both $A_{\rm SP}$ and $B_{\rm SP}$ are odd cycle and we choose $A=A_{\rm SP}$ and $B=A_{\rm SP} +B_{\rm SP}$ for the two 1-cycles to have different oddness/evenness. In that choice, the field theory associated to the 1-cusped 3-manifold is 
 \begin{align}
 \begin{split}
 &\CL_{\CT[m006,A_{\rm SP};A_{\rm SP}+B_{\rm SP}]} [\vec{\Phi}, v, w; W_X, W_C] 
 \\
 &=  \int d^4\theta (\Phi_1^\dagger e^{-w}\Phi_1 +\Phi_2^\dagger e^{-v+w} \Phi_2 +\Phi_3^\dagger e^{v+w} \Phi_3 ) + \frac{1}{2\pi } \int d^4 \theta (\Sigma_v W_A +\Sigma_w W_C)
 \\
 & + \frac{1}{4\pi}\int d^4 \theta \left(-\Sigma_v  v-\frac{3}2 \Sigma_w  w\right) + \left(\int d^2 \theta \Phi_1^2 \Phi_2 \Phi_3 +(c.c)\right) \ .
 \end{split}
 \end{align}
 Here $\{\Phi_i\}_{i=1}^3$ are chiral superfields. $v$ and $w$ are dynamical vector multiplets while $W_A$ and $W_C$ are background vector multiplets coupled to $U(1)_A \times U(1)_C$ flavor symmetry.  
 The theory is nothing but $U(1)_{-1}\times U(1)_{-\frac{3}2}$ gauge theory coupled to 3 chirals with gauge charges $(0,-1),(-1,1),(1,1)$ with superpotential $\Phi_1^2 \Phi_2 \Phi_3$. The theory has $U(1)_A \times U(1)_C$ flavor symmetries which are topological symmetries associated to the two $U(1)$s. Since the $A_{\rm SP}$  is non-closable, the $U(1)_A$ symmetry is expected to be enhanced to $SU(2)_A$ symmetry at IR according to the prediction of 3D-3D correspondence in  \eqref{su(2)A}. Below,  we will  confirm the symmetry enhancement from the superconformal index computation. Then, the field theories for the Dehn filled manifolds are
 \begin{align}
 	\begin{split}
 		&\CT[(m006)_{pA_{\rm SP}+B_{\rm SP}}]  =  \frac{\CT[ m006, A_{\rm SP} ;A_{\rm SP}+B_{\rm SP}]}{``SO(3)"_{p-1}} 
 		\\
 		&=\begin{cases} 
 			\left( \frac{\CT[ m006, A_{\rm SP} ;A_{\rm SP}+B_{\rm SP}]}{SU(2)_{p-1}} \right)/\mathbb{Z}_2\; , \quad \textrm{even $(p-1)$} 
 			\\
 			\left( \frac{\CT[ m006, A_{\rm SP} ;A_{\rm SP}+B_{\rm SP}]}{SU(2)_{p-1}} \otimes U(1)_{-2} \right)/\mathbb{Z}_2\;, \quad \textrm{odd $(p-1)$} \ .
 		\end{cases}
 	\end{split}
 	\label{Field theory  for m006p}
 \end{align} 
 
\subsubsection{SUSY partition functions}
\paragraph{Squashed 3-sphere partition function} The $S^3_b$ partition function of the theory, \\$\frac{\CT[m006, A_{\rm SP} ;A_{\rm SP}+B_{\rm SP}]}{SU(2)_{p-1}} $, is given by ($\hbar:=2\pi i b^2$)
 \begin{align}
 	\begin{split}
 		&\mathcal{Z}_b^{(m006)_p} \left(W= m + (i \pi +\frac{\hbar}2)\nu \right)
 		\\
 		&= \frac{1}2 \int \frac{dXdZ_1 dZ_2}{(2\pi \hbar)^{3/2}} \left( 4 \sinh (X) \sinh(\frac{2\pi i X}{\hbar})\right) \psi_\hbar (-Z_2 +2\pi i +\hbar) \psi_\hbar (-Z_1+Z_2 -2\pi i -\hbar) 
 		\\
 		& \quad   \times   \psi_\hbar (Z_1+Z_2)  \exp \left((p-1)\frac{X^2}\hbar +\frac{2X (Z_1+i \pi +\frac{\hbar}2 )+ Z_2 (W- (i\pi +\frac{\hbar}2))}\hbar \right) \;.
 	\end{split}
 \end{align}
 The geometrical R-charge corresponds to $\nu=1$.
 
 \paragraph{Superconformal index} The index is given by 
 \begin{align}
 	\begin{split}
 		\CI_q^{(m006)_p} (u,\nu) &= \sum_{m_x}\oint_{|u_x|=1} \frac{du_x}{2\pi i u_x}     \Delta (m_x, u_x)  u_x^{2(p-1)m_x} \CI^{m006} (m_x, u_x; u, \nu)\;, \;\textrm{where}
 		\\
 		\CI^{m006} (m_x, u_x; u,\nu)& =\sum_{m_1, m_2 \in \mathbb{Z}} \oint_{|z_1|=1}\frac{dz_1}{2\pi i z_1}  \oint_{|z_2|=1} \frac{dz_2}{2\pi i z_2}   \CI_\Delta  \left(-m_2,\frac{q}{z_2}\right) \CI_\Delta  \left(m_2-m_1,\frac{z_2}{z_1 q}\right)
 		\\
 		&  \qquad \times  \CI_\Delta  (m_1+m_2,z_1 z_2)    \left(-q^{1/2}z_1\right)^{2m_x} u_x^{2m_1} \left(u(-q^{1/2})^{\nu-1}\right)^{m_2} \ .
 		\label{Index for m006p}
 	\end{split}
 \end{align}
 $\CI^{m006}(m_x, u_x;u, \nu)$ is the generalized superconformal index for $\CT[m006, A_{\rm SP}; A_{\rm SP}+B_{\rm SP}]$ where $(m_x,u_x)$ are background (monopole flux, fugacity)  coupled to the $U(1)_A$ while $u$ is the fugacity for the $U(1)_C$. $\nu$ parametrizes the mixing between $U(1)$ R-symmetry and the $U(1)_C$. The $U(1)_A$ symmetry will be enhanced to $SU(2)_A$ at IR and cannot be mixed the R-symmetry. To see the enhancement, we compute the index 
 \begin{align}
 	\begin{split}
 &\CI_q^{m006}(m_x=0,u_x;u, \nu) 
 \\
 &= 1+ \left(-2 -  u_x^2 -\frac{1}{u_x^2} - \frac{1}u - \frac{1}{u u_x^2}  - \frac{u_x^2}{u}\right) q - \left(-2-\frac{1}u +u +\frac{1}{u u_x^4}+ \frac{u_x^4}{u}\right)q^2 +O(q^3)\bigg{|}_{u \rightarrow (-q^{1/2})^{\nu} u }
 \end{split}
 \end{align}
 The term $\left(-2-u_x^2-\frac{1}{u_x^2}\right)q$ can be identified with the contributions from conserved current multiplets for $U(1)_C$ and $SU(2)_A$.  The conserved current multiplet of the 3D $\CN=2$ SCFT contains a conformal primary with $r=1, j_3=\frac{1}2$ and $\Delta = \frac{3}2$. The term exists regardless of the mixing parameter $\nu$. Furthermore, the index respects the Weyl $\mathbb{Z}_2$ symmetry 
 \begin{align}
 \CI_q^{m006} (m_x, u_x;u, \nu ) =  \CI_q^{m006} (-m_x, 1/u_x;u, \nu ) \;.
 \end{align}
 The summation ranges of $m_x$ and $m_{1,2}$ are
 \begin{align}
 	& \begin{cases}
 		m_x \in  \mathbb{Z}_{\geq 0 },\;m_1,m_2 \in \mathbb{Z}\;, \quad \textrm{for }\frac{\CT[m006, A_{\rm SP} ;A_{\rm SP}+B_{\rm SP}]}{SU(2)_{p-1}}
 		\\
 		m_x \in  \frac{1}2\mathbb{Z}_{\geq 0 },\;m_1,m_2 \in \mathbb{Z}\;, \quad \textrm{for }\CT[(m006)_{pA_{\rm SP} +B_{\rm SP} }]=\frac{\CT[m006, A_{\rm SP} ;A_{\rm SP}+B_{\rm SP}]}{``SO(3)"_p} \ .
 		\label{Index for m006p-2}
 	\end{cases}
 \end{align}
 
\paragraph{Bethe-vacua and handle-gluing/fibering operators} Expanding the integrand of $\mathcal{Z}_b^{(\textrm{m006})_p}$ in the limit in which $\hbar \to 0$, we get
\begin{equation}
\begin{aligned}
&\mathcal{W}_0^{(\textrm{m006})_p} (X,Z_1,Z_2;m,\nu) \\
&= \textrm{Li}_2 (e^{Z_2})+\textrm{Li}_2 (e^{Z_1-Z_2})+\textrm{Li}_2 (e^{-Z_1-Z_2}) + (p-1) X^2 + 2X(Z_1 + \pi i) + (m+(\nu-1)\pi i) Z_2
\\
&  \quad \pm 2\pi i X\;,
\\
&\mathcal{W}_1^{(\textrm{m006})_p} (X,Z_1,Z_2;m,\nu) \\
&=\frac{1}{2} \left( \log(1-e^{Z_2}) -3 \log(1-e^{Z_1-Z_2})-\log(1-e^{-Z_1-Z_2}) \right) + X + \frac{\nu-1}{2} Z_2 +\log (2\sinh X)\ .
\end{aligned}
\end{equation}
Extremizing the twisted superpotential $\mathcal{W}_0^{(\textrm{m006})_p}$, we obtain	
\begin{equation}
\begin{aligned}
\left(\mathcal{S}_{\textrm{BE}} \textrm{ of } \frac{\CT[m006, A_{\rm SP} ;A_{\rm SP}+B_{\rm SP}]}{SU(2)_{p-1}} \right) =&\bigg{ \{} (x,z_1,z_2) \;\; : \;\; z_1^2 x^{2 p-2}=1\ , \;  \frac{x^2 (1-z_1z_2)}{z_1 (z_1-z_2)}=1\ ,
\\ 
&\;  -\frac{e^{m+i \pi  \nu} (z_1-z_2) (z_1 z_2-1)}{z_1 (z_2-1) z_2^2}=1\ , \; x^2 \neq 1 \bigg{\}} / \mathbb{Z}^{\rm Weyl}_2\ . \label{Bethe-in-m006}
\end{aligned}
\end{equation}
Here $x=e^X, z_{1,2}=e^{Z_{1,2}}$, and the Weyl group $\mathbb{Z}^{\rm Weyl}_2$ acts as
\begin{equation}
\mathbb{Z}^{\rm Weyl}_2 \;\; : \;\; (x,z_1,z_2) \; \leftrightarrow \; (1/x,1/z_1,z_2)\ .
\end{equation}
Then, the handle-gluing and fibering operators are given by
\begin{equation}\label{f-m006}\hspace{-0.2cm}
\begin{aligned}
&\left(\mathcal{H}_\alpha \textrm{ of } \frac{\CT[m006, A_{\rm SP} ;A_{\rm SP}+B_{\rm SP}]}{SU(2)_{p-1}}\right)
\\
& =\left. \frac{2 \left(z_1-z_2\right){}^2 \left((p+3) \left(z_1^2+1\right) z_2^2-2 z_2 \left(3 p z_1+z_1^2+1\right)+4 p z_1-2 z_1 z_2^3\right) z_2^{-\nu
   -3}}{\left(x^2-1\right)^2 z_1 \left(z_2-1\right){}^2} \right|_{(x,z_{1,2}) = (x^{(\alpha)},z_{1,2}^{(\alpha)})}\ , \\
&\left(\mathcal{F}_\alpha \textrm{ of } \frac{\CT[m006, A_{\rm SP} ;A_{\rm SP}+B_{\rm SP}]}{SU(2)_{p-1}}\right) 
\\
&= \left. \exp \left(\frac{i (\CW^{(\textrm{m006})_{p}}_0-2\pi i n^{(\a)}_x X -2\pi i n^{(\a)}_{z_1} Z_1-2\pi i n^{(\a)}_{z_2} Z_2 - m Z_2)}{2\pi }\right) \right|_{(X,Z_{1,2}) = (\log x^{(\a)}, \log z_{1,2}^{(\a)})}\ . \nonumber
\end{aligned}
\end{equation}
Three integers $(n_x^{(\a)}, n_{z_1}^{(\a)}, n_{z_2}^{(\a)})$ for each Bethe-vacuum $\a$ are chosen  as in \eqref{vecn}.

 \subsubsection{IR phases}
 According to \cite{2018arXiv181211940D}, the Dehn filled manifolds are
 \begin{align}
 	(m006)_{p A_{\rm SP}+ B_{\rm SP}}   = 
 	\begin{cases}
 		\textrm{Lens space $L(15,4)$}\;, \quad p=0
 		\\
 		\textrm{Atoroidal SFS $S^2((2,1),(2,1),(3,2))$}\;, \quad p= 1
 		\\
 		\textrm{Atoroidal SFS $S^2((2,1),(3,2),(4,-3))$}\;, \quad p= -1
 		\\
 		\textrm{Atoroidal SFS $S^2((2,1),(3,1),(7,-5))$}\;, \quad p= -2
 		\\
 		 \textrm{Graph}\;, \quad p= -3
         \\
         \textrm{Hyperbolic}\;, \quad |p+1|>2 \ .
 	\end{cases}
 \end{align}
 According to \eqref{unitary/non-unitary 3-manifolds},   $\textrm{TFT}[S^2 ((2,1),(2,1) (3,2))]$ and $\textrm{TFT}[S^2 ((2,1),(3,2) (4,-3))]$ are   unitary while  $\textrm{TFT}[S^2 ((2,1),(3,1) (7,-5))]$ is non-unitary.  From the our proposal in \eqref{proposal for IR phases}, we expect that
 \begin{align}
 	\mathcal{T}[(m006)_{p A_{\rm SP}+ B_{\rm SP}} ] =
 	\begin{cases}
 		 \textrm{SUSY is spontaneously broken}\;, \quad p= 0 
 		 \\
 		\textrm{Unitary  TQFT}\;, \quad p=-1, 1
 		\\
 		\textrm{$\CN=4$ rank-0 SCFT}\;, \quad p= -2
 		\\
 		\textrm{$\CN=2$ SCFT with $U(1)$ flavor symmetry and CPOs}\;, \; p=-3
        \\
        \textrm{$\CN=2$ SCFT}\;, \; |p+1|>2 \ .
 	\end{cases} \label{IR phases of m006p}
 \end{align}

Computing the superconformal indices \eqref{Index for m006p}, \eqref{Index for m006p-2}, we obtain
\begin{equation}\label{I-m006} \hspace{0cm}
\begin{aligned}
&\left(\CI_q (u;\nu=0)\textrm{ of } \frac{\CT[m006, A_{\rm SP} ;A_{\rm SP}+B_{\rm SP} ]}{SU(2)_{p-1=-1}} \textrm{ and } \frac{\CT[m006, A_{\rm SP} ;A_{\rm SP}+B_{\rm SP} ]}{``SO(3)"_{p-1=-1}}\right)  =0\ , \\
&\left(\CI_q (u;\nu=0)\textrm{ of } \frac{\CT[m006, A_{\rm SP} ;A_{\rm SP}+B_{\rm SP} ]}{SU(2)_{p-1=0,-2}} \textrm{ and } \frac{\CT[m006, A_{\rm SP} ;A_{\rm SP}+B_{\rm SP} ]}{``SO(3)"_{p-1=0,-2}}\right)  =1\ ,  \\
&\left(\CI_q (u;\nu=0)\textrm{ of } \frac{\CT[m006, A_{\rm SP} ;A_{\rm SP}+B_{\rm SP} ]}{SU(2)_{p-1=-3}} \textrm{ and } \frac{\CT[m006, A_{\rm SP} ;A_{\rm SP}+B_{\rm SP} ]}{``SO(3)"_{p-1=-3}}\right)\\
&=1-q - \left(u+\frac{1}u \right)q^{3/2}-2q^2-u q^{5/2}+\left(\frac{1}{u^2}-1\right) q^3  + \left(\frac{1}{u}-u\right)q^{7/2} + O(q^{4}) \ , \\
&\left(\CI_q (u;\nu=0)\textrm{ of } \frac{\CT[m006, A_{\rm SP} ;A_{\rm SP}+B_{\rm SP} ]}{SU(2)_{p-1=-4}}\right) \\
&=1+\left(\frac{1}{u^2} -1\right)q -u q^{3/2} +\left(\frac{1}{u^4} -2\right)q^2 +\left(\frac{2}u-u\right)q^{5/2}+\left(1+\frac{1}{u^6}\right)q^3+ O(q^{7/2})\ , \\
&\left(\CI_q (u;\nu=0)\textrm{ of } \frac{\CT[m006, A_{\rm SP} ;A_{\rm SP}+B_{\rm SP} ]}{``SO(3)"_{p-1=-4}}\right)\\
&=1- \frac{q^{1/2}}{u}+\left(\frac{1}{u^2} -2\right)q -\left(u+\frac{1}{u^3}\right) q^{3/2} +\left(\frac{1}{u^4} -3\right)q^2 + O(q^{5/2})\ .
\end{aligned}
\end{equation}
The results  for $p=0,\pm 1$ are compatible with our expectation in \eqref{IR phases of m006p}.\footnote{When $p=1$, the UV 1-form $\mathbb{Z}_2$ symmetry decouples at IR. Thus, we only need to sum over $m_x \in \mathbb{Z}$ even for the $``SO(3)"$ gauging as explained below \eqref{Index for Np-2}.  } For $p=-2$ case, one can check that the $\nu=0$  corresponds to the superconformal point, i.e. $\nu_{\rm IR}=0$, from F-maximization. The superconformal index contains the term $-(u+\frac{1}u) q^{2/3}$, which is compatible with the expected SUSY enhancement, see \eqref{SUSY enhancement from index}.   When $p=0$, $\left(\mathcal{S}_{\textrm{BE}} \textrm{ of } \frac{\CT[m006, A_{\rm SP} ;A_{\rm SP}+B_{\rm SP}]}{SU(2)_{p-1=-1}} \right)$ is an empty set as expected from the spontaneous supersymmetry breaking.

Now, let us compute the handle-gluing and fibering operators to probe the topological field theories when $p=\pm 1, -2$.

\paragraph{Unitary topological field theory  when $p=\pm 1$} 
The handle-gluing and fibering operators are given by\footnote{There is a tricky Bethe-vacuum with $(z_1, z_2 , x) = (1,1,\pm i )$. To see it, we first deform the equation $z_1^2 x^{2p-2}=1$ in \eqref{Bethe-in-m006} by $z_1^2 x^{2p-2} = e^{\epsilon}$, and obtain Bethe-vacua and their $(\CH_\a,\CF_\alpha)$, and then take $\epsilon \rightarrow 0$. }
\begin{align}
	\begin{split}
		&\left(\{\CH^{-1}_\alpha (m=0, \nu =\pm  1)\} \textrm{ of } \frac{\CT[m006, A_{\rm SP} ;A_{\rm SP}+B_{\rm SP}]}{SU(2)_{p-1=0}}\right) =\bigg{ \{} 1  \bigg{\}} \;,
		\\
		&\left(\{\CF_\alpha (m=0, \nu = \pm 1)\} \textrm{ of }\frac{\CT[m006, A_{\rm SP} ;A_{\rm SP}+B_{\rm SP}]}{SU(2)_{p-1=0}}\right) =  \bigg{ \{} e^{\frac{i \pi}4}\bigg{ \}} \;,
        \\
        &\left(\{\CH^{-1}_\alpha (m=0, \nu =\pm  1)\} \textrm{ of } \frac{\CT[m006, A_{\rm SP} ;A_{\rm SP}+B_{\rm SP}]}{SU(2)_{p-1=-2}}\right) 
		= \bigg{ \{}\frac{1}{4},\frac{1}{4}, \frac{1}2 \bigg{ \}} \;,
		\\
		&\left(\{\CF_\alpha (m=0, \nu = \pm 1)\} \textrm{ of } \frac{\CT[m006, A_{\rm SP} ;A_{\rm SP}+B_{\rm SP}]}{SU(2)_{p-1=-2}}\right) 
		=\left\{e^{-\frac{9 i \pi }{8}},e^{-\frac{i \pi }{8}}, e^{\frac{i \pi}2}\right\}\;.
	\end{split}
\end{align}
Further note that $|\CF_\alpha |=1$  and $\sum_\alpha (\mathcal{H}_\alpha)^{-1}=1$, which also imply the emergence of TQFT at IR, see \eqref{F, H in TQFT} and \eqref{sum of 1/H in TQFT}. When $p=1$, there is a single Bethe-vacuum with $x=\pm i $ (modulo the Weyl $\mathbb{Z}_2$) on which the 1-form $\mathbb{Z}_2$ symmetry trivially acts. It is another evidence that the 1-form symmetry decouples at IR.

\paragraph{SUSY enhancement  when $p=-2$}
We have already observed strong evidences for the SUSY enhancement at the superconformal R-charge $\nu=0$ when $p=-2$. Here, we shall analyze the emergent non-unitary TQFTs in the degenerate limits. In the degenerate limits $u=1$, $\nu \rightarrow \pm 1$, the superconformal index becomes trivial:
\begin{align}\hspace{0cm}
&\left(\CI_q (u=1;\nu=\pm 1)\textrm{ of } \frac{\CT[m006, A_{\rm SP} ;A_{\rm SP}+B_{\rm SP} ]}{SU(2)_{p-1=-3}} \textrm{ and } \frac{\CT[m006, A_{\rm SP} ;A_{\rm SP}+B_{\rm SP} ]}{``SO(3)"_{p-1=-3}}\right)=1\;.
\end{align}
This is a strong evidence for emergence of the non-unitary TQFTs in the degenerate limits.  
To probe the emergent non-unitary TQFTs, one may compute the handle-gluing and fibering operators as following:
\begin{align}
	\begin{split}
    		&\left(\{\CH^{-1}_\alpha (m=0, \nu =  1)\} \textrm{ of } \frac{\CT[m006, A_{\rm SP} ;A_{\rm SP}+B_{\rm SP}]}{SU(2)_{p-1=-3}} \right) =\bigg{\{}(\zeta_5^3)^2,(\zeta_5^1)^2,(\zeta_5^2)^2\bigg{\}}\times \bigg{\{}1, 1 \bigg{\}}\;,
		\\
		&\left(\{\CF_\alpha (m=0, \nu =  1)\} \textrm{ of } \frac{\CT[m006, A_{\rm SP} ;A_{\rm SP}+B_{\rm SP}]}{SU(2)_{p-1=-3}} \right) =\left\{e^{-\frac{5 i \pi }{42}},e^{-\frac{41 i \pi }{42}},e^{-\frac{17i \pi }{42}}\right\}  \times \left\{1,e^{\frac{\pi i}2}\right\}  \;.\,
        \\
		&\left(\{\CH^{-1}_\alpha (m=0, \nu =-  1)\} \textrm{ of } \frac{\CT[m006, A_{\rm SP} ;A_{\rm SP}+B_{\rm SP}]}{SU(2)_{p-1=-3}} \right) = \bigg{\{}(\zeta_5^2)^2,(\zeta_5^1)^2,(\zeta_5^3)^2\bigg{\}}\times \bigg{\{}1, 1 \bigg{\}}\;,
		\\
		&\left(\{\CF_\alpha (m=0, \nu = - 1)\} \textrm{ of } \frac{\CT[m006, A_{\rm SP} ;A_{\rm SP}+B_{\rm SP}]}{SU(2)_{p-1=-3}} \right) =\left\{e^{-\frac{41 i \pi }{42}},e^{-\frac{5 i \pi }{42}},e^{-\frac{17i \pi }{42}}\right\}  \times \left\{1,e^{\frac{\pi i}2}\right\}  \;,
	\end{split}
\end{align}
where $\zeta_k^m$ is defined in \eqref{zetakm}. From the above results, one can check that $(\CH_\alpha,\CF_\alpha)$ satisfy the conditions in \eqref{properties of ran0 SCFT-1}, which are another evidences for the emergence of non-unitary TQFTs in the degenerate limits, $m=0$ and  $\nu \rightarrow \pm 1$. Here, the Bethe-vacuum $\alpha=0$ is chosen  to satisfy  the following relation:
\begin{align}
|\CZ_{b}(m=0, \nu=\pm 1)| = \bigg{|}\sum_\a \frac{\CF_\a (m=0, \nu=\pm 1) }{\CH_\a (m=0, \nu=\pm 1)} \bigg{|} = \frac{1}{\sqrt{\CH_{\a=0} (m=0,\nu=\pm 1)}}\;.
\end{align}
In addition, one can obtain the value of round 3-sphere partition function at $m=\nu=0$ as following:
\begin{align}
	|\CZ_{b=1}(m=0, \nu=0)| =  \textrm{min}_\alpha[\mathcal{H}_{\alpha}(m=0,\nu=\pm1)^{-1/2}]  = \sqrt{\frac{2}{7}} \sin \left(\frac{\pi}{7}\right)\ ,
\end{align}
which is expected from \eqref{properties of ran0 SCFT-3}. Using the relation in  \eqref{S,T from H,F}, one can compute $|S_{0\alpha}|$ for the topological field theories in the degenerate limits and show that the unitarity condition \eqref{unitarity of TQFT} is violated. It implies that the emergent TQFTs are indeed non-unitary.

\paragraph{$U(1)$ flavor symmetry and chiral primary operators when $p=-3$} The superconformal indices \eqref{I-m006} for $p=-3$ at geometric R-charge $\nu=1$ are given by
\begin{align}\hspace{-0.7cm}
	\begin{split}
&\left(\CI_q (u;\nu=1)\textrm{ of } \frac{\CT[m006, A_{\rm SP} ;A_{\rm SP}+B_{\rm SP} ]}{SU(2)_{p-1=-4}}\right)  
\\
&=\left(\sum_{n=0}^\infty u^{-2n}\right)\times q^0-q+\left(-2-\frac{2}{u}+u\right) q^2+\left(1-\frac{2}{u^2}-\frac{3}{u}+u\right) q^3+ O(q^4)\ ,
\\
&\left(\CI_q (u;\nu=1)\textrm{ of } \frac{\CT[m006, A_{\rm SP} ;A_{\rm SP}+B_{\rm SP} ]}{``SO(3)_{p-1=-4}"}\right) 
\\
& =\left(\sum_{n=0}^\infty u^{-n}\right)\times q^0-2 q-\left(3+\frac{4}{u}-u\right)q^2 +\left(2-\frac{4}{u^2}-\frac{6}{u}+2u\right)q^3+O\left(q^{4}\right)\ .
	\end{split}
\end{align}
As expected, the indices diverge at $\nu=1$ and $u=1$. The infinity is regularized by the fugacity $u$. This infinity   is just  due to the bad choice of R-charge, $\nu=1$, which is different from the correct IR superconformal R-charge.  The correct R-charge should be determined from the F-maximization principle \eqref{F-maximization}, and we numerically find that
\begin{align}
\nu_{\rm IR}\simeq -0.706\;, \quad |\CZ_{b=1}(m=0,\nu=\nu_{\rm IR})| \simeq 0.0642483\ .
\end{align}
The  term $\sum_{n=0}^\infty u^{-n}$ comes from  $``SO(3)"$ monopole operators, which are chiral primaries with $R_{\rm geo}=0$.

\subsection{Example : $M=(m007)_{pA_{\rm SP}+B_{\rm SP}}$} $A_{\rm SP}$ is a non-closable cycle since $(m006)_{A_{\rm SP}} = L(3,1) $. Both $A_{\rm SP}$ and $B_{\rm SP}$ are odd and we choose $A = A_{\rm SP}$ and $B= A_{\rm SP}+B_{\rm SP}$. 

\subsubsection{Field theory} \label{subsec : field theory for m007p}
The field theory is \cite{Gang:2018wek} 
\begin{align}
\begin{split}
&\CT[(m007)_{pA_{\rm SP}+B_{\rm SP}}] = \frac{\CT[m007,A_{\rm SP};A_{\rm SP}+B_{\rm SP}]}{``SO(3)"_{p-1}}
\\
&=\begin{cases} 
\frac{(U(1)_{\frac{3}2}\times SU(2)_{p-1} \textrm{\;coupled to chirals $\Phi$ in  $\textrm{(Adj)}_1$   and $M$ in $\mathbf{1}_{-2}$ with  $\CW_{\rm m007}$)} }{\mathbb{Z}_2}\;, \quad \textrm{odd $p$}
\\
\frac{(U(1)_{\frac{3}2}\times SU(2)_{p-1} \textrm{\;coupled to chirals $\Phi$ in  $\textrm{(Adj)}_1$   and $M$ in $\mathbf{1}_{-2}$ with  $\CW_{\rm m007}$)}\otimes U(1)_{-2}}{\mathbb{Z}^{\rm diag}_2}, \quad \textrm{even $p$} 
\end{cases}
\end{split}\ .
\label{Field theory  for m007}
\end{align}
Here, $\mathbf{1}_q$ means neutral under $SU(2)$ and having charge $q$ under $U(1)$. The superpotential is given by
\begin{align}
\CW_{m007} = M \times \textrm{Tr}(\Phi^2)\;.
\end{align}
Also, the theory has a $U(1)_{\rm top}$ symmetry.

\subsubsection{SUSY partition function}
\paragraph{Squashed 3-sphere partition function} The $S^3_b$ partition function of the theory, \\$\frac{\CT[m007,A_{\rm SP};A_{\rm SP}+B_{\rm SP}]}{SU(2)_{p-1}} = U(1)_{\frac{3}2}\times SU(2)_{p-1}$ + $\Phi$ (in Adj$_{+1}$)+$M$ (in $\mathbf{1}_{-2}$), is given by \\($\hbar:=2\pi i b^2$)
\begin{align}
\begin{split}
&\mathcal{Z}_b^{(\textrm{m007})_p} \left( W = m +(i \pi +\frac{\hbar}2)\nu \right)
\\
&= \frac{1}{2} \int \frac{dXdZ}{2\pi \hbar} \left( 4 \sinh (X) \sinh(\frac{2\pi i X}{\hbar})\right)  \psi_\hbar (Z+2X) \psi_\hbar (Z-2X)   \psi_\hbar (Z)   \psi_\hbar (-2Z+2\pi i +\hbar)  
\\
& \qquad \times \exp \left(\frac{5Z^2}{2\hbar} + (p+1)\frac{X^2}\hbar -\frac{Z(W+2\pi i +\hbar)}{\hbar}+\frac{(2\pi i +\hbar)^2}{4\hbar} \right).
\end{split}
\end{align}
Here, $m$ is the real mass for the $U(1)_{\rm top}$ symmetry, i.e. FI parameter and $\nu$ parametrizes the mixing between $U(1)_R$ and $U(1)_{\rm top}$. In the above expression,  the R-symmetry at $\nu=1$ corresponds to the geometrical R-symmetry in \eqref{R-geo}, i.e. $R_{\nu=1} = R_{\rm geo}$.

 \paragraph{Superconformal index} The index is given by
 \begin{equation}\label{I-m007}
 \begin{aligned}
& \mathcal{I}_q^{(\textrm{m007})_p}(u,\nu) = \sum_{(m_x,m_z)} \oint_{|u_x|=1} \frac{du_x}{2\pi i u_x}\oint_{|u_z|=1} \frac{du_z}{2\pi i u_z} \Delta(m_x,u_x) \CI_\Delta (m_z+2m_x,u_zu_x^2) \\
 &\times \CI_\Delta (m_z-2m_x,u_zu_x^{-2})\CI_\Delta (m_z,u_z)\CI_\Delta (-2m_z,u_z^{-2}q) u_z^{5m_z} u_x^{(2p+2)m_x} (u(-q^{1/2})^{(\nu+2)})^{-m_z}\;.
 \end{aligned}
 \end{equation}
Here, $(m_z,u_z)$ and $(m_x,u_x)$ are (monopole flux, fugacity) coupled to the $U(1)$ and $SU(2)$ gauge symmetry respectively. The summation range for the monopole fluxes is
\begin{align}
& \begin{cases}
m_x \in  \mathbb{Z}_{\geq 0 },\;m_z \in \mathbb{Z}\;, \quad \textrm{for }\frac{\CT[m007,A_{\rm SP};A_{\rm SP}+B_{\rm SP}]}{SU(2)_{p-1}}
\\
m_x \in  \frac{1}2\mathbb{Z}_{\geq 0 },\;m_z \in \mathbb{Z}\;, \quad \textrm{for }\CT[(m007)_{pA_{\rm SP} +B_{\rm SP} }]=\frac{\CT[m007, A_{\rm SP} ;A_{\rm SP}+ B_{\rm SP}]}{``SO(3)"_{p-1}} \ .
\label{I-m007-2}
\end{cases}
\end{align}

\paragraph{Bethe-vacua and handle-gluing/fibering operators} Expanding the integrand of $\mathcal{Z}_b^{(\textrm{m007})_p}$ in the limit in which $\hbar \to 0$, we get
\begin{equation}
\begin{aligned}
&\mathcal{W}_0^{(\textrm{m007})_p} (X,Z;m,\nu) \\
&= \textrm{Li}_2(e^{-Z-2X})+\textrm{Li}_2(e^{-Z+2X})+\textrm{Li}_2(e^{-Z})+\textrm{Li}_2(e^{2Z}) +(p+1) X^2+\frac{5}{2}Z^2-(m+i \pi  (\nu+2)) Z
\\
&  \quad - \pi^2 \pm 2\pi i X  \ ,\\
&\mathcal{W}_1^{(\textrm{m007})_p} (X,Z;m,\nu) \\
&=-\frac{1}{2}(\log (1-e^{-Z-2X})+\log (1-e^{-Z+2X})+\log (1-e^{-Z})-\log (1-e^{2Z})) - \frac{\nu+2}{2} Z 
\\
& \quad + \pi i +\log (2\sinh X)\ .
\end{aligned}
\end{equation}
Extremizing the twisted superpotential $\mathcal{W}_0^{(\textrm{m007})_p}$, we obtain	
\begin{equation}
\begin{aligned}
&\left(\mathcal{S}_{\textrm{BE}} \textrm{ of } \frac{\CT[m007,A_{\rm SP};A_{\rm SP}+B_{\rm SP}]}{SU(2)_{p-1}}\right) \\
&= \left\{ (x,z) \;\; : \;\; \frac{x^{2 p-2} \left(x^2 z-1\right)^2}{\left(x^2-z\right)^2}=1\ , \; -\frac{z^2 e^{-m-i \pi  \nu} \left(x^2-z\right) \left(x^2 z-1\right)}{x^2 (z-1) (z+1)^2} =1\ , \; x^2 \neq 1 \right\} / \mathbb{Z}^{\rm Weyl}_2\ ,
\end{aligned}
\end{equation}
where $x=e^X, z=e^Z$ and the Weyl group $\mathbb{Z}_2$ acts as
\begin{equation}
\mathbb{Z}^{\rm Weyl}_2 \;\; : \;\; (x,z) \; \leftrightarrow \; (1/x,z)\ .
\end{equation}
Then, the handle-gluing and fibering operators are given by
\begin{equation}\label{f-m007}
\hspace{-1cm}
\begin{aligned}
&\left(\mathcal{H}_\alpha \textrm{ of } \frac{\CT[m007,A_{\rm SP};A_{\rm SP}+B_{\rm SP}]}{SU(2)_{p-1}}\right) \\
&=\frac{2 z^{\nu-1} \left(x^4 z \left(-p (z-3)+2 z^2+z-5\right)+x^2 \left(p \left(z^4+z^3-5 z^2+z-2\right)+z \left(z
   \left(z^2+z-1\right)-3\right)+6\right)\right)}{\left(x^2-1\right)^2 (z-1) (z+1)^2}\\
&\quad +\frac{2 z^{\nu-1} z \left(-p (z-3)+2 z^2+z-5\right)}{\left(x^2-1\right)^2 (z-1) (z+1)^2}\bigg{|}_{(x,z) = (x^{(\alpha)},z^{(\alpha)})} \\
&\left(\mathcal{F}_\alpha \textrm{ of } \frac{\CT[m007,A_{\rm SP};A_{\rm SP}+B_{\rm SP}]}{SU(2)_{p-1}}\right) 
\\
&= \left. \exp \left(\frac{i (\CW^{(\textrm{m007})_{p}}_0-2\pi i n^{(\a)}_x X -2\pi i n^{(\a)}_z Z + m Z)}{2\pi }\right) \right|_{(X,Z) = (\log x^{(\a)}, \log z^{(\a)})}\; . \\
\end{aligned}
\end{equation}
Two integers $(n_x^{(\a)}, n_z^{(\a)})$ for each Bethe-vacuum $\a$ are chosen as in \eqref{vecn}.

\subsubsection{IR phases}
According to \cite{2018arXiv181211940D}, the Dehn filled manifolds are
\begin{align}
	(m007)_{p A_{\rm SP} + B{\rm SP}}   = 
	\begin{cases}
		\textrm{Atoroidal SFS $S^2 ((2,1),(3,2),(3,-1))$}\;, \quad p=2
		\\
		\textrm{Atoroidal SFS $S^2 ((2,1),(2,1),(5,-2))$}\;, \quad p=1
		\\
		\textrm{Atoroidal SFS $S^2 ((3,1),(3,1),(3,-1))$}\;, \quad p=0
		\\
        \textrm{Atoroidal SFS $S^2 ((2,1),(4,1),(5,-3))$}\;, \quad p=-1
		\\
        \textrm{Atoroidal SFS $S^2 ((2,1),(3,1),(9,-7))$}\;, \quad p=-2
		\\
		\textrm{Graph}\;, \quad p=-3
		\\
		\textrm{Hyperbolic}\;, \quad p>2 \textrm{ or } p<-3 \ .
	\end{cases}\ . \label{topological types of m007p}
\end{align}
According to \eqref{unitary/non-unitary 3-manifolds}, the  topological field theories $\textrm{TFT}[M]$ for the above Seifert fibered manifolds are unitary when $p=0,2$ while they are non-unitary when $p=1,-1,-2$.
Combined with the mathematical facts, our proposal \eqref{proposal for IR phases} predicts that
\begin{align}
	\begin{split}
	&\CT[(m007)_{pA_{\rm SP}+B_{\rm SP}}]  
	\\
	&=
	\begin{cases}
		\textrm{Unitary TQFT}\;, \quad p=0,2
		\\
		\textrm{$\mathcal{N}=4$ rank-0 SCFT}\;, \quad p=1,-1,-2
		\\
		\textrm{$\CN=2$ SCFT with $U(1)$ flavor symmetry and CPOs}\;, \quad p= -3
		\\
		\textrm{$\mathcal{N}=2$ SCFT}\;, \quad p>2 \textrm{ or } p<-3
	\end{cases}
\end{split}\ .
\end{align}
Now, we will  confirm the expected IR phases using the general methods outlined in Section \ref{sec : strategy}. From now on, we redefine $\nu$ as
\begin{equation}
\begin{aligned}
\nu \leftarrow \left\{ \begin{array}{ll}
\nu\ , & p=0,1,2 \\
2-\nu\ , & p=-1,-2,-3
\end{array} \right.\ .
\end{aligned}
\end{equation}

\paragraph{Unitary topological field theory  when $p=0,2$} The superconformal indices \eqref{I-m007}, \eqref{I-m007-2} for $p=0,2$ are given by
\begin{equation}
\left(\CI_q (u,\nu)\textrm{ of } \frac{\CT[m007, A_{\rm SP} ; A_{\rm SP}+B_{\rm SP}]}{SU(2)_{p-1=-1,1}}\right) = \left(\CI_q (u,\nu)\textrm{ of } \frac{\CT[m007, A_{\rm SP} ; A_{\rm SP} +B_{\rm SP}]}{``SO(3)"_{p-1=-1,1}}\right) =  1\ ,
\end{equation}
thus implying the gapped phase. To probe the topological field theories, let us compute the handle-gluing and fibering operators \eqref{f-m007}. They are given by
\begin{align}
	\begin{split}
		&\left(\{\CH^{-1}_\alpha (m=0, \nu = \pm 1)\}\textrm{ of } \frac{\CT[m007,A_{\rm SP};A_{\rm SP}+B_{\rm SP}]}{SU(2)_{p-1=-1}}\right) = \bigg{ \{}  \frac{1}{4},\frac{1}{4},\frac{1}{4},\frac{1}{4}\bigg{ \}} \;.
		\\
		&\left(\{\CF_\alpha (m=0, \nu = \pm 1)\} \textrm{ of } \frac{\CT[m007,A_{\rm SP};A_{\rm SP}+B_{\rm SP} ]}{SU(2)_{p-1=-1}}\right) =  \bigg{ \{}  e^{\frac{\pi i}3},e^{-\frac{2\pi i}3}, e^{-\frac{\pi i}6}, e^{-\frac{\pi i}6} \bigg{ \}}  \;.
        \\
        &\left(\{\CH^{-1}_\alpha (m=0, \nu =\pm  1)\} \textrm{ of } \frac{\CT[m007,A_{\rm SP};A_{\rm SP}+B_{\rm SP}]}{SU(2)_{p-1=1}}\right) 
		= \bigg{ \{}\frac{1}{2},\frac{1}{2} \bigg{ \}} \;.
		\\
		&\left(\{\CF_\alpha (m=0, \nu = \pm 1)\} \textrm{ of } \frac{\CT[m007,A_{\rm SP}; A_{\rm SP}+B_{\rm SP} ]}{SU(2)_{p-1=1}}\right) 
		=e^{\frac{\pi i }{12}}  \bigg{ \{}1,e^{\frac{\pi i }{2}}  \bigg{ \}}  \;.
	\end{split}
\end{align}
From the above, one can check that
\begin{align}
	&\CH_\alpha  \in \mathbb{R}_+, \;\;\sum_\alpha\frac{1}{\CH_\alpha } =1 \; \textrm{ and } \;
	|\CF_\alpha  |=1\;. \nonumber
\end{align}
These are another evidences for the emergence of TQFTs at IR, see \eqref{F, H in TQFT} and \eqref{sum of 1/H in TQFT}.
 Turning on the real mass parameter $m$, we have 
\begin{align}\hspace{0cm}
	\begin{split}
		&\left(\{\CH^{-1}_\alpha (m, \nu=1)\} \textrm{ of } \frac{\CT[m007,A_{\rm SP};A_{\rm SP}+B_{\rm SP}]}{SU(2)_{p-1=-1}}\right)= \left\{\frac{1}{4} ,\frac{1}{4} ,\frac{1}{4} ,\frac{1}{4} \right\}\;,
           \\
                &\left(\{\CF_\alpha (m, \nu =1)\} \textrm{ of } \frac{\CT[m007,A_{\rm SP};A_{\rm SP}+B_{\rm SP}]}{SU(2)_{p-1=-1}}\right) 
		=e^{\frac{i m^2}{8\pi}}   \left\{e^{\frac{\pi i}3},e^{-\frac{2\pi i}3}, e^{-\frac{\pi i}6}, e^{-\frac{\pi i}6}\right\} \;,
		\\
        &\left(\{\CH^{-1}_\alpha (m, \nu =1)\} \textrm{ of } \frac{\CT[m007,A_{\rm SP};A_{\rm SP}+B_{\rm SP}]}{SU(2)_{p-1=1}}\right) 
		=  e^{-m\nu}\left\{\frac{1}{2} ,\frac{1}{2}\right\} \;,
        \\
                &\left(\{\CF_\alpha (m, \nu=1)\} \textrm{ of } \frac{\CT[m007,A_{\rm SP};A_{\rm SP}+B_{\rm SP}]}{SU(2)_{p-1=1}}\right) 
		=e^{\frac{im^2}{4\pi}}\left\{e^{\frac{\pi i}{12}},e^{\frac{7\pi i}{12}}\right\} \;.
	\end{split}
\end{align}
Note that they are independent of the continuous parameter $m$, modulo ambiguity in \eqref{ambiguity in H and F},  which also imply the emergence of TQFT at IR.

\paragraph{SUSY enhancement  when $p=1,-1,-2$} The superconformal indices \eqref{I-m007}, \eqref{I-m007-2} for $p=1,-1,-2$ are given by\footnote{For $p=1$, the $\mathbb{Z}_2$ 1-form symmetry of $\frac{\CT[m007, A_{\rm SP};A_{\rm SP}+B_{\rm SP}]}{SU(2)_{p-1}}$ decouples at IR. In this case, $\frac{\CT[m007, A_{\rm SP} ;A_{\rm SP} +B_{\rm SP} ]}{``SO(3)"_{p-1=0}}$ is identical to $\frac{\CT[m007, A_{\rm SP} ;A_{\rm SP} +B_{\rm SP} ]}{``SU(2)"_{p-1=0}}$ and we only need to sum over $m_x \in \mathbb{Z}_{\geq 0}$ in  \eqref{I-m007-2} even for the $``SO(3)"$ gauging.}
\begin{align}
	\begin{split}
		& \left(\CI_q (u;\nu=0)\textrm{ of } \frac{\CT[m007, A_{\rm SP} ;A_{\rm SP}+ B_{\rm SP} ]}{SU(2)_{p-1=0}}=\frac{\CT[m007, A_{\rm SP} ;A_{\rm SP} +B_{\rm SP} ]}{``SO(3)"_{p-1=0}}\right)  \\
		 &=1-q-\left(u+\frac{1}{u}\right) q^{3/2}-2 q^2-\left(u+\frac{1}{u}\right) q^{5/2}-2 q^3-\left(u+\frac{1}{u}\right) q^{7/2}-2
		q^4+O\left(q^{5}\right)\ ,
		\\
        &= \left(\CI_q (u;\nu=0)\textrm{ of } \frac{\CT[m007, A_{\rm SP} ; A_{\rm SP}+B_{\rm SP}]}{SU(2)_{p-1=-2}} \textrm{ and } \frac{\CT[m007, A_{\rm SP} ;A_{\rm SP}+ B_{\rm SP}]}{``SO(3)"_{p-1=-2}}\right)  
		\\
		 &=1-q-\left(u+\frac{1}{u}\right) q^{3/2}-2 q^2-\left(u+\frac{1}{u}\right) q^{5/2}-2 q^3-\left(u+\frac{1}{u}\right) q^{7/2}-2
   q^4+O\left(q^{5}\right)\ ,
		\\
		&\left(\CI_q (u;\nu=0)\textrm{ of } \frac{\CT[m007, A_{\rm SP} ; A_{\rm SP}+B_{\rm SP} ]}{SU(2)_{p-1=-3}} \textrm{ and } \frac{\CT[m007, A_{\rm SP} ; A_{\rm SP}+B_{\rm SP} ]}{``SO(3)"_{p-1=-3}}\right)  
		\\
		 &= 1-q-\left(u+\frac{1}{u}\right) q^{3/2}-2 q^2-\frac{q^{5/2}}{u}-\left(1-u^2\right) q^3-\left(\frac{1}{u}-u\right) q^{7/2}+O\left(q^{9/2}\right)\ .
	\end{split}
\end{align}
The index at general R-charge mixing parameter $\nu$ can be obtained by the relation \eqref{SCI under mixing}. To compute the correct IR superconformal index, one needs to determine the  IR superconformal R-symmetry, parametrized by $\nu_{\rm IR}$, using the F-maximization \eqref{F-maximization}. Utilizing the Bethe-sum formula of the round 3-sphere partition function \eqref{round 3-sphere from Bethe-sum}, one can confirm that
\begin{align}
	|\mathcal{Z}_{b=1}^{(m007)_{p}} (m=0, \nu=0)| <|\mathcal{Z}_{b=1}^{(m007)_{p}} (m=0, \nu\neq 0)|\quad \textrm{for $p=1,-1,-2$}\ .
\end{align}
Therefore, $\nu= 0$ corresponds to the correct IR superconformal R-charge according to the F-maximization, i.e. $\nu_{\rm IR}=0$. Then, one finds that the indices at superconformal R-charge satisfy the necessary condition \eqref{SUSY enhancement from index} for the SUSY enhancement:
\begin{align}
&\left(\CI_q (u,\nu=0)\textrm{ of } \frac{\CT[m007, A_{\rm SP} ; A_{\rm SP}+B_{\rm SP}]}{SU(2)_{p-1=0,-2,-3}} \textrm{ and } \frac{\CT[m007, A_{\rm SP} ;A_{\rm SP}+ B_{\rm SP}]}{``SO(3)"_{p-1=0,-2,-3}}\right)   \nonumber
\\ 
&\ni -\left(u+\frac{1}{u}\right)q^{3/2} \ . \nonumber
\end{align}
In the degenerate limits $u=1$, $\nu \rightarrow \pm 1$, the superconformal indices become 1, i.e. 
\begin{align}
	\begin{split}
	&\left(\CI_q (u=1;\nu=\pm 1)\textrm{ of } \frac{\CT[m007,A_{\rm SP};A_{\rm SP}+B_{\rm SP}]}{SU(2)_{p-1=0,-2,-3}} \textrm{ and } \frac{\CT[m007,A_{\rm SP};A_{\rm SP}+B_{\rm SP}]}{``SO(3)"_{p-1=0,-2,-3}}\right)
	\\
	&=1\;.
	\end{split}
\end{align}
This is a strong evidence for emergence of the non-unitary TQFTs in the degenerate limits.  
To probe the emergent non-unitary TQFTs, one may compute the handle-gluing and fibering operators as following: 
\begin{align}\hspace{-0.5cm}
	\begin{split}
    		&\left(\{\CH^{-1}_\alpha (m=0, \nu =  1)\} \textrm{ of } \frac{\CT[m007,A_{\rm SP};A_{\rm SP}+B_{\rm SP}]}{SU(2)_{p-1=0}}\right) = \left\{2(\zeta_3^2)^2,2 (\zeta_3^4)^2 \right\}\;.
		\\
		&\left(\{\CF_\alpha (m=0, \nu =  1)\} \textrm{ of } \frac{\CT[m007,A_{\rm SP};A_{\rm SP}+B_{\rm SP}]}{SU(2)_{p-1=0}}\right) =\left\{  e^{-\frac{\pi i}{5}},e^{\frac{\pi i}{5}}\right\} \;.
        \\
		&\left(\{\CH^{-1}_\alpha (m=0, \nu =-  1)\}\textrm{ of } \frac{\CT[m007,A_{\rm SP};A_{\rm SP}+B_{\rm SP}]}{SU(2)_{p-1=0}}\right) = \left\{ 2(\zeta_3^2)^2,2 (\zeta_3^4)^2 \right\} \;.
		\\
		&\left(\{\CF_\alpha (m=0, \nu = - 1)\} \textrm{ of } \frac{\CT[m007,A_{\rm SP};A_{\rm SP}+B_{\rm SP}]}{SU(2)_{p-1=0}}\right) =\left\{   e^{\frac{\pi i}{5}},e^{-\frac{\pi i}{5}} \right\} \;.
        \\
		&\left(\{\CH^{-1}_\alpha (m=0, \nu =  1)\}\textrm{ of } \frac{\CT[m007,A_{\rm SP};A_{\rm SP}+B_{\rm SP}]}{SU(2)_{p-1=-2}}\right) = \left\{\frac{(\zeta_3^2)^2}{2},\frac{(\zeta_3^2)^2}{2},\frac{(\zeta_3^4)^2}{2},\frac{(\zeta_3^4)^2}{2},(\zeta_3^2)^2,(\zeta_3^4)^2\right\} \;.
		\\
		&\left(\{\CF_\alpha (m=0, \nu =  1)\}\textrm{ of } \frac{\CT[m007,A_{\rm SP};A_{\rm SP}+B_{\rm SP}]}{SU(2)_{p-1=-2}}\right) =\left\{e^{-\frac{7 i \pi }{40}},e^{\frac{33 i \pi }{40}},e^{-\frac{23 i \pi }{40}},e^{\frac{17 i
   \pi }{40}},e^{-\frac{i \pi }{20}},e^{-\frac{9 i \pi }{20}}\right\} \;.
        \\
        &\left(\{\CH^{-1}_\alpha (m=0, \nu =-  1)\} \textrm{ of } \frac{\CT[m007,A_{\rm SP};A_{\rm SP}+B_{\rm SP}]}{SU(2)_{p-1=-2}}\right) = \left\{\frac{(\zeta_3^2)^2}{2},\frac{(\zeta_3^2)^2}{2},\frac{(\zeta_3^4)^2}{2},\frac{(\zeta_3^4)^2}{2},(\zeta_3^2)^2,(\zeta_3^4)^2\right\} \;.
		\\
		&\left(\{\CF_\alpha (m=0, \nu = - 1)\} \textrm{ of } \frac{\CT[m007,A_{\rm SP};A_{\rm SP}+B_{\rm SP}]}{SU(2)_{p-1=-2}}\right) =\left\{e^{-\frac{23 i \pi }{40}},e^{\frac{17 i
   \pi }{40}},e^{-\frac{7 i \pi }{40}},e^{\frac{33 i \pi }{40}},e^{-\frac{9 i \pi }{20}},e^{-\frac{i \pi }{20}}\right\} \;.
        \\
       &\left(\{\CH^{-1}_\alpha (m=0, \nu =  1)\} \textrm{ of } \frac{\CT[m007,A_{\rm SP};A_{\rm SP}+B_{\rm SP}]}{SU(2)_{p-1=-3}}\right)= \left\{ (\zeta_7^4)^2,(\zeta_7^1)^2, (\zeta_7^2)^2 ,(\zeta_7^3)^2\right\}\times \bigg{\{}1, 1 \bigg{\}}\;.  
		\\
		&\left(\{\CF_\alpha (m=0, \nu =  1)\} \textrm{ of } \frac{\CT[m007,A_{\rm SP};A_{\rm SP}+B_{\rm SP}]}{SU(2)_{p-1=-3}}\right) =\left\{e^{-\frac{7 i \pi }{36}},e^{\frac{17 i \pi }{36}},e^{-\frac{31 i \pi }{36}},e^{-\frac{5 i \pi }{12}}\right\}  \times  \{1, e^{\frac{\pi  i}2}\}\;.
        \\
        		&\left(\{\CH^{-1}_\alpha (m=0, \nu =-  1)\}\textrm{ of } \frac{\CT[m007,A_{\rm SP};A_{\rm SP}+B_{\rm SP}]}{SU(2)_{p-1=-3}}\right)= \left\{ (\zeta_7^2)^2,(\zeta_7^1)^2,(\zeta_7^3)^2, (\zeta_7^4)^2 \right\}\times \bigg{\{}1, 1 \bigg{\}}\;.  
		\\
		&\left(\{\CF_\alpha (m=0, \nu = - 1)\} \textrm{ of } \frac{\CT[m007,A_{\rm SP};A_{\rm SP}+B_{\rm SP}]}{SU(2)_{p-1=-3}}\right) =\left\{e^{\frac{17 i \pi }{36}},e^{-\frac{7 i \pi }{36}},e^{-\frac{5 i \pi }{12}},e^{-\frac{31 i \pi }{36}}\right\} \times \{1, e^{\frac{\pi i }2}\}\;.
	\end{split}
\end{align}
Note that the above $(\CH_\alpha,\CF_\alpha)$ satisfy the conditions in \eqref{properties of ran0 SCFT-1}, which are another evidence for the SUSY enhancement.
Here, the Bethe-vacuum $\alpha=0$ is chosen  to satisfy  the following relation:
\begin{align}
|\CZ_{b}(m=0, \nu=\pm 1)| = \bigg{|}\sum_\a \frac{\CF_\a (m=0, \nu=\pm 1) }{\CH_\a (m=0, \nu=\pm 1)} \bigg{|} = \frac{1}{\sqrt{\CH_{\a=0} (m=0,\nu=\pm 1)}}\;.
\end{align}
In addition, one can obtain the value of round 3-sphere partition function at $m=\nu=0$ as following:
\begin{align}
	|\CZ_{b=1}(m=0, \nu=0)| =  \textrm{min}_\alpha[\mathcal{H}_{\alpha}(m=0,\nu=\pm1)^{-1/2}]  = \begin{cases}
		\frac{2}{\sqrt{5}} \sin \left(\frac{4\pi}{5}\right),  \quad p=1
		\\
		\frac{1}{\sqrt{5}} \sin \left(\frac{4\pi}{5}\right), \quad p=-1
		\\
		\frac{\sqrt{2}}{3} \sin \left(\frac{\pi}{9}\right), \quad p=-2
	\end{cases} \ .
\end{align}
Note that this is compatible with \eqref{properties of ran0 SCFT-3}. 
Using the relation in  \eqref{S,T from H,F}, one can compute $|S_{0\alpha}|$ for the topological field theories in the degenerate limits and show that the unitarity condition \eqref{unitarity of TQFT} is violated. It implies that the emergent TQFTs are indeed non-unitary. 

\paragraph{$U(1)$ flavor symmetry and chiral primary operators when $p=-3$} The superconformal indices \eqref{I-m007}, \eqref{I-m007-2} for $p=-3$ are given by
\begin{align}\hspace{0cm}
	\begin{split}
&   \left(\CI_q (u;\nu=0)\textrm{ of } \frac{\CT[m007,A_{\rm SP};A_{\rm SP}+B_{\rm SP}]}{SU(2)_{p-1=-4}}\right) \\
   & =1-\left(1-u^2\right) q-\frac{q^{3/2}}{u}-\left(2-u^4\right) q^2-\left(\frac{1}{u}-u\right) q^{5/2}+u^6 q^3-\left(\frac{1}{u}-2 u\right)
   q^{7/2}+O\left(q^{4}\right)\ ,
   \\
&\left(\CI_q (u;\nu=0)\textrm{ of } \frac{\CT[m007,A_{\rm SP};A_{\rm SP}+B_{\rm SP}]}{``SO(3)"_{p-1=-4}}\right)  
		\\
		& =1-u q^{1/2}-\left(2-u^2\right) q-\left(\frac{1}{u}+u^3\right) q^{3/2}-\left(3-u^4\right) q^2-\left(\frac{2}{u}-2 u+u^5\right)
   q^{5/2}+O\left(q^3\right)\ .
	\end{split}
\end{align}
Then, the indices at  $\nu=1$ (geometric R-charge) become
\begin{align}\hspace{0cm}
	\begin{split}
&\left(\CI_q (u;\nu=1)\textrm{ of } \frac{\CT[m007,A_{\rm SP};A_{\rm SP}+B_{\rm SP}]}{SU(2)_{p-1=-4}}\right)  
\\
&=\left(\sum_{n=0}^\infty u^{2n}\right)\times q^0-q-\left(2-\frac{1}{u}+u\right) q^2+\left(\frac{1}{u}-2 u-2 u^2\right) q^3+O\left(q^4\right)\ ,
   \\
&\left(\CI_q (u;\nu=1)\textrm{ of } \frac{\CT[m007,A_{\rm SP};A_{\rm SP}+B_{\rm SP}]}{``SO(3)"_{p-1=-4}}\right) 
		\\
		& =\left(\sum_{n=0}^\infty u^n\right)\times q^0 -2 q-\left(3-\frac{1}{u}+2 u\right) q^2+\left(\frac{2}{u}-4u-4u^2\right) q^3+O\left(q^4\right)\ .
	\end{split}
\end{align}
As expected, the indices diverge at $\nu=1$ and $u=1$. The infinity is regularized by the fugacity $u$. This infinity   is just  due to the bad choice of R-charge, $\nu=1$, which is different from the correct IR superconformal R-charge.  The correct R-charge should be determined from the F-maximization principle \eqref{F-maximization}, and we numerically find that
\begin{align}
\nu_{\rm IR}\simeq 0.052\;, \quad |\CZ_{b=1}(m=0,\nu=\nu_{\rm IR})| \simeq 0.107532\ .
\end{align}
The  term $\left(\sum_{n=0}^\infty u^{n} \right)$ comes from  $``SO(3)"$ monopole operators which are chiral primaries with $R_{\rm geo}=0$.

\subsection{Example : $M=(m009)_{pA_{\rm SP}+B_{\rm SP}}$} 
\subsubsection{Field theory}  \label{subsec : field theory for m009p}
The corresponding 3D theory is given as
\begin{align}
\begin{split}
&\CT[(m009)_{pA_{\rm SP} +B_{\rm SP}}] =  \frac{\CT[m009,A_{\rm SP};B_{\rm SP}-A_{\rm SP}]}{``SO(3)"_{p+1}}\;, 
\\
&\CT[m009,A_{\rm SP};B_{\rm SP}-A_{\rm SP}] = \left(\U(1)_{-1/2} + 3\Phi s \textrm{ of charge }(+1,+1,-1) \right) + \CW_{\rm sup} \ . \label{T[m009]}
\end{split}
\end{align}
Both of $A_{\rm SP}$ and $B_{\rm Sp}$ are odd cycle and we choose the $B$ cycle as $B_{\rm SP}-A_{\rm SP}$. $A_{\rm SP}$ is a non-closable cycle since $(m009)_{A_{\rm SP}} =L(2,1)  = \mathbb{RP}^3$ \cite{2018arXiv181211940D}. 
The superpotential is 
\begin{align}
\CW_{\rm sup}  =  \Phi_1 \Phi_2 \Phi_3^2 + (V_{-1})^2   \ .
\end{align}
Here, $V_{q}$ denotes the 1/2 BPS bare monopole operator with flux $q$, which is gauge-invariant when $q=-1$. 
Charge assignment of the theory is given in Table \ref{T[m009] : charge table}.
\begin{table}[h]
	\begin{center}
		\begin{tabular}{|c|c|c|c|}
			\hline
			& $ \U(1)_{\rm gauge}$ &  $\U(1)_{x}$ & $\U(1)_{R_{\nu_{\rm geom}}}$ \\
			\hline
			$(\Phi_1, \Phi_2, \Phi_3)$ &  $(+1, +1, -1)$ &  $(+2, -2, 0)$ & $(0,0,+1)$
			\\
			\hline
			$V_q$  & $- \frac{q}2 -\frac{|q|}2$ & $0$  & $|q|$
			\\
			\hline
		\end{tabular}
	\end{center}
	\caption{Charge of elementary fields and local operators in $\CT[m009, A_{\rm SP};B_{\rm SP}-A_{\rm SP}]$}
	\label{T[m009] : charge table}
\end{table}
\\
Since the $A_{\rm SP}$ is a non-closable cycle, the $U(1)_x$ in $\CT[m009, A_{\rm SP};B_{\rm SP}-A_{\rm SP}]$ is expected to be enhanced to $SU(2)$ at IR.  Superconformal index of the theory is
\begin{align}
	\begin{split}
	&\CI^{m009} (m_x, u_x)
	\\
	& =\sum_{m \in  \mathbb{Z}} \oint_{|u|=1}\frac{du}{2\pi i u}  u_x^{4 m_x} u^m \CI_\Delta  \left(m+2m_x, -q^{1/2}u u_x^2\right)  \CI_\Delta  \left(m-2m_x, -q^{1/2}\frac{u} {u_x^2}\right)  \CI_\Delta  \left(-m, \frac{1}{u}\right) \label{I-m009}
	\end{split}
\end{align}
$(m_x,u_x)$ is the $(\textrm{background monopole flux}, \textrm{fugacity})$ for the $U(1)_x$ flavor symmetry. At $m_x=0$, the index is given by
\begin{align}
\begin{split}
&\CI^{m009} \big{(}m_x=0, u_x\big{)} 
\\
&= 1+ \left(-u_x^2-\frac{1}{u_x^2}-1\right)q^{1/2} + \left(-u_x^2-\frac{1}{u_x^2}-1\right) q +  \left(-u_x^2-\frac{1}{u_x^2}-2\right) q^{3/2}-q^2 +\cdots\ .
\end{split}
\end{align}
 The $q^{1/2}$-term comes from the chiral primary operators 
\begin{align}
\begin{split}
&\Phi_1 \Phi_3 \; : \; -q^{1/2} u_x^2\;, \quad  \Phi_2 \Phi_3 \; :\; -q^{1/2} u_x^{-2}\;, \quad  V_- \;: \; -q^{1/2} \;. 
\end{split}
\end{align}
The $q^{1}$-term comes from following operators\footnote{Dressed monopole operators, $\phi_1 V_{+1}$ and $\phi_2 V_{+1}$, are dyonic $1/4$ BPS operators and have spin $j=1/2$. }
\begin{align}
\phi_1 \psi^*_{1+} - \phi_2 \psi^*_{2+} \; :\; -q \;, \quad  \phi_1  V_{+1} \; :\; -q u_x^2\;, \quad \phi_2 V_{+1} \; :\; -q u_x^{-2}\;.
\end{align}
These operators have $r=1, j_3 = 1/2$ and  $\Delta= \frac{3}2$, which can be regarded as conformal primaries in  conserved current multiplets. It  is consistent with the expected IR symmetry enhancement, $U(1)_x \rightarrow SO(3)$. 
For $\CT[(m009)_{pA_{\rm SP}+B{\rm SP}}]$,  we gauge the  enhanced $SO(3)$ flavor symmetry with an additional CS level $(p+1)$. 

\subsubsection{SUSY partition functions}
\paragraph{Squashed 3-sphere partition function} The $S^3_b$ partition function of the theory, \\$\frac{\CT[m009, A_{\rm SP} ;B_{\rm SP}-A{\rm SP}]}{SU(2)_{p+1}} $, is given by 
\begin{align}
	\begin{split}
		&\mathcal{Z}_b^{(m009)_p}= \frac{1}2 \int \frac{dXdZ}{(2\pi \hbar)} \left( 4 \sinh (X) \sinh(\frac{2\pi i X}{\hbar})\right) \exp \left((p+3)\frac{X^2}\hbar +\frac{Z^2}{2\hbar} \right)  
		\\
		&\qquad \qquad \qquad \qquad  \qquad \times \psi_\hbar (Z+2X+\pi i +\frac{\hbar}2 )  \psi_\hbar (Z-2X+\pi i +\frac{\hbar}2 )  \psi_\hbar (-Z)\;.
	\end{split}
\end{align}
\paragraph{Superconformal index} The superconformal index is
\begin{align}
\CI_q^{(m009)_p} &= \sum_{m_x}\oint_{|u_x|=1} \frac{du_x}{2\pi i u_x}     \Delta (m_x, u_x)  u_x^{2(p+1)m_x} \CI^{m009} (m_x, u_x)\;.  \label{I-m009-2}
\end{align}
Here, $(m_x,u_x)$ are (monopole flux, fugacity) for $SU(2)$ gauge symmetry. The summation range for the monopole flux is given by
\begin{align}
 	& \begin{cases}
 		m_x \in  \mathbb{Z}_{\geq 0 }\;, \quad \textrm{for }\frac{\CT[m009, A_{\rm SP} ;B_{\rm SP}-A{\rm SP}]}{SU(2)_{p+1}}
 		\\
 		m_x \in  \frac{1}2\mathbb{Z}_{\geq 0 }\;, \quad \textrm{for }\CT[(m009)_{pA_{\rm SP} +B_{\rm SP} }]=\frac{\CT[m009, A_{\rm SP} ;B_{\rm SP}-A{\rm SP}]}{``SO(3)"_{p+1}} \ .
 		\label{Index for m009p-2}
 	\end{cases}
 \end{align}

\paragraph{Bethe-vacua and handle-gluing/fibering operators}
Expanding the integrand of $\mathcal{Z}_b^{(\textrm{m009})_p}$ in the limit in which $\hbar \to 0$, we get
\begin{equation}
\begin{aligned}
&\mathcal{W}_0^{(\textrm{m009})_p} (X,Z) = \textrm{Li}_2(-e^{-Z-2X}) +\textrm{Li}_2(-e^{-Z+2X}) + \textrm{Li}_2(e^{Z}) + (p+3)X^2 + \frac{1}{2}Z^2 \pm 2 \pi i X  \ ,\\
&\mathcal{W}_1^{(\textrm{m009})_p} (X,Z) =-\frac{1}{2} \log(1-e^{Z}) + \log(2\sinh X)\ .
\end{aligned}
\end{equation}
Extremizing the twisted superpotential $\mathcal{W}_0^{(\textrm{m009})_p}$, we obtain	
\begin{equation}
\begin{aligned}
&\left(\mathcal{S}_{\textrm{BE}} \textrm{ of } \frac{\CT[m009, A_{\rm SP} ;B_{\rm SP}-A_{\rm SP}]}{SU(2)_{p+1}} \right) \\
&= \left\{ (x,z) \;\; : \;\; \frac{x^{2 p+2} \left(x^2 z+1\right)^2}{\left(x^2+z\right)^2}=1\ , \; \frac{\left(\frac{1}{x^2 z}+1\right) \left(x^2+z\right)}{1-z} =1\ , \; x^2 \neq 1 \right\} / \mathbb{Z}^{\rm Weyl}_2\ ,
\end{aligned}
\end{equation}
where $x=e^X, z=e^{Z}$, and the Weyl group $\mathbb{Z}_2$ acts as
\begin{equation}
\mathbb{Z}^{\rm Weyl}_2 \;\; : \;\; (x,z) \; \leftrightarrow \; (1/x,z)\ .
\end{equation}
Then, the handle-gluing and fibering operators are given by
\begin{equation}\label{f-m009}\hspace{0cm}
\begin{aligned}
&\left(\mathcal{H}_\alpha \textrm{ of } \frac{\CT[m009, A_{\rm SP} ;B_{\rm SP}-A_{\rm SP}]}{SU(2)_{p+1}}\right) 
\\
&=\left. \frac{2 x^2 \left((p-1) x^2-(p+3) \left(x^4+x^2+1\right) z^2+2 z \left(-(p-1) x^2+x^4+1\right)\right)}{\left(x^2-1\right)^2 \left(x^2+z\right)
   \left(x^2 z+1\right)} \right|_{(x,z) = (x^{(\alpha)},z^{(\alpha)})}\ , \\
&\left(\mathcal{F}_\alpha \textrm{ of } \frac{\CT[m009, A_{\rm SP} ;B_{\rm SP}-A_{\rm SP}]}{SU(2)_{p+1}}\right) 
\\
&= \left. \exp \left(\frac{i (\CW^{(\textrm{m009})_{p}}_0-2\pi i n^{(\a)}_x X -2\pi i n^{(\a)}_{z} Z)}{2\pi }\right) \right|_{(X,Z) = (\log x^{(\a)}, \log z^{(\a)})}\ . \\
\end{aligned}
\end{equation}
Three integers $(n_x^{(\a)}, n_{z}^{(\a)})$ for each Bethe-vacuum $\a$   are chosen as in \eqref{vecn}.

\subsubsection{IR phases}
According to  \cite{2018arXiv181211940D}, the Dehn filled manifolds are
\begin{align}
	(m009)_{p A_{\rm SP} + B_{\rm SP}}   = 
	\begin{cases}
		\textrm{Atoroidal SFS $S^2 ((3,1),(3,1),(5,-4))$}\;, \quad p=-2
		\\
		\textrm{Atoroidal SFS $S^2 ((2,1),(4,1),(6,-5))$}\;, \quad p=-1
		\\
		\textrm{Atoroidal SFS $S^2 ((2,1),(3,1),(8,-7))$}\;, \quad p=0
		\\
		\textrm{Torus bundle  with } \varphi = \begin{pmatrix}  3 & 2 \\ 1 & 1 \end{pmatrix}\;, \quad p=1
		\\
        \textrm{Atoroidal SFS $S^2 ((2,1),(4,1),(5,-4))$}\;, \quad p=2
		\\
		\textrm{Graph}\;, \quad p= \pm 3
		\\
		\textrm{Hyperbolic}\;, \quad |p|>3\ .
	\end{cases}\ . \label{topological types of m009p}
\end{align}
According to \eqref{unitary/non-unitary 3-manifolds}, the  topological field theories $\textrm{TFT}[M]$ for the above Seifert fibered manifolds are all unitary.
Combined with the mathematical facts, our proposal \eqref{proposal for IR phases} predicts that
\begin{align}
	\begin{split}
	&\CT[(m009)_{pA_{\rm SP}+B_{\rm SP}}]  
	\\
	&=
	\begin{cases}
		\textrm{Unitary TQFT}\;, \quad p=0,\pm 1,\pm 2
		\\
		\textrm{$\CN=2$ SCFT with $U(1)$ flavor symmetry and CPOs}\;, \quad p= \pm 3
		\\
		\textrm{$\mathcal{N}=2$ SCFT}\;, \quad |p|>3 \ .
	\end{cases}
\end{split}\ .
\end{align}

\paragraph{Unitary topological field theory  when $p=0,\pm 1, \pm 2$} The superconformal indices \eqref{I-m009}, \eqref{I-m009-2} for $p=0,\pm 1, \pm 2$ are given by
\begin{equation}
\left(\CI_q\textrm{ of }  \frac{\CT[m009,A_{\rm SP};B_{\rm SP}-A_{\rm SP}]}{SU(2)_{p+1}} \right)=\left(\CI_q\textrm{ of }  \frac{\CT[m009,A_{\rm SP};B_{\rm SP}-A_{\rm SP}]}{``SO(3)"_{p+1}} \right) =  1\ ,
\end{equation}
thus implying the gapped phase. To probe the topological field theories, let us compute the handle-gluing and fibering operators \eqref{f-m009}. They are given by\footnote{For $p=\pm 1$, there is a tricky Bethe-vacuum with $(x,z) = (\pm  i , 1)$. The solution is  visible  only after deforming the Bethe-vacua equation,$\frac{x^{2 p+2} \left(x^2 z+1\right)^2}{\left(x^2+z\right)^2}=1$, by $\frac{x^{2 p+2} \left(x^2 z+1\right)^2}{\left(x^2+z\right)^2}=e^{\epsilon}$ with arbitrary small (but non-zero) $\epsilon$.  }
\begin{align}
	\begin{split}
		&\left(\{\CH^{-1}_\alpha\}\textrm{ of } \frac{\CT[m009, A_{\rm SP} ;B_{\rm SP}-A_{\rm SP}]}{SU(2)_{p+1=1}}\right) = \bigg{ \{}2(\zeta_6^1)^2, 2(\zeta_6^3)^2\bigg{ \}}\times  \bigg{\{} 1,1 \bigg{\}}\;.
		\\
		&\left(\{\CF_\alpha\} \textrm{ of } \frac{\CT[m009, A_{\rm SP} ;B_{\rm SP}-A_{\rm SP}]}{SU(2)_{p+1=1}}\right) =  \left\{e^{\frac{43 i \pi }{48}},e^{-\frac{29 i \pi }{48}}\right\} \times \bigg{\{} 1,e^{\frac{\pi i }2} \bigg{\}}\;.
        \\
        		&\left(\{\CH^{-1}_\alpha\} \textrm{ of } \frac{\CT[m009, A_{\rm SP} ;B_{\rm SP}-A_{\rm SP}]}{SU(2)_{p+1=2}}\right) = \bigg{ \{} \frac{1}{12},\frac{1}{12},\frac{1}{2}, \frac{1}3 \bigg{ \}} \;.
		\\
		&\left(\{\CF_\alpha\} \textrm{ of } \frac{\CT[m009, A_{\rm SP} ;B_{\rm SP}-A_{\rm SP}]}{SU(2)_{p+1=2}}\right) = \left\{e^{\frac{11 i \pi }{12}},e^{-\frac{i \pi }{12}},e^{\frac{19 i \pi }{24}},\pm e^{\frac{i \pi}4}\right\} \;.
        \\
        		&\left(\{\CH^{-1}_\alpha\} \textrm{ of } \frac{\CT[m009, A_{\rm SP} ;B_{\rm SP}-A_{\rm SP}]}{SU(2)_{p+1=0}}\right) = \bigg{ \{} \frac{1}{12},\frac{1}{12},\frac{1}{3},\frac{1}2 \bigg{ \}} \;.
		\\
		&\left(\{\CF_\alpha\} \textrm{ of } \frac{\CT[m009, A_{\rm SP} ;B_{\rm SP}-A_{\rm SP}]}{SU(2)_{p+1=0}}\right) =  \left\{e^{\frac{7 i \pi }{8}},e^{\frac{7 i \pi }{8}},e^{\frac{13 i \pi }{24}}, \pm e^{i \pi }\right\}  \;.
        \\
        		&\left(\{\CH^{-1}_\alpha\} \textrm{ of } \frac{\CT[m009, A_{\rm SP} ;B_{\rm SP}-A_{\rm SP}]}{SU(2)_{p+1=3}}\right) = \bigg{ \{} (\zeta_3^1)^2, (\zeta_3^2)^2\bigg{ \}}  \times \bigg{\{} 1,1 \bigg{\}} \;.
		\\
		&\left(\{\CF_\alpha\} \textrm{ of } \frac{\CT[m009, A_{\rm SP} ;B_{\rm SP}-A_{\rm SP}]}{SU(2)_{p+1=3}}\right) = \left\{e^{\frac{53 i \pi }{120}},e^{-\frac{43 i \pi }{120}}\right\} \bigg{\{} 1,e^{\frac{\pi i}2} \bigg{\}}\;.  
        \\
        		&\left(\{\CH^{-1}_\alpha\} \textrm{ of } \frac{\CT[m009, A_{\rm SP} ;B_{\rm SP}-A_{\rm SP}]}{SU(2)_{p+1=-1}}\right) = \bigg{ \{} (\zeta_3^1)^2, (\zeta_3^2)^2\bigg{ \}} \times \bigg{\{} 1,1 \bigg{\}} \;.
		\\
		&\left(\{\CF_\alpha\} \textrm{ of } \frac{\CT[m009, A_{\rm SP} ;B_{\rm SP}-A_{\rm SP}]}{SU(2)_{p+1=-1}}\right) = \left\{e^{\frac{7 i \pi }{20}},e^{-\frac{17 i \pi }{20}}\right\} \times \bigg{\{} 1,e^{\frac{\pi i }2} \bigg{\}} \;. 
	\end{split}
\end{align}
One can easily see that
\begin{equation}
\CH_\a \in \mathbb{R}_+\  , \quad \sum_\alpha \frac{1}{\CH_\a} = 1\ , \quad |\CF_\alpha|= 1\ , 
\end{equation}
which also imply the emergence of TQFT at IR, see \eqref{F, H in TQFT} and \eqref{sum of 1/H in TQFT}. For $p=\pm 1$ case, the $\CF_\alpha$ associated with the tricky Bethe-vacuum, $(x,z)_\alpha=(\pm i , 1)$, has a sign ambiguity and the sign depends on the phase factor of the small parameter  $\epsilon$. It means that $(\CF_\alpha)^2$ (instead of $\CF_\alpha$) is well-defined and  $ \frac{\CT[m009, A_{\rm SP} ;B_{\rm SP}-A_{\rm SP}]}{SU(2)_{p+1=0}}|_{p=\pm 1} $ actually flow to fermionic TQFTs at IR.

\paragraph{$U(1)$ flavor symmetry and chiral primary operators when $p= \pm 3$} The superconformal indices \eqref{I-m009}, \eqref{I-m009-2} for $p=\pm 3$ are given by\footnote{In this case, the UV $\mathbb{Z}_2$ 1-form symmetry decouples at IR. Thus, we only need to sum over $m_x \in \mathbb{Z}$ even for the ``SO(3)" gauging as discussed below \eqref{Index for Np-2}.}
\begin{align}
	\begin{split}
		& \left(\CI_q \textrm{ of }  \frac{\CT[m009,A_{\rm SP};B_{\rm SP}-A_{\rm SP}]}{SU(2)_{p+1}} =  \frac{\CT[m009,A_{\rm SP};B_{\rm SP}-A_{\rm SP}]}{``SO(3)"_{p+1}}\right) 
		\\
		& =\left(\sum_{n=0}^\infty 1\right) \times q^0 - q^{3/2}- q^2- q^{5/2}- q^3-2 q^{7/2}-2 q^4-2 q^{9/2}+O\left(q^{5}\right)\ .
	\end{split}
\end{align}
As expected, the index diverges. The  term $\left(\sum_{n=0}^\infty 1 \right)$ comes from  $``SO(3)"$ monopole operators, which are chiral primaries with $R_{\rm geo}=0$. This strongly suggests that there is an accidental symmetry at IR and the correct superconformal R-symmetry is a mixing between the $U(1)_{\rm geo}$ and the accidental symmetry. Under the correct IR R-symmetry, the unitarity bound $R_{\rm IR}> \frac{1}2$ should be met for the 1/2 BPS monopole operators.

\section*{Acknowledgements}
We would like to thank Zhihao Duan, Dongwook Ghim, Seok Kim, Ki-Hong Lee, Kimyeong Lee, Sungjay Lee, Sakura Schafer-Nameki, Jaewon Song, Yuji Tachikawa and Jingxiang Wu for useful discussion.
The work was presented at the workshop "Geometry, Representation Theory and Quantum Fields" and the webinar series "Symmetry Seminar." We thank the organizers and  participants.
SC thanks SNU for kind hospitality during his visiting, where part of this work was done.
The work of SC is supported by a KIAS Individual Grant (PG081601) at Korea Institute for Advanced Study.
The work of DG is supported in part by the National Research Foundation of Korea grant NRF-2021R1G1A1095318 and NRF-2022R1C1C1011979. DG also acknowledges support by Creative-Pioneering Researchers Program through Seoul National University.
The research of HK is supported by Samsung Science and Technology Foundation under Project Number SSTF-BA2002-05 and by the National Research Foundation of Korea (NRF) grant funded by the Korea government (MSIT) (No. 2018R1D1A1B07042934).

\appendix

\section{3D indices for $M= N_{PA_{\rm SP}+QB_{\rm SP}}$ with $N=m003,m004,\ldots$} \label{App : 3D index examples}

In the appendix \ref{App : 3D index examples}, we test our proposals given in Fig.~\ref{fig: census of non-hyperbolics} and Eqn.~\eqref{3D index pattern} for the 3D index for a large number of (179) closed non-hyperbolic 3-manifolds by explicit computations. Our examples cover all non-hyperbolic 3-manifolds constructed by exceptional Dehn fillings from 1-cusped hyperbolic manifolds whose number of tetrahedra in the ideal triangulation equals 2 (m003, m004), and 3 (m006, m007, m009, m010, m011, m015, m016, m017, m019). Several examples with 4 (m022, m023, m026, m027, m029, m030, m036, m043, m130, m135, m136, m160, m207), 5 (m247, m249, m294, m410), and 7 (v2050, v2051, v2099, v2274, v2334) ideal tetrahedra will also be carried out to construct the atoroidal connected sum (such as $\mathbb{RP}^3 \, \sharp \, \mathbb{RP}^3$), toroidal SFS, and hyperbolic piece in JSJ, in the left of Fig.~\ref{fig: census of non-hyperbolics}, by exceptional Dehn fillings. 
We will not analyze the toroidal connected sum, such as $\mathbb{T}^3 \, \sharp \, L(3,1)$, here.
All these non-hyperbolic 3-manifolds and their topological types, corresponding IR phases based on our proposal in \eqref{proposal for IR phases} are listed in Table~\ref{Appen-tab}. 
For simplicity, we shall only compute the 3D index. According to the  3D-3D relation in \eqref{3D-3D relations}, this invariant is equal to (possibly with an overall factor $2$) the superconformal index of class R theory at geometric R-charge with all the flavor fugacities  turned off. 

	\begin{longtable}{ |c|c|c|c| }
		\hline
		 $M$ & Topological type & IR phase
		\\ 
		\hline
		$m003_{-2,1}$  & Finite/Real/Unitary (atoroidal SFS) & Gapped/TQFT
		\\
		$m003_{-1,1}$  & Empty (lens space) & SUSY broken
		\\
		$m003_{0,1}$  & Empty (lens space) & SUSY broken
		\\
		$m003_{1,1}$  & Finite/Real/Unitary (atoroidal SFS) & Gapped/TQFT
		\\
		$m003_{1,0}$  & Empty (lens space) & SUSY broken
		\\
		$m003_{-3,2}$  & Infinite (graph) & $\CN=2$ SCFT with CPOs
		\\
		$m003_{-1,2}$  & Finite/Real/Unitary (SOL) & Gapped/TQFT
		\\
		$m003_{1,2}$  & Infinite (graph) & $\CN=2$ SCFT with CPOs
		\\
		\hline
		$m004_{\pm 4,1}$  & Infinite (graph) & $\CN=2$ SCFT with CPOs
		\\
		$m004_{\pm 3,1}$  & Finite/Real/Unitary (atoroidal SFS) & Gapped/TQFT
		\\
		$m004_{\pm 2,1}$  & Finite/Real/Unitary (atoroidal SFS) & Gapped/TQFT
		\\
		$m004_{\pm 1,1}$  & Finite/Real/Unitary (atoroidal SFS) & Gapped/TQFT
		\\
		$m004_{0,1}$  & Finite/Real/Unitary (SOL) & Gapped/TQFT
		\\
		$m004_{1,0}$  & Empty (lens space $S^3 \cong L(1,1)$) & SUSY broken
		\\
		\hline
		$m006_{-3,1}$  &  Infinite (graph) & $\CN=2$ SCFT with CPOs
		\\
		$m006_{-2,1}$  & Finite/Real/Non-unitary (atoroidal SFS) & $\CN=4$ rank-0 SCFT
		\\
		$m006_{-1,1}$  & Finite/Real/Unitary (atoroidal SFS) & Gapped/TQFT
		\\
		$m006_{0,1}$  & Empty (lens space) & SUSY broken
		\\
		$m006_{1,1}$  & Finite/Real/Unitary (atoroidal SFS) & Gapped/TQFT
		\\
		$m006_{1,0}$  & Empty (lens space) & SUSY broken
		\\
		$m006_{1,2}$  & Infinite (graph) & $\CN=2$ SCFT with CPOs
		\\
		\hline
		$m007_{-3,1}$   & Infinite (graph) & $\CN=2$ SCFT with CPOs
		\\
		$m007_{-2,1}$  & Finite/Real/Non-unitary (atoroidal SFS) & $\CN=4$ rank-0 SCFT
		\\
		$m007_{-1,1}$  & Finite/Real/Non-unitary (atoroidal SFS) & $\CN=4$ rank-0 SCFT
		\\
		$m007_{0,1}$  & Finite/Real/Unitary (atoroidal SFS) & Gapped/TQFT
		\\
		$m007_{1,1}$  & Finite/Real/Non-unitary (atoroidal SFS) & $\CN=4$ rank-0 SCFT
		\\
		$m007_{2,1}$  & Finite/Real/Unitary (atoroidal SFS) & Gapped/TQFT
		\\
		$m007_{1,0}$  & Empty (lens space) & SUSY broken
		\\
		\hline
		$m009_{-3,1}$  & Infinite (graph) & $\CN=2$ SCFT with CPOs
		\\
		$m009_{-2,1}$  & Finite/Real/Unitary (atoroidal SFS) & Gapped/TQFT
		\\
		$m009_{-1,1}$  & Finite/Real/Unitary (atoroidal SFS) & Gapped/TQFT
		\\
		$m009_{0,1}$  & Finite/Real/Unitary (atoroidal SFS) & Gapped/TQFT
		\\
		$m009_{1,1}$  & Finite/Real/Unitary (SOL) & Gapped/TQFT
		\\
		$m009_{2,1}$ & Finite/Real/Unitary (atoroidal SFS) & Gapped/TQFT
		\\
		$m009_{3,1}$  & Infinite (graph) & $\CN=2$ SCFT with CPOs
		\\
		$m009_{1,0}$  & Empty (lens space $\mathbb{RP}^3 \cong L(2,1)$) & SUSY broken
		\\
		\hline
		$m010_{-2,1}$  & Finite/Real/Unitary (SOL) & Gapped/TQFT
		\\
		$m010_{-1,1}$  & Infinite (atoroidal connected sum) & $\CN=2$ SCFT with CPOs
		\\
		$m010_{0,1}$  & Finite/Real/Unitary (atoroidal SFS) & Gapped/TQFT
		\\
		$m010_{1,1}$  & Finite/Real/Unitary (atoroidal SFS) & Gapped/TQFT
		\\
		$m010_{2,1}$  & Infinite (graph) & $\CN=2$ SCFT with CPOs
		\\
		$m010_{1,0}$  & Empty (lens space) & SUSY broken
		\\
        \hline
        $m011_{-1,1}$  & Finite/Real/Unitary (atoroidal SFS) & Gapped/TQFT
        \\
        $m011_{0,1}$  & Empty (lens space) & SUSY broken
        \\        
        $m011_{1,1}$  & Finite/Real/Unitary (atoroidal SFS) & Gapped/TQFT
        \\
        $m011_{2,1}$  & Finite/Real/Non-unitary (atoroidal SFS) & $\CN=4$ rank-0 SCFT
        \\
        $m011_{1,0}$  & Empty (lens space) & SUSY broken
        \\
        $m011_{-1,2}$  & Infinite (graph) & $\CN=2$ SCFT with CPOs
        \\
        \hline    
        $m015_{-2,1}$  & Infinite (graph) & $\CN=2$ SCFT with CPOs
        \\
        $m015_{-1,1}$  & Finite/Real/Non-unitary (atoroidal SFS) & $\CN=4$ rank-0 SCFT
        \\
        $m015_{0,0}$  & Finite/Real/Non-unitary (atoroidal SFS) & $\CN=4$ rank-0 SCFT
        \\        
        $m015_{1,1}$  & Finite/Real/Non-unitary (atoroidal SFS) & $\CN=4$ rank-0 SCFT
        \\
        $m015_{2,1}$  & Infinite (graph) & $\CN=2$ SCFT with CPOs
        \\
        $m015_{1,0}$  & Empty (lens space $S^3 \cong L(1,1)$) & SUSY broken
        \\
        \hline
        $m016_{-2,1}$  & Infinite (graph) & $\CN=2$ SCFT with CPOs
        \\
        $m016_{-1,1}$  & Empty (lens space) & SUSY broken
        \\
        $m016_{0,1}$  & Empty (lens space) & SUSY broken
        \\        
        $m016_{1,1}$  & Finite/Real/Non-unitary (atoroidal SFS) & $\CN=4$ rank-0 SCFT
        \\
        $m016_{2,1}$  & Infinite (graph) & $\CN=2$ SCFT with CPOs
        \\
        $m016_{1,0}$  & Empty (lens space $S^3 \cong L(1,1)$) & SUSY broken
        \\
        $m016_{-1,2}$  & Infinite (graph) & $\CN=2$ SCFT with CPOs
        \\
        \hline
        $m017_{-2,1}$  & Infinite (graph) & $\CN=2$ SCFT with CPOs
        \\
        $m017_{-1,1}$  & Empty (lens space) & SUSY broken
        \\
        $m017_{0,1}$  & Empty (lens space) & SUSY broken
        \\        
        $m017_{1,1}$  & Finite/Real/Non-unitary (atoroidal SFS) & $\CN=4$ rank-0 SCFT
        \\
        $m017_{2,1}$  & Infinite (graph) & $\CN=2$ SCFT with CPOs
        \\
        $m017_{1,0}$  & Empty (lens space) & SUSY broken
        \\
        $m017_{-1,2}$  & Infinite (graph) & $\CN=2$ SCFT with CPOs
        \\
        \hline
        $m019_{-2,1}$  & Infinite (graph) & $\CN=2$ SCFT with CPOs
        \\
        $m019_{-1,1}$  & Finite/Real/Non-unitary (atoroidal SFS) & $\CN=4$ rank-0 SCFT
        \\
        $m019_{0,1}$  & Empty (lens space) & SUSY broken
        \\        
        $m019_{1,1}$  & Empty (lens space) & SUSY broken
        \\
        $m019_{1,0}$  & Empty (lens space) & SUSY broken
        \\
        $m019_{1,2}$  & Infinite (graph) & $\CN=2$ SCFT with CPOs
        \\
        \hline
        $m022_{-2,1}$  & Infinite (graph) & $\CN=2$ SCFT with CPOs
        \\
        $m022_{-1,1}$ & Finite/Real/Unitary (atoroidal SFS)  & Gapped/TQFT
        \\
        $m022_{0,1}$  & Finite/Real/Unitary (atoroidal SFS)  & Gapped/TQFT
        \\
        $m022_{1,1}$  & Empty (lens space) & SUSY broken
        \\
        $m022_{2,1}$  & Finite/Real/Unitary (SOL)  & Gapped/TQFT
        \\
        $m022_{1,0}$  & Empty (lens space) & SUSY broken
        \\        
        \hline
        $m023_{-2,1}$ & Finite/Real/Unitary (atoroidal SFS)  & Gapped/TQFT
        \\
        $m023_{-1,1}$ & Finite/Real/Unitary (SOL)  & Gapped/TQFT
        \\
        $m023_{0,1}$  & Finite/Real/Unitary (atoroidal SFS)  & Gapped/TQFT
        \\
        $m023_{1,1}$  & Finite/Real/Unitary (atoroidal SFS)  & Gapped/TQFT
        \\
        $m023_{2,1}$  & Finite/Real/Unitary (atoroidal SFS)  & Gapped/TQFT
        \\
        $m023_{3,1}$ &  Infinite (graph) & $\CN=2$ SCFT with CPOs
        \\
        $m023_{1,0}$  & Empty (lens space) & SUSY broken
        \\
        \hline 
        $m026_{-2,1}$  & Finite/Real/Non-unitary (atoroidal SFS) & $\CN=4$ rank-0 SCFT
        \\
        $m026_{-1,1}$  & Finite/Real/Unitary (atoroidal SFS)  & Gapped/TQFT
        \\
        $m026_{0,1}$  & Empty (lens space) & SUSY broken
        \\
        $m026_{1,1}$ & Finite/Real/Unitary (atoroidal SFS) & Gapped/TQFT
        \\
        $m026_{1,0}$ & Empty (lens space) & SUSY broken
        \\
        $m026_{1,2}$ &  Infinite (graph) & $\CN=2$ SCFT with CPOs
        \\
        \hline      
        $m027_{-1,1}$  & Finite/Real/Unitary (atoroidal SFS)  & Gapped/TQFT
        \\
        $m027_{0,1}$  & Empty (lens space) & SUSY broken
        \\        
        $m027_{1,1}$ & Finite/Real/Unitary (atoroidal SFS)  & Gapped/TQFT
        \\
        $m027_{2,1}$  & Finite/Real/Non-unitary (atoroidal SFS) & $\CN=4$ rank-0 SCFT
        \\
        $m027_{1,0}$ & Empty (lens space) & SUSY broken
        \\
        $m027_{-1,2}$ &  Infinite (graph) & $\CN=2$ SCFT with CPOs
        \\
        \hline          
        $m029_{-1,1}$  & Finite/Real/Non-unitary (atoroidal SFS) & $\CN=4$ rank-0 SCFT
        \\
        $m029_{0,1}$ & Finite/Real/Unitary (atoroidal SFS)  & Gapped/TQFT
        \\
        $m029_{1,1}$ & Finite/Real/Non-unitary (atoroidal SFS) & $\CN=4$ rank-0 SCFT
        \\
        $m029_{2,1}$  & Finite/Real/Non-unitary (atoroidal SFS) & $\CN=4$ rank-0 SCFT
        \\
        $m029_{3,1}$ &  Infinite (graph) & $\CN=2$ SCFT with CPOs
        \\
        $m029_{1,0}$  & Empty (lens space) & SUSY broken
        \\        
        \hline      
        $m030_{-2,1}$ &  Infinite (graph) & $\CN=2$ SCFT with CPOs
        \\   
        $m030_{-1,1}$  & Finite/Real/Non-unitary (atoroidal SFS) & $\CN=4$ rank-0 SCFT
        \\
        $m030_{0,1}$ & Finite/Real/Non-unitary (atoroidal SFS) & $\CN=4$ rank-0 SCFT
        \\
        $m030_{1,1}$  & Finite/Real/Unitary (atoroidal SFS)  & Gapped/TQFT
        \\        
        $m030_{2,1}$ & Finite/Real/Non-unitary (atoroidal SFS) & $\CN=4$ rank-0 SCFT
        \\
        $m030_{1,0}$  & Empty (lens space) & SUSY broken
        \\
        \hline      
        $m036_{-2,1}$  & Infinite (toroidal SFS)  & $\CN=2$ SCFT with CPOs
        \\
        $m036_{-1,1}$  & Empty (lens space) & SUSY broken
        \\
        $m036_{0,1}$ & Finite/Real/Unitary (atoroidal SFS)  & Gapped/TQFT
        \\
        $m036_{1,1}$ & Finite/Real/Non-unitary (atoroidal SFS) & $\CN=4$ rank-0 SCFT
        \\
        $m036_{2,1}$ &  Infinite (graph) & $\CN=2$ SCFT with CPOs
        \\
        $m036_{1,0}$  & Empty (lens space) & SUSY broken
        \\        
        \hline            
        $m043_{-1,1}$  & Empty (lens space) & SUSY broken
        \\
        $m043_{0,1}$  & Empty (lens space) & SUSY broken
        \\        
        $m043_{1,1}$ & Finite/Real/Non-unitary (atoroidal SFS) & $\CN=4$ rank-0 SCFT
        \\
        $m043_{2,1}$  &  Infinite (graph) & $\CN=2$ SCFT with CPOs
        \\
        $m043_{1,0}$ & Empty (lens space $S^2\times S^1 \cong L(0,1)$)    & SUSY broken
        \\
        $m043_{-1,2}$ &  Infinite (graph) & $\CN=2$ SCFT with CPOs
        \\
        \hline     
        $m130_{-2,1}$  &  Infinite (graph) & $\CN=2$ SCFT with CPOs
        \\
        $m130_{-1,1}$ & Infinite (toroidal SFS)  & $\CN=2$ SCFT with CPOs
        \\
        $m130_{0,1}$  & Empty (lens space) & SUSY broken
        \\
        $m130_{1,1}$ & Finite/Real/Unitary (atoroidal SFS)  & Gapped/TQFT
        \\
        $m130_{1,0}$  & Empty (lens space) & SUSY broken
        \\        
        $m130_{1,2}$ &  Infinite (graph) & $\CN=2$ SCFT with CPOs
        \\
        \hline       
        $m135_{-1,1}$ & Finite/Real/Unitary (SOL)  & Gapped/TQFT
        \\
        $m135_{0,1}$  & Infinite (atoroidal connected sum)  & $\CN=2$ SCFT with CPOs
        \\
        $m135_{1,1}$ &  Infinite (graph) & $\CN=2$ SCFT with CPOs
        \\
        $m135_{1,0}$   & Infinite (atoroidal connected sum)  & $\CN=2$ SCFT with CPOs
        \\
        \hline       
        $m136_{\pm 2,1}$ &  Infinite (graph) & $\CN=2$ SCFT with CPOs
        \\
        $m136_{\pm 1,1}$& Finite/Real/Unitary (atoroidal SFS)  & Gapped/TQFT
        \\
        $m136_{0,1}$   & Finite/Real/Unitary (SOL)  & Gapped/TQFT
        \\        
        $m136_{1,0}$  & Infinite (atoroidal connected sum $\mathbb{RP}^3 \, \sharp \, \mathbb{RP}^3$)  & $\CN=2$ SCFT with CPOs
        \\
        \hline    
        $m160_{-2,1}$ & Finite/Real/Non-unitary (atoroidal SFS) & $\CN=4$ rank-0 SCFT
        \\
        $m160_{-1,1}$ & Finite/Real/Non-unitary (atoroidal SFS) & $\CN=4$ rank-0 SCFT
        \\
        $m160_{0,1}$  & Infinite (atoroidal connected sum) & $\CN=2$ SCFT with CPOs
        \\
        $m160_{1,0}$   & Finite/Real/Unitary (atoroidal SFS)  & Gapped/TQFT
        \\        
        \hline    
        $m207_{-1,1}$ & Finite/Real/Unitary (SOL)  & Gapped/TQFT
        \\
        $m207_{0,1}$ & Infinite (atoroidal connected sum)  & $\CN=2$ SCFT with CPOs
        \\
        $m207_{1,0}$  & Infinite (atoroidal connected sum) & $\CN=2$ SCFT with CPOs
        \\        
        \hline     
        $m247_{-1,1}$ & Infinite (toroidal SFS)  & $\CN=2$ SCFT with CPOs
        \\
        $m247_{0,1}$ & Empty (lens space) & SUSY broken
        \\
        $m247_{1,0}$   & Infinite (atoroidal connected sum)  & $\CN=2$ SCFT with CPOs
        \\        
        \hline      
        $m249_{-1,1}$ & Finite/Real/Unitary (atoroidal SFS)  & Gapped/TQFT
        \\
        $m249_{0,1}$    & Infinite (toroidal SFS)  & $\CN=2$ SCFT with CPOs
        \\        
        $m249_{1,0}$ & Empty (lens space) & SUSY broken
        \\
        \hline          
        $m294_{0,1}$  & Finite/Real/Non-unitary (atoroidal SFS) & $\CN=4$ rank-0 SCFT 
        \\         
        $m294_{1,1}$    & Finite/Real/Unitary (toroidal SFS)  & Gapped/TQFT
        \\        
        $m294_{1,0}$  & Infinite (atoroidal connected sum)  & $\CN=2$ SCFT with CPOs
        \\
        $m294_{1,2}$ & Infinite (graph)  & $\CN=2$ SCFT with CPOs
        \\
        \hline           
        $m410_{0,1}$  & Finite/Real/Non-unitary (atoroidal SFS) & $\CN=4$ rank-0 SCFT 
        \\
        $m410_{1,1}$ & Infinite (graph)  & $\CN=2$ SCFT with CPOs
        \\  
        $m410_{1,0}$  & Infinite (atoroidal connected sum $\mathbb{RP}^3 \, \sharp \, S^2 \times S^1$)  & $\CN=2$ SCFT with CPOs
        \\
        \hline    
        $v2050_{-2,1}$  & Infinite (hyperbolic piece in JSJ)  & $\CN=2$ SCFT with CPOs
        \\
        $v2050_{1,0}$  & Empty (lens space) & SUSY broken
        \\         
        \hline    
        $v2051_{-2,1}$  & Infinite (hyperbolic piece in JSJ)  & $\CN=2$ SCFT with CPOs
        \\
        $v2051_{1,1}$  & Infinite (graph)  & $\CN=2$ SCFT with CPOs
        \\
        $v2051_{1,0}$   & Finite/Real/Unitary (atoroidal SFS)  & Gapped/TQFT
        \\
        \hline    
        $v2099_{-2,1}$  & Infinite (hyperbolic piece in JSJ)  & $\CN=2$ SCFT with CPOs
        \\
        $v2099_{0,1}$  & Infinite (graph)  & $\CN=2$ SCFT with CPOs
        \\
        $v2099_{1,1}$  & Finite/Real/Non-unitary (atoroidal SFS) & $\CN=4$ rank-0 SCFT 
        \\
        $v2099_{1,0}$   & Finite/Real/Unitary (atoroidal SFS)  & Gapped/TQFT
        \\
        \hline
        $v2274_{-2,1}$   & Infinite (hyperbolic piece in JSJ)  & $\CN=2$ SCFT with CPOs
        \\
        $v2274_{2,1}$   & Infinite (hyperbolic piece in JSJ)  & $\CN=2$ SCFT with CPOs
        \\
        $v2274_{1,0}$ & Finite/Real/Unitary (atoroidal SFS)  & Gapped/TQFT
        \\
        \hline
        $v2334_{-2,1}$   & Infinite (hyperbolic piece in JSJ)  & $\CN=2$ SCFT with CPOs
        \\
        $v2334_{0,1}$  & Finite/Real/Non-unitary (atoroidal SFS) & $\CN=4$ rank-0 SCFT 
        \\
        $v2334_{1,1}$   & Finite/Real/Non-unitary (atoroidal SFS) & $\CN=4$ rank-0 SCFT 
        \\    
        $v2334_{1,0}$ & Finite/Real/Unitary (atoroidal SFS)  & Gapped/TQFT
        \\
        \hline
        \caption{Topological types of non-hyperbolic 3-manifolds obtained by exceptional Dehn fillings and the IR phases for the associated class R theories. Here $N_{P,Q}$ denotes the 3-manifold  $M=N_{PA_{\rm SP}+QB_{\rm SP}}$.  Refer to \cite{2018arXiv181211940D} for the  topological types of the non-hyperbolic 3-manifolds in the conventional mathematical classification scheme.}
        \label{Appen-tab}
	\end{longtable}

The 3D indices for the class R theories associated to non-hyperbolic 3-manifolds listed in Table \ref{Appen-tab} are given as follows. Below, $\CI_q(N_{P,Q})$ denotes the 3D index $\CI_{M} (q)$ for $M=N_{PA_{\rm SP}+Q B_{\rm SP}}$.
\begingroup
\allowdisplaybreaks
\begin{align}
        &\CI_q(m003_{-2,1}) = 1\ , \nonumber\\
        &\CI_q(m003_{-1,1}) = 0\ , \nonumber\\
        &\CI_q(m003_{0,1}) = 0\ , \nonumber\\
        &\CI_q(m003_{1,1}) = 1\ , \nonumber\\
        &\CI_q(m003_{1,0}) = 0\ , \nonumber\\
        &\CI_q(m003_{-3,2}) = \infty \times q^0 -2q -4q^2+ O(q^3)\ , \nonumber\\
        &\CI_q(m003_{-1,2}) = 1\ , \nonumber\\
        &\CI_q(m003_{1,2}) = \infty \times q^0 -2q -4q^2+ O(q^3)\ , \nonumber\\
        &\CI_q(m004_{-4,1}) = \infty \times q^0 -2q^2 -2q^3 -4q^4 -4q^5 -6q^6 +O(q^7)\ , \nonumber\\
        &\CI_q(m004_{-3,1}) = 1\ , \nonumber\\
        &\CI_q(m004_{-2,1}) = 1\ , \nonumber\\
        &\CI_q(m004_{-1,1}) = 1\ , \nonumber\\
        &\CI_q(m004_{0,1}) = 1\ , \nonumber\\
        &\CI_q(m004_{1,1}) = 1\ , \nonumber\\
        &\CI_q(m004_{2,1}) = 1\ , \nonumber\\
        &\CI_q(m004_{3,1}) = 1\ , \nonumber\\
        &\CI_q(m004_{4,1}) = \infty \times q^0 -2q^2 -2q^3 -4q^4 -4q^5 -6q^6 +O(q^7)\ , \nonumber\\
        &\CI_q(m004_{1,0}) = 0\ , \nonumber\\
        &\CI_q(m006_{-3,1}) = \infty \times q^0 -2q -6q^2 -6q^3 + O(q^4)\ , \nonumber\\
        &\CI_q(m006_{-2,1}) = 1\ , \nonumber\\
        &\CI_q(m006_{-1,1}) = 1\ , \nonumber\\
        &\CI_q(m006_{0,1}) = 0\ , \nonumber\\
        &\CI_q(m006_{1,1}) = 2\ , \nonumber\\
        &\CI_q(m006_{1,0}) = 0\ , \nonumber\\
        &\CI_q(m006_{1,2}) = \infty \times q^0 -2q -4q^2+O(q^3)\ , \nonumber\\
        &\CI_q(m007_{-3,1}) = \infty \times q^0 -2q -4q^2 -6q^3 + O(q^4)\ , \nonumber\\
        &\CI_q(m007_{-2,1}) = 1\ , \nonumber\\
        &\CI_q(m007_{-1,1}) = 1\ , \nonumber\\
        &\CI_q(m007_{0,1}) = 1\ , \nonumber\\
        &\CI_q(m007_{1,1}) = 2\ , \nonumber\\
        &\CI_q(m007_{2,1}) = 1\ , \nonumber\\
        &\CI_q(m007_{1,0}) = 0\ , \nonumber\\
        &\CI_q(m009_{-3,1}) = \infty \times q^0 -2q^{3/2}-2q^2 -2q^{5/2}-2q^3 -4q^{7/2} -4q^4 -4q^{9/2} + O(q^5)\ , \nonumber\\
        &\CI_q(m009_{-2,1}) = 1\ , \nonumber\\
        &\CI_q(m009_{-1,1}) = 1\ , \nonumber\\
        &\CI_q(m009_{0,1}) = 1\ , \nonumber\\
        &\CI_q(m009_{1,1}) = 1\ , \nonumber\\
        &\CI_q(m009_{2,1}) = 1\ , \nonumber\\
        &\CI_q(m009_{3,1}) = \infty \times q^0 -2q^{3/2}-2q^2 -2q^{5/2}-2q^3 -4q^{7/2} -4q^4 -4q^{9/2} + O(q^5)\ , \nonumber\\
        &\CI_q(m009_{1,0}) = 0\ , \nonumber\\
        &\CI_q(m010_{-2,1}) = 1\ , \nonumber\\
        &\CI_q(m010_{-1,1}) = \cdots + q^{-2}-q^{-3/2}+q^{-1}-q^{-1/2}+1-q^{1/2}+q-q^{3/2}+q^2+O(q^{5/2})\ , \nonumber\\
        &\CI_q(m010_{0,1}) = 2\ , \nonumber\\
        &\CI_q(m010_{1,1}) = 1\ , \nonumber\\
        &\CI_q(m010_{2,1}) = \infty \times q^0 -q^{1/2}-q-2q^{3/2}-3q^2-2q^{5/2}-2q^3- 2q^{7/2} -5q^4 + O(q^{9/2})\ , \nonumber\\
        &\CI_q(m010_{1,0}) = 0\ , \nonumber\\
        &\CI_q(m011_{-1,1}) = 1\ , \nonumber\\
        &\CI_q(m011_{0,1}) = 0\ , \nonumber\\
        &\CI_q(m011_{1,1}) = 2\ , \nonumber\\
        &\CI_q(m011_{2,1}) = 1\ , \nonumber\\
        &\CI_q(m011_{1,0}) = 0\ , \nonumber\\
        &\CI_q(m011_{-1,2}) = \infty \times q^0 -q -3q^2 + O(q^3)\ , \nonumber\\
        &\CI_q(m015_{-2,1}) = \infty \times q^0 -2q -4q^2 -4q^3 +O(q^4)\ , \nonumber\\
        &\CI_q(m015_{-1,1}) = 1\ , \nonumber\\
        &\CI_q(m015_{0,1}) = 1\ , \nonumber\\
        &\CI_q(m015_{1,1}) = 1\ , \nonumber\\
        &\CI_q(m015_{2,1}) = \infty \times q^0 -2q -4q^2 -4q^3 + O(q^4)\ , \nonumber\\
        &\CI_q(m015_{1,0}) = 0\ , \nonumber\\
        &\CI_q(m016_{-2,1}) = \infty \times q^0 -2q -4q^2 -4q^3 -6q^4 -4q^5 -8q^6 -4q^7+ O(q^8)\ , \nonumber\\
        &\CI_q(m016_{-1,1}) = 0\ , \nonumber\\
        &\CI_q(m016_{0,1}) = 0\ , \nonumber\\
        &\CI_q(m016_{1,1}) = 1\ , \nonumber\\
        &\CI_q(m016_{2,1}) = \infty \times q^0 -4q -6q^2 -4q^3 -8q^4 -4q^5 -12q^6 -4q^7 + O(q^8)\ , \nonumber\\
        &\CI_q(m016_{1,0}) = 0\ , \nonumber\\
        &\CI_q(m016_{-1,2}) = \infty \times q^0 -2q -4q^2 +O(q^3)\ , \nonumber\\
        &\CI_q(m017_{-2,1}) = \infty \times q^0 -2q -4q^2 -4q^3 -6q^4 -4q^5 -8q^6 -4q^7  +O(q^8)\ , \nonumber\\
        &\CI_q(m017_{-1,1}) = 0\ , \nonumber\\
        &\CI_q(m017_{0,1}) = 0\ , \nonumber\\
        &\CI_q(m017_{1,1}) = 1\ , \nonumber\\
        &\CI_q(m017_{2,1}) = \infty \times q^0 -4q -6q^2 -4q^3 -8q^4 -4q^5 +O(q^6)\ , \nonumber\\
        &\CI_q(m017_{1,0}) = 0\ , \nonumber\\
        &\CI_q(m017_{-1,2}) = \infty \times q^0 -2q -4q^2 +O(q^3)\ , \nonumber\\
        &\CI_q(m019_{-2,1}) = \infty \times q^0 -2q -3q^2 -3q^3 -5q^4 -3q^5 -4q^6+O(q^7)\ , \nonumber\\
        &\CI_q(m019_{-1,1}) = 1\ , \nonumber\\
        &\CI_q(m019_{0,1}) = 0\ , \nonumber\\
        &\CI_q(m019_{1,1}) = 0\ , \nonumber\\
        &\CI_q(m019_{1,0}) = 0\ , \nonumber\\
        &\CI_q(m019_{1,2}) = \infty \times q^0 -2q -4q^2 + O(q^3)\ , \nonumber\\
        &\CI_q(m022_{-2,1}) = \infty \times q^0 -q -3q^2 -3q^3 +O(q^4)\ , \nonumber\\
        &\CI_q(m022_{-1,1}) = 1\ , \nonumber\\
        &\CI_q(m022_{0,1}) = 1\ , \nonumber\\
        &\CI_q(m022_{1,1}) = 0\ , \nonumber\\
        &\CI_q(m022_{2,1}) = 1\ , \nonumber\\
        &\CI_q(m022_{1,0}) = 0\ , \nonumber\\
        &\CI_q(m023_{-2,1}) = 1\ , \nonumber\\
        &\CI_q(m023_{-1,1}) = 1\ , \nonumber\\
        &\CI_q(m023_{0,1}) = 1\ , \nonumber\\
        &\CI_q(m023_{1,1}) = 1\ , \nonumber\\
        &\CI_q(m023_{2,1}) = 1\ , \nonumber\\
        &\CI_q(m023_{3,1}) = \infty \times q^0 -2q^2 -2q^3 +O(q^4)\ , \\
        &\CI_q(m023_{1,0}) = 0\ , \nonumber\\
        &\CI_q(m026_{-2,1}) = 1\ , \nonumber\\
        &\CI_q(m026_{-1,1}) = 1\ , \nonumber\\
        &\CI_q(m026_{0,1}) = 0\ , \nonumber\\
        &\CI_q(m026_{1,1}) = 1\ , \nonumber\\
        &\CI_q(m026_{1,0}) = 0\ , \nonumber\\
        &\CI_q(m026_{1,2}) = \infty \times q^0-q^{1/2}-q-2q^{3/2}-3 q^2 + O(q^{5/2})\ , \nonumber\\
        &\CI_q(m027_{-1,1}) = 1\ , \nonumber\\
        &\CI_q(m027_{0,1}) = 0\ , \nonumber\\
        &\CI_q(m027_{1,1}) = 1\ , \nonumber\\
        &\CI_q(m027_{2,1}) = 1\ , \nonumber\\
        &\CI_q(m027_{1,0}) = 0\ , \nonumber\\
        &\CI_q(m027_{-1,2}) = \infty \times q^0-q-q^{3/2}-3 q^2-q^{5/2} + O(q^{3})\ , \nonumber\\
        &\CI_q(m029_{-1,1}) = 2\ , \nonumber\\
        &\CI_q(m029_{0,1}) = 1\ , \nonumber\\
        &\CI_q(m029_{1,1}) = 1\ , \nonumber\\
        &\CI_q(m029_{2,1}) = 1\ , \nonumber\\
        &\CI_q(m029_{3,1}) = \infty \times q^0 -2q -4q^2 +O(q^3)\ , \nonumber\\
        &\CI_q(m029_{1,0}) = 0\ , \nonumber\\
        &\CI_q(m030_{-2,1}) = \infty \times q^0 -2q -4q^2 -6q^3 +O(q^4)\ , \nonumber\\
        &\CI_q(m030_{-1,1}) = 1\ , \nonumber\\
        &\CI_q(m030_{0,1}) = 1\ , \nonumber\\
        &\CI_q(m030_{1,1}) = 1\ , \nonumber\\
        &\CI_q(m030_{2,1}) = 2\ , \nonumber\\
        &\CI_q(m030_{1,0}) = 0\ , \nonumber\\
        &\CI_q(m036_{-2,1}) = \infty \times q^0\ , \nonumber\\
        &\CI_q(m036_{-1,1}) = 0\ , \nonumber\\
        &\CI_q(m036_{0,1}) = 1\ , \nonumber\\
        &\CI_q(m036_{1,1}) = 1\ , \nonumber\\
        &\CI_q(m036_{2,1}) = \infty \times q^0 -4q -8q^2 -6q^3 +O(q^4)\ , \nonumber\\
        &\CI_q(m036_{1,0}) = 0\ , \nonumber\\
        &\CI_q(m043_{-1,1}) = 0\ , \nonumber\\
        &\CI_q(m043_{0,1}) = 0\ , \nonumber\\
        &\CI_q(m043_{1,1}) = 1\ , \nonumber\\
        &\CI_q(m043_{2,1}) = \infty \times q^0 -2q -3q^2 +O(q^3)\ , \nonumber\\
        &\CI_q(m043_{1,0}) = 0\ , \nonumber\\
        &\CI_q(m043_{-1,2}) = \infty \times q^0-q^{1/2}-q-2q^{3/2}-3 q^2 + O(q^{5/2})\ , \nonumber\\
        &\CI_q(m130_{-2,1}) = \infty \times q^0 -4q -6q^2 +O(q^3)\ , \nonumber\\
        &\CI_q(m130_{-1,1}) = \infty \times q^0 \ , \nonumber\\
        &\CI_q(m130_{0,1}) = 0\ , \nonumber\\
        &\CI_q(m130_{1,1}) = 1\ , \nonumber\\
        &\CI_q(m130_{1,0}) = 0\ , \nonumber\\
        &\CI_q(m130_{1,2}) = \infty \times q^0-2q-6 q^2 + O(q^3)\ ,  \nonumber\\
        &\CI_q(m135_{-1,1}) = 1\ , \nonumber\\
        &\CI_q(m135_{0,1}) = \cdots+q^{-1}-q^{-1/2}+1 -q^{1/2}+q+O(q^{3/2})\ ,  \nonumber\\
        &\CI_q(m135_{1,1}) = \infty \times q^0\ , \nonumber\\
        &\CI_q(m135_{1,0}) = \cdots+q^{-1}-q^{-1/2}+1 -q^{1/2}+q+O(q^{3/2})\ ,  \nonumber\\
        &\CI_q(m136_{-2,1}) = \infty \times q^0 -2q -4q^{3/2} -4q^2 +O(q^{5/2})\ , \nonumber\\
        &\CI_q(m136_{-1,1}) = 1\ , \nonumber\\
        &\CI_q(m136_{0,1}) = 1\ , \nonumber\\
        &\CI_q(m136_{1,1}) = 1\ , \nonumber\\
        &\CI_q(m136_{2,1}) = \infty \times q^0 -2q -4q^{3/2} -4q^2 +O(q^{5/2})\ , \nonumber\\
        &\CI_q(m136_{1,0}) = \cdots+2q^{-1} -2q^{-1/2} +2 -2q^{1/2} +O(q)\ , \nonumber\\
        &\CI_q(m160_{-2,1}) = 1\ , \nonumber\\
        &\CI_q(m160_{-1,1}) = 2\ , \nonumber\\
        &\CI_q(m160_{0,1}) = \cdots +q^{-3}+q^{-2}+q^{-1}+1 + O(q)\ , \nonumber\\
        &\CI_q(m160_{1,0}) = 1\ , \nonumber\\
        &\CI_q(m207_{-1,1}) = 1\ , \nonumber\\
        &\CI_q(m207_{0,1}) = \cdots +q^{-3}+q^{-2}+q^{-1}+1 + O(q)\ ,   \nonumber\\
        &\CI_q(m207_{1,0}) = \cdots +q^{-3}+q^{-2}+q^{-1}+1 + O(q)\ ,  \nonumber\\
        &\CI_q(m247_{-1,1}) = \infty \times q^0 +2q +O(q^2)\ , \nonumber\\
        &\CI_q(m247_{0,1}) = 0\ , \nonumber\\
        &\CI_q(m247_{1,0}) = \cdots + q^{-2}+ q^{-1}+1+O(q)\ , \nonumber\\
        &\CI_q(m249_{-1,1}) = 1\ , \nonumber\\
        &\CI_q(m249_{0,1}) = \infty \times q^0 + 2 q^2+O(q^3)\ , \nonumber\\
        &\CI_q(m249_{1,0}) = 0\ , \nonumber\\
        &\CI_q(m294_{0,1}) = 2 \ , \nonumber\\
        &\CI_q(m294_{1,1}) = 1\ , \nonumber\\
        &\CI_q(m294_{1,0}) = \cdots + q^{-2}+ q^{-1}+1+O(q)\ , \nonumber\\
        &\CI_q(m294_{1,2}) = \infty\times q^0 - 4q +O(q^2)\ , \nonumber\\
        &\CI_q(m410_{0,1}) = 1\ , \nonumber\\
        &\CI_q(m410_{1,1}) = \infty \times q^0 -q^{1/2}-2q^{3/2}-2q^2-2q^{5/2}-2q^3+O(q^{7/2})\ , \nonumber\\
        &\CI_q(m410_{1,0}) = \cdots  -\infty \times q^{-1/2}+ \infty \times q^0 - \infty \times q^{1/2} +\cdots\ , \nonumber\\
        &\CI_q(v2050_{-2,1}) = \infty \times q^0 - \infty \times q^1 + \cdots\ , \nonumber\\
        &\CI_q(v2050_{1,0}) = 0\ , \nonumber\\ 
        &\CI_q(v2051_{-2,1}) = \infty \times q^0 - \infty \times q^1 + \cdots\ , \nonumber\\
        &\CI_q(v2051_{1,1}) = \infty \times q^0 -2q-2q^2+ O(q^3)\ , \nonumber\\
        &\CI_q(v2051_{1,0}) = 1\ , \nonumber\\
        &\CI_q(v2099_{-2,1}) = \infty \times q^0 - 2 q^{1/2}- \infty \times q + \cdots\ , \nonumber\\
        &\CI_q(v2099_{0,1}) =   \infty \times q^0 - 2q^{1/2} - 6q^{3/2}+O(q^2)\ , \nonumber\\
        &\CI_q(v2099_{1,1}) =1\ ,  \nonumber\\
        &\CI_q(v2099_{1,0}) =1\ , \nonumber\\
        &\CI_q(v2274_{-2,1}) =\infty \times q^0 - \infty \times q^1 +\cdots \ , \nonumber\\
        &\CI_q(v2274_{2,1}) =\infty \times q^0 - \infty \times q^1 +\cdots \ , \nonumber\\
        &\CI_q(v2274_{1,0}) =2\ , \nonumber\\
        &\CI_q(v2334_{-2,1}) =\infty \times q^0 - \infty \times q^1 +\cdots \ , \nonumber\\
        &\CI_q(v2334_{0,1}) =1\ , \nonumber\\
        &\CI_q(v2334_{1,1}) =1\ , \nonumber\\
        &\CI_q(v2334_{1,0}) =1\ . \nonumber
\end{align}
\endgroup
Comparing Table \ref{Appen-tab} and the above 3D indices, one can indeed see that our proposals in Fig. \ref{fig: census of non-hyperbolics} (along with  \eqref{3D index pattern}) and \eqref{proposal for IR phases} for the 3D index and IR phase are consistent for all the above closed non-hyperbolic 3-manifolds. Among the above atoroidal Seifert fibered spaces, followings have non-trivial $\cap_{\rho} \textrm{Inv}(\rho)  = \mathbb{Z}_2$ and thus their 3D indices become 2 instead of 1 as discussed around \eqref{Inv-rho}.
\begin{align}
\begin{split}
&m006_{1,1} = S^2 ((2,1),(2,1),(3,2)) \;, \quad m007_{1,1} = S^2 ((2,1),(2,1),(5,-2))\;, 
\\
& m010_{0,1} = S^2 ((2,1),(2,1),(4,-1))\;, \quad m011_{1,1} =S^2 ((2,1),(2,1),(3,-2)) \;, 
\\
& m029_{-1,1} = S^2 ((2,1),(2,1),(7,-2)) \;, \quad m030_{2,1} = S^2 ((2,1),(2,1),(5,2))  \;, 
\\
& m160_{-1,1} =S^2 ((2,1),(2,1),(7,-4))  \;, \quad m294_{0,1} = S^2 ((2,1),(2,1),(10,-7))\;, 
\\
& v2274_{1,0} =S^2 ((2,1),(2,1),(3,-2)) \;.
\end{split}
\end{align}

\bibliographystyle{ytphys}
\bibliography{ref}

\providecommand{\href}[2]{#2}\begingroup\raggedright\begin{thebibliography}{10}

\bibitem{Gaiotto:2009we}
D.~Gaiotto, ``{N=2 dualities},''
  \href{http://dx.doi.org/10.1007/JHEP08(2012)034}{{\em JHEP} {\bfseries 08}
  (2012) 034}, \href{http://arxiv.org/abs/0904.2715}{{\ttfamily arXiv:0904.2715
  [hep-th]}}.

\bibitem{Gaiotto:2009hg}
D.~Gaiotto, G.~W. Moore, and A.~Neitzke, ``{Wall-crossing, Hitchin Systems, and
  the WKB Approximation},'' \href{http://arxiv.org/abs/0907.3987}{{\ttfamily
  arXiv:0907.3987 [hep-th]}}.

\bibitem{Dimofte:2011ju}
T.~Dimofte, D.~Gaiotto, and S.~Gukov, ``{Gauge Theories Labelled by
  Three-Manifolds},'' \href{http://dx.doi.org/10.1007/s00220-013-1863-2}{{\em
  Commun. Math. Phys.} {\bfseries 325} (2014) 367--419},
\href{http://arxiv.org/abs/1108.4389}{{\ttfamily arXiv:1108.4389 [hep-th]}}.

\bibitem{Dimofte:2011py}
T.~Dimofte, D.~Gaiotto, and S.~Gukov, ``{3-Manifolds and 3d Indices},''
  \href{http://dx.doi.org/10.4310/ATMP.2013.v17.n5.a3}{{\em Adv. Theor. Math.
  Phys.} {\bfseries 17} no.~5, (2013) 975--1076},
\href{http://arxiv.org/abs/1112.5179}{{\ttfamily arXiv:1112.5179 [hep-th]}}.

\bibitem{Cho:2020ljj}
G.~Y. Cho, D.~Gang, and H.-C. Kim, ``{M-theoretic Genesis of Topological
  Phases},'' \href{http://dx.doi.org/10.1007/JHEP11(2020)115}{{\em JHEP}
  {\bfseries 11} (2020) 115}, \href{http://arxiv.org/abs/2007.01532}{{\ttfamily
  arXiv:2007.01532 [hep-th]}}.

\bibitem{Gang:2021hrd}
D.~Gang, S.~Kim, K.~Lee, M.~Shim, and M.~Yamazaki, ``{Non-unitary TQFTs from 3D
  $ \mathcal{N} $ = 4 rank 0 SCFTs},''
  \href{http://dx.doi.org/10.1007/JHEP08(2021)158}{{\em JHEP} {\bfseries 08}
  (2021) 158}, \href{http://arxiv.org/abs/2103.09283}{{\ttfamily
  arXiv:2103.09283 [hep-th]}}.

\bibitem{2018arXiv181211940D}
N.~M. {Dunfield}, ``{A census of exceptional Dehn fillings},'' {\em arXiv
  e-prints} (Dec., 2018) arXiv:1812.11940,
  \href{http://arxiv.org/abs/1812.11940}{{\ttfamily arXiv:1812.11940
  [math.GT]}}.

\bibitem{SnapPy}
M.~Culler, N.~M. Dunfield, M.~Goerner, and J.~R. Weeks, ``Snap{P}y, a computer
  program for studying the geometry and topology of $3$-manifolds.'' Available
  at \url{http://snappy.computop.org}.

\bibitem{thurston2014three}
W.~P. Thurston, ``Three-dimensional geometry and topology, volume 1,'' in {\em
  Three-Dimensional Geometry and Topology, Volume 1}.
\newblock Princeton university press, 2014.

\bibitem{jaco1979seifert}
W.~Jaco and P.~B. Shalen, ``Seifert fibered spaces in 3-manifolds,'' in {\em
  Geometric topology}, pp.~91--99.
\newblock Elsevier, 1979.

\bibitem{johannson2006homotopy}
K.~Johannson, {\em Homotopy equivalences of 3-manifolds with boundaries},
  vol.~761.
\newblock Springer, 2006.

\bibitem{Gang:2018wek}
D.~Gang and K.~Yonekura, ``{Symmetry enhancement and closing of knots in 3d/3d
  correspondence},''
\href{http://arxiv.org/abs/1803.04009}{{\ttfamily arXiv:1803.04009 [hep-th]}}.

\bibitem{Gang:2018gyt}
D.~Gang, ``{Quantum Approach to Dehn Surgery Problem},''
  \href{http://arxiv.org/abs/1803.11143}{{\ttfamily arXiv:1803.11143
  [math.GT]}}.

\bibitem{Cui:2021lyi}
S.~X. Cui, Y.~Qiu, and Z.~Wang, ``{From Three Dimensional Manifolds to Modular
  Tensor Categories},'' \href{http://arxiv.org/abs/2101.01674}{{\ttfamily
  arXiv:2101.01674 [math.QA]}}.

\bibitem{Cui:2021yes}
S.~X. Cui, P.~Gustafson, Y.~Qiu, and Q.~Zhang, ``{From Torus Bundles to
  Particle-Hole Equivariantization},''
  \href{http://arxiv.org/abs/2106.01959}{{\ttfamily arXiv:2106.01959
  [math.QA]}}.

\bibitem{Garoufalidis:2013axa}
S.~Garoufalidis, C.~D. Hodgson, J.~H. Rubinstein, and H.~Segerman,
  ``{1-efficient triangulations and the index of a cusped hyperbolic
  3-manifold},'' \href{http://arxiv.org/abs/1303.5278}{{\ttfamily
  arXiv:1303.5278 [math.GT]}}.

\bibitem{Garoufalidis:2016ckn}
S.~Garoufalidis, C.~Hodgson, N.~Hoffman, and H.~Rubinstein, ``{The 3D-index and
  normal surfaces},'' \href{http://arxiv.org/abs/1604.02688}{{\ttfamily
  arXiv:1604.02688 [math.GT]}}.

\bibitem{to-apper}
D.~Gang, H.~Kang, G.~Kim, and H.-C. Kim, ``To appear,''.

\bibitem{Dimofte:2012qj}
T.~D. Dimofte and S.~Garoufalidis, ``{The Quantum content of the gluing
  equations},'' {\em Geom. Topol.} {\bfseries 17} (2013) 1253--1316,
\href{http://arxiv.org/abs/1202.6268}{{\ttfamily arXiv:1202.6268 [math.GT]}}.

\bibitem{Gang:2017cwq}
D.~Gang, M.~Romo, and M.~Yamazaki, ``{All-Order Volume Conjecture for Closed
  3-Manifolds from Complex Chern-Simons Theory},''
\href{http://arxiv.org/abs/1704.00918}{{\ttfamily arXiv:1704.00918 [hep-th]}}.

\bibitem{Gadde:2013wq}
A.~Gadde, S.~Gukov, and P.~Putrov, ``{Walls, Lines, and Spectral Dualities in
  3d Gauge Theories},'' \href{http://dx.doi.org/10.1007/JHEP05(2014)047}{{\em
  JHEP} {\bfseries 05} (2014) 047},
  \href{http://arxiv.org/abs/1302.0015}{{\ttfamily arXiv:1302.0015 [hep-th]}}.

\bibitem{Beem:2012mb}
C.~Beem, T.~Dimofte, and S.~Pasquetti, ``{Holomorphic Blocks in Three
  Dimensions},'' \href{http://dx.doi.org/10.1007/JHEP12(2014)177}{{\em JHEP}
  {\bfseries 12} (2014) 177}, \href{http://arxiv.org/abs/1211.1986}{{\ttfamily
  arXiv:1211.1986 [hep-th]}}.

\bibitem{Chung:2014qpa}
H.-J. Chung, T.~Dimofte, S.~Gukov, and P.~Su\l{}kowski, ``{3d-3d Correspondence
  Revisited},'' \href{http://dx.doi.org/10.1007/JHEP04(2016)140}{{\em JHEP}
  {\bfseries 04} (2016) 140}, \href{http://arxiv.org/abs/1405.3663}{{\ttfamily
  arXiv:1405.3663 [hep-th]}}.

\bibitem{Pei:2015jsa}
D.~Pei and K.~Ye, ``{A 3d-3d appetizer},''
  \href{http://dx.doi.org/10.1007/JHEP11(2016)008}{{\em JHEP} {\bfseries 11}
  (2016) 008}, \href{http://arxiv.org/abs/1503.04809}{{\ttfamily
  arXiv:1503.04809 [hep-th]}}.

\bibitem{Gukov:2017kmk}
S.~Gukov, D.~Pei, P.~Putrov, and C.~Vafa, ``{BPS spectra and 3-manifold
  invariants},'' \href{http://dx.doi.org/10.1142/S0218216520400039}{{\em J.
  Knot Theor. Ramifications} {\bfseries 29} no.~02, (2020) 2040003},
  \href{http://arxiv.org/abs/1701.06567}{{\ttfamily arXiv:1701.06567
  [hep-th]}}.

\bibitem{Eckhard:2019jgg}
J.~Eckhard, H.~Kim, S.~Schafer-Nameki, and B.~Willett, ``{Higher-Form
  Symmetries, Bethe Vacua, and the 3d-3d Correspondence},''
  \href{http://dx.doi.org/10.1007/JHEP01(2020)101}{{\em JHEP} {\bfseries 01}
  (2020) 101}, \href{http://arxiv.org/abs/1910.14086}{{\ttfamily
  arXiv:1910.14086 [hep-th]}}.

\bibitem{Dimofte:2010tz}
T.~Dimofte, S.~Gukov, and L.~Hollands, ``{Vortex Counting and Lagrangian
  3-manifolds},'' \href{http://dx.doi.org/10.1007/s11005-011-0531-8}{{\em Lett.
  Math. Phys.} {\bfseries 98} (2011) 225--287},
  \href{http://arxiv.org/abs/1006.0977}{{\ttfamily arXiv:1006.0977 [hep-th]}}.

\bibitem{Kim:2009wb}
S.~Kim, ``{The Complete superconformal index for N=6 Chern-Simons theory},''
  \href{http://dx.doi.org/10.1016/j.nuclphysb.2012.07.015,
  10.1016/j.nuclphysb.2009.06.025}{{\em Nucl. Phys.} {\bfseries B821} (2009)
  241--284}, \href{http://arxiv.org/abs/0903.4172}{{\ttfamily arXiv:0903.4172
  [hep-th]}}.
[Erratum: Nucl. Phys.B864,884(2012)].

\bibitem{Imamura:2011su}
Y.~Imamura and S.~Yokoyama, ``{Index for three dimensional superconformal field
  theories with general R-charge assignments},''
  \href{http://dx.doi.org/10.1007/JHEP04(2011)007}{{\em JHEP} {\bfseries 04}
  (2011) 007}, \href{http://arxiv.org/abs/1101.0557}{{\ttfamily arXiv:1101.0557
  [hep-th]}}.

\bibitem{Jafferis:2010un}
D.~L. Jafferis, ``{The Exact Superconformal R-Symmetry Extremizes Z},''
  \href{http://dx.doi.org/10.1007/JHEP05(2012)159}{{\em JHEP} {\bfseries 05}
  (2012) 159},
\href{http://arxiv.org/abs/1012.3210}{{\ttfamily arXiv:1012.3210 [hep-th]}}.

\bibitem{Razamat:2014pta}
S.~S. Razamat and B.~Willett, ``{Down the rabbit hole with theories of class $
  \mathcal{S} $},'' \href{http://dx.doi.org/10.1007/JHEP10(2014)099}{{\em JHEP}
  {\bfseries 10} (2014) 099}, \href{http://arxiv.org/abs/1403.6107}{{\ttfamily
  arXiv:1403.6107 [hep-th]}}.

\bibitem{Closset:2017zgf}
C.~Closset, H.~Kim, and B.~Willett, ``{Supersymmetric partition functions and
  the three-dimensional A-twist},''
  \href{http://dx.doi.org/10.1007/JHEP03(2017)074}{{\em JHEP} {\bfseries 03}
  (2017) 074},
\href{http://arxiv.org/abs/1701.03171}{{\ttfamily arXiv:1701.03171 [hep-th]}}.

\bibitem{Closset:2018ghr}
C.~Closset, H.~Kim, and B.~Willett, ``{Seifert fibering operators in 3d
  $\mathcal{N}=2$ theories},''
  \href{http://dx.doi.org/10.1007/JHEP11(2018)004}{{\em JHEP} {\bfseries 11}
  (2018) 004}, \href{http://arxiv.org/abs/1807.02328}{{\ttfamily
  arXiv:1807.02328 [hep-th]}}.

\bibitem{Gukov:2015sna}
S.~Gukov and D.~Pei, ``{Equivariant Verlinde formula from fivebranes and
  vortices},'' \href{http://dx.doi.org/10.1007/s00220-017-2931-9}{{\em Commun.
  Math. Phys.} {\bfseries 355} no.~1, (2017) 1--50},
  \href{http://arxiv.org/abs/1501.01310}{{\ttfamily arXiv:1501.01310
  [hep-th]}}.

\bibitem{Benini:2015noa}
F.~Benini and A.~Zaffaroni, ``{A topologically twisted index for
  three-dimensional supersymmetric theories},''
  \href{http://dx.doi.org/10.1007/JHEP07(2015)127}{{\em JHEP} {\bfseries 07}
  (2015) 127}, \href{http://arxiv.org/abs/1504.03698}{{\ttfamily
  arXiv:1504.03698 [hep-th]}}.

\bibitem{Benini:2016hjo}
F.~Benini and A.~Zaffaroni, ``{Supersymmetric partition functions on Riemann
  surfaces},'' {\em Proc. Symp. Pure Math.} {\bfseries 96} (2017) 13--46,
  \href{http://arxiv.org/abs/1605.06120}{{\ttfamily arXiv:1605.06120
  [hep-th]}}.

\bibitem{Closset:2016arn}
C.~Closset and H.~Kim, ``{Comments on twisted indices in 3d supersymmetric
  gauge theories},'' \href{http://dx.doi.org/10.1007/JHEP08(2016)059}{{\em
  JHEP} {\bfseries 08} (2016) 059},
  \href{http://arxiv.org/abs/1605.06531}{{\ttfamily arXiv:1605.06531
  [hep-th]}}.

\bibitem{Kim:2010mr}
H.-C. Kim and S.~Kim, ``{Supersymmetric vacua of mass-deformed M2-brane
  theory},'' \href{http://dx.doi.org/10.1016/j.nuclphysb.2010.06.002}{{\em
  Nucl. Phys. B} {\bfseries 839} (2010) 96--111},
  \href{http://arxiv.org/abs/1001.3153}{{\ttfamily arXiv:1001.3153 [hep-th]}}.

\bibitem{Intriligator:2013lca}
K.~Intriligator and N.~Seiberg, ``{Aspects of 3d N=2 Chern-Simons-Matter
  Theories},'' \href{http://dx.doi.org/10.1007/JHEP07(2013)079}{{\em JHEP}
  {\bfseries 07} (2013) 079}, \href{http://arxiv.org/abs/1305.1633}{{\ttfamily
  arXiv:1305.1633 [hep-th]}}.

\bibitem{Hama:2011ea}
N.~Hama, K.~Hosomichi, and S.~Lee, ``{SUSY Gauge Theories on Squashed
  Three-Spheres},'' \href{http://dx.doi.org/10.1007/JHEP05(2011)014}{{\em JHEP}
  {\bfseries 05} (2011) 014}, \href{http://arxiv.org/abs/1102.4716}{{\ttfamily
  arXiv:1102.4716 [hep-th]}}.

\bibitem{faddeev1994quantum}
L.~D. Faddeev and R.~M. Kashaev, ``Quantum dilogarithm,'' {\em Modern Physics
  Letters A} {\bfseries 9} no.~05, (1994) 427--434.

\bibitem{Gang:2019jut}
D.~Gang and M.~Yamazaki, ``{Expanding 3d $ \mathcal{N} $ = 2 theories around
  the round sphere},'' \href{http://dx.doi.org/10.1007/JHEP02(2020)102}{{\em
  JHEP} {\bfseries 02} (2020) 102},
  \href{http://arxiv.org/abs/1912.09617}{{\ttfamily arXiv:1912.09617
  [hep-th]}}.

\bibitem{witten1989quantum}
E.~Witten, ``Quantum field theory and the jones polynomial,'' {\em
  Communications in Mathematical Physics} {\bfseries 121} no.~3, (1989)
  351--399.

\bibitem{Cordova:2016emh}
C.~Cordova, T.~T. Dumitrescu, and K.~Intriligator, ``{Multiplets of
  Superconformal Symmetry in Diverse Dimensions},''
  \href{http://dx.doi.org/10.1007/JHEP03(2019)163}{{\em JHEP} {\bfseries 03}
  (2019) 163}, \href{http://arxiv.org/abs/1612.00809}{{\ttfamily
  arXiv:1612.00809 [hep-th]}}.

\bibitem{Benini:2018umh}
F.~Benini and S.~Benvenuti, ``{$ \mathcal{N} $ = 1 dualities in 2+1
  dimensions},'' \href{http://dx.doi.org/10.1007/JHEP11(2018)197}{{\em JHEP}
  {\bfseries 11} (2018) 197}, \href{http://arxiv.org/abs/1803.01784}{{\ttfamily
  arXiv:1803.01784 [hep-th]}}.

\bibitem{Gaiotto:2018yjh}
D.~Gaiotto, Z.~Komargodski, and J.~Wu, ``{Curious Aspects of Three-Dimensional
  ${\cal N}=1$ SCFTs},'' \href{http://dx.doi.org/10.1007/JHEP08(2018)004}{{\em
  JHEP} {\bfseries 08} (2018) 004},
  \href{http://arxiv.org/abs/1804.02018}{{\ttfamily arXiv:1804.02018
  [hep-th]}}.

\bibitem{Jafferis:2011ns}
D.~Jafferis and X.~Yin, ``{A Duality Appetizer},''
  \href{http://arxiv.org/abs/1103.5700}{{\ttfamily arXiv:1103.5700 [hep-th]}}.

\bibitem{Gang:2018huc}
D.~Gang and M.~Yamazaki, ``{Three-dimensional gauge theories with supersymmetry
  enhancement},'' \href{http://dx.doi.org/10.1103/PhysRevD.98.121701}{{\em
  Phys. Rev. D} {\bfseries 98} no.~12, (2018) 121701},
  \href{http://arxiv.org/abs/1806.07714}{{\ttfamily arXiv:1806.07714
  [hep-th]}}.

\end{thebibliography}\endgroup

\end{document}